\documentclass[11pt]{article}

\usepackage{tabularx}
\usepackage{booktabs}

\usepackage{diagbox}
\usepackage{amssymb}

\usepackage{amsfonts}
\usepackage{amsthm}
\usepackage{amsmath}
\usepackage{bbm}
\usepackage{bm}
\usepackage[inline]{enumitem}
\usepackage{amsbsy}
\usepackage[utf8]{inputenc}
\usepackage[margin=1in]{geometry}
\usepackage{setspace}
\usepackage{color,xcolor}
\usepackage{float}
\usepackage[subrefformat=parens,labelformat=parens,justification = centering]{subfig}
\usepackage[justification = justified]{caption}
\usepackage{tikz}
\usetikzlibrary{positioning}
\usetikzlibrary{arrows.meta, matrix}
\usetikzlibrary{shapes,arrows,chains}
\usetikzlibrary{arrows,positioning,calc,topaths}

\usepackage{titlesec}
\titleformat*{\section}{\large\bfseries}
\titleformat*{\subsection}{\normalsize\bfseries}
\titleformat{\subsubsection}[runin]
  {\normalfont\normalsize\bfseries}{\thesubsubsection}{1em}{}
\titleformat*{\paragraph}{\normalsize\bfseries}
\titleformat*{\subparagraph}{\normalsize\bfseries}

\usepackage{multirow}
\usepackage{makecell}
\usepackage{natbib}
\usepackage[T1]{fontenc}

\usepackage{chngcntr}
\usepackage{apptools}
\usepackage{algorithm}
\usepackage{algpseudocode}

\newtheorem{theorem}{Theorem}[section]
\newtheorem{proposition}{Proposition}[section]

\newtheorem{lemma}[theorem]{Lemma}

\theoremstyle{remark}
\newtheorem{case}{Case}
\newcommand{\dt}{\frac{d}{dt}}
\newcommand{\dd}{\,\text{d}}

\newcommand{\ig}{\textrm{IG}}
\newcommand{\ga}{\textrm{Gamma}}
\newcommand{\sss}{\bm{s}}
\newcommand{\ttheta}{\bm{\theta}}

\usepackage[toc,page,header]{appendix}
\usepackage{minitoc}

\usepackage[colorlinks=true,allcolors=blue]{hyperref}
\usepackage[capitalize,noabbrev]{cleveref}
\crefformat{equation}{#2\textup{(#1)}#3}
\noptcrule 

\title{
Bayesian inference for aggregated Hawkes processes
}
\date{}
\author{Lingxiao Zhou and Georgia Papadogeorgou 
\thanks{This material is based upon work supported by the National Science Foundation under Grant No. 2124124.} \\[-5pt]
{\normalsize Department of Statistics, University of Florida} \\[-20pt]}

\usepackage{xargs, xstring}
\newcommandx{\history}[1][1=t]{\mathcal{H}_{#1}}

\allowdisplaybreaks

\begin{document}
\maketitle

\begin{abstract}
The Hawkes process, a self-exciting point process, has a wide range of applications in modeling earthquakes, social networks and stock markets. The established estimation process requires that researchers have access to the exact time stamps and spatial information. However, available data are often rounded or aggregated. We develop a Bayesian estimation procedure for the parameters of a Hawkes process based on aggregated data. Our approach is developed for temporal, spatio-temporal, and mutually exciting Hawkes processes where data are available over discrete time periods and regions. We show theoretically that the parameters of the Hawkes process are identifiable from aggregated data under general specifications. We demonstrate the method on simulated data under various model specifications in the presence of one or more interacting processes, and under varying coarseness of data aggregation. Finally, we examine the internal and cross-excitation effects of airstrikes and insurgent violence events from February 2007 to June 2008, with some data aggregated by day.
\end{abstract}

Understanding the dynamics generating point pattern data is the goal of point pattern modeling. A dynamic often of interest is self-excitement, where the occurrence of an event triggers subsequent events. A common model for studying excitement in point pattern data is the Hawkes process \citep{hawkes1971spectra}. Initially applied in seismology \citep{ogata1998space}, Hawkes processes have been used to study terrorist attacks \citep{porter2012self,lewis2012self}, financial markets \citep{lapham2014hawkes}, and epidemiology \citep{chiang2022hawkes}, among others. Their wide applicability has spurred extensive research on precise and efficient parameter estimation, within both frequentist \citep{veen2008estimation} and Bayesian \citep{Rasmussen2013bayesian} frameworks.
Beyond an event's timing, researchers often have access to its location. Including spatial information in point process modeling can offer further insights into the process' self-excitement dynamics.
Besides studying {\it self}-excitement, the Hawkes process can model cross-excitement among multiple point processes, each representing different phenomena. Mutually-exciting Hawkes processes provide a framework to study interaction dynamics where an event of one type can trigger subsequent events of the same or another type.


Standard estimation techniques for the Hawkes process require that the exact time and location of the events is known. However, real data are often only available at aggregated levels across time and space. For example, we might only know the date that an event occurred without knowing its exact time, and some published earthquake data are aggregated over towns. Since data of this form are available within bins of time and space they could be referred to as {\it binned} or {\it discretized}, though we maintain the term {\it aggregated} for the remainder of this manuscript.
When data are aggregated, standard estimation and inference techniques for the underlying continuous time and space Hawkes point process are not applicable.
In these cases, it is common practice to model the discretized point pattern dynamics using, for example, autoregressive models in time or space \citep{aldor2016spatio,darolles2019bivariate}.
These models could in principle perform adequately for forecasting event counts, but they can only do so over intervals that are multiples of the data's aggregation size. More importantly, since they ignore the possible dependencies among events within the discrete bins, they are not useful for learning the process dynamics at the scale at which they naturally occur, which is often of explicit interest. 
Even though there exists some work on inferring the underlying Hawkes process model parameters from discrete-time models based on temporal, aggregated point pattern data \citep{kirchner2016hawkes, kirchner2017estimation}, the equivalence is only established for autoregressive models with infinite lags (which are impossible to fit), and it is not clear whether the results are sensitive to the scale of data aggregation, or how the method can be extended to incorporate space, other marks, or multiple processes.
\cite{shlomovich2022parameter} proposed a 
Monte Carlo Expectation-Maximization (MC-EM) algorithm to estimate the parameters of a Hawkes process based on aggregated temporal data, and \cite{Shlomovich2022parameterb} extended it to multivariate Hawkes processes. Their approach, which is only developed for temporal point patterns, is based on an expectation-maximization algorithm which cannot be used directly for performing inference over model parameters.

In this manuscript, we propose an estimation technique for modeling point pattern data using Hawkes processes when the available data are aggregated across time, space, or both. We review the Hawkes process in \cref{sec:methods}. We turn our attention to aggregated point pattern data in \cref{sec:aggregated}.
We show theoretically that the parameters of a Hawkes process are identifiable from aggregated data under general model conditions, and provide examples for when these conditions hold.
We introduce our approach, which is positioned within the Bayesian paradigm, and it is based on considering the real (continuous) time of event occurrence as a latent variable. We propose a Markov Chain Monte Carlo (MCMC) scheme for learning the parameters of the Hawkes process which involves straightforward imputation of these latent variables.
Our approach improves on the existing methodology in a number of ways. First, by placing our approach within the Bayesian framework, our estimation strategy leads to straightforward inference procedures that propagate all sources of uncertainty.
Second, it allows for aggregated point pattern data with spatial information, which can itself be available at coarse spatial regions. Lastly, it allows for modeling multiple, mutually-exciting Hawkes processes, each of which can be aggregated across time, space, or both, with the same or different granularity.
We evaluate the performance of the proposed approach in terms of estimation and inference over simulated data sets under different coarseness of aggregation in time and space (\cref{sec:simulations}). We use the proposed method to study the dynamics of airstrikes and insurgent violence in Iraq (\cref{sec:application}). We conclude with a discussion and possible extensions in \cref{sec: discussion}.

\section{The Hawkes process}
\label{sec:methods}

We review the temporal Hawkes process in \cref{subsec:continuous_hp}, the spatio-temporal Hawkes process in \cref{subsec:continuous_hp_st}, and the multivariate Hawkes process in \cref{subsec:continuous_hp_multi}.

\subsection{The temporal Hawkes process}
\label{subsec:continuous_hp}

Hawkes processes are a class of point processes which exhibit a {\it self-exciting behavior}, meaning that the occurrence of an event can lead to the occurrence of additional, subsequent events. Consider a temporal point process $\bm X = \{t_1, t_2, \dots\}$ where $t_i$ can be interpreted as the time of occurrence of the $i^{th}$ event, and $0\leq t_i<t_{i+1}$ for all $i\in\mathbb{Z}^+$. We use $\history$ to denote the process' history up to time $t$, i.e. $\history = \{ t^*\in\bm X: t^*<t \}.$ Consider a right-continuous counting measure $N(A)$ defined as the number of event times in $A.$ Then a  point process can be specified by its conditional intensity, defined as the limit of the expected number of points in an infinitesimal time window following $t$ given the realized history \citep{daley2003introduction},
\begin{equation}\label{eq: intensity_temporal}
\lambda^*(t) = \lim_{\Delta t \downarrow 0}
\frac{\textrm{E}[\ \textrm{N}\{(t,t+\Delta t)\} \mid \history]}{\Delta t}.
\end{equation}
A temporal point process is a Hawkes process if its conditional intensity is of the form
\begin{equation}\label{eq: intensity}
  \lambda^*(t) = \mu(t)+ \sum_{i:t_i<t}\alpha \ g(t-t_i).
\end{equation}
The superscript $^*$ denotes that the intensity is defined conditional on the history. The function $\mu(t)$ is a non-negative function which is often assumed to be constant, and $\alpha$ is a positive value with $\alpha<1$ to ensure stationarity. In turn, $g(t)$ is a density function on $[0,\infty)$, and common choices are the exponential and Lomax densities (see \cref{tab:specification}), the latter of which specifies the Epidemic Type Aftershock Sequences (ETAS) model in seismology \citep{ogata1998space}.

An alternative formalization of the Hawkes process views the Hawkes process as a Poisson cluster process \citep{hawkes1974cluster}. From this point of view, an event is either an immigrant that occurred in the system separately from previous events, or an offspring of a previous event which occurred due to the process' self-excitement. The set of these relations is called the process' branching structure. For $i\in\mathbb{Z}^+$, we  denote event $i$'s (latent) branching information as $Y_i$, where $Y_i = 0$ if $t_i$ is an immigrant, and $Y_i = j$ if $t_i$ is an offspring of event $t_j$. Let $\bm x = \{t_1, t_2, \dots,t_n\}$ be a realization of a Hawkes process on the temporal window $[0,T)$ with $n\in\mathbb{Z}^+$ number of events. 
Let $\bm Y = \{Y_1, Y_2, \dots, Y_n\}$ be the branching structure. Based on the branching structure, we can partition the set of event indices $\{1, 2, \dots, n\}$ into $I$, $O_1,O_2,\dots,O_n$, representing the immigrant events, and the offspring of event $1, 2, \dots, n$, respectively. Therefore, $I = \{i: Y_i = 0, 1\leq i\leq n\}$ and $O_j = \{i:Y_i = j, 1\leq i\leq n\}$ for $j\in\mathbb{Z}^+$. Moreover, we use $\bm{x}_I = \{t_i : i\in I\}$ for the set of immigrant events, and $\bm{x}_{O_j} = \{t_i : i\in O_j\}$ for the set of offspring events of event $t_j$. An immigrant event $t_j$ might lead to no offspring (in which case $O_j = \varnothing$ and $\bm x_{O_j} = \varnothing$), or one or more offspring with event times in $\bm{x}_{O_j}$. By definition, the last event has an empty set of offspring, and $O_n = \varnothing.$

Under the Poisson cluster representation of the Hawkes process, $\bm x$ can be thought of as occurring in the following manner. 
First, the set of all immigrants events (or ``generation 0'') are generated from a Poisson process with $\lambda_I(t)= \mu(t)$ on $[0,T)$. Then, each of the immigrant events $t_i\in \bm{x}_I$ leads to offspring events in generation 1 as a draw from a Poisson process with intensity $\lambda_{O_i}(t) = \alpha g(t-t_i)$ on $[0,T)$.  At the next step, each event in generation 1 leads to events in generation 2 in the same manner, and that continues on.
 Due to this representation, $\mu(t)$ and $\alpha g(t)$ are called background and offspring intensity, respectively. This representation also illustrates that $\alpha$ represents the mean number of offspring of an event, and $g(t)$ is the density function specifying the length of the time interval between an offspring and its parent. Therefore, the parameter $\alpha$ needs to satisfy that $\alpha\in(0,1)$ for the stationarity of the process.

For our discussion in \cref{sec:aggregated}, it is useful to introduce the likelihood of an observed point pattern $\bm{x} = \{t_1,t_2,\dots,t_n\}$ with $n$ events and corresponding branching structure $\bm Y = \{Y_1,Y_2,\dots,Y_n\}$ during an observed time period $[0, T)$ under the Hawkes model in \cref{eq: intensity}. If $\bm \theta$ is the vector of all parameters, then  
\begin{align}
\label{eq:likelihood}
p(\bm{x},\bm{Y}|\bm \theta) &= p(\bm{x}_I|\bm \theta)\prod_{j=1}^np(\bm{x}_{O_j}|\bm \theta),
\end{align}
where
\(
p(\bm{x}_I|\bm \theta) = {\mathrm{e}}^{-M(T)}\prod_{i\in I}\mu(t_{i}),\)
and
\(
p(\bm{x}_{O_j}|\bm \theta)= {\mathrm{e}}^{-\alpha G(T-t_j)}\prod_{i\in O_j}\alpha g(t_i-t_j)
\)
for
$M(t) = \int_0^t \mu(u)\, \mathrm{d} u$,
and $G(t) = \int_{0}^{t}g(u)\,\mathrm{d} u$. Integrating the branching structure $\bm Y$ out of the joint likelihood in \cref{eq:likelihood} returns the likelihood of the point pattern $\bm x$. From \cref{eq:likelihood}, we see that working with the joint likelihood of ($\bm x, \bm Y$) leads to a convenient factorization of the likelihood, where $\bm x$ can be viewed as independent Poisson processes $\bm x_I,\bm x_{O_1}, \dots,\bm x_{O_n}$, which can lead to computational gains \citep[e.g.][]{zhou2020efficient}.

\subsection{The spatio-temporal Hawkes process}
\label{subsec:continuous_hp_st}

When the locations of the events are available, a point process is of the form $\bm{X} = \{(t_1,\bm{s}_1),(t_2,\bm{s}_2),\dots\}$, where $t_i$ and $\bm{s}_i$ denote the time and location of the $i^{th}$ event, and $(t_i,\sss_i)\in[0,\infty)\times\mathbb{R}^2$ for $i\in\mathbb{Z}^+.$ 
The conditional intensity is defined in a similar way to that for temporal processes in \cref{eq: intensity_temporal} as 
\begin{equation}
  \lambda^*(t,\bm s) = \lim_{\Delta \bm s ,\Delta t \downarrow 0}
\frac{\textrm{E}[\ \textrm{N}\{(t,t+\Delta t)\times B(\bm s,\Delta \bm s)\} \mid \history]}{\Delta t |B(\bm s,\Delta \bm s)|},  
\end{equation}
where $N$ is now the counting measure over $ [0,\infty)\times \mathbb{R}^2$, $|B(\bm s,\Delta \bm s)|$ is the Lebesgue measure of the ball $B(\bm s,\Delta \bm s)$, and  $\history = \{(t^*,\bm{s}^*)\in\bm X: t^*<t\}.$ A spatio-temporal Hawkes process is one whose conditional intensity is of the form
\begin{align}\label{eq: general_intensity_st}
     \lambda^*(t,\bm s) = u(t,\bm s) + \ \sum_{i:t_i<t}\alpha \ g(t-t_i,\bm{s}-\bm{s}_i),
\end{align}
where $g(t,\bm{s})$ is a density function on $[0,\infty)\times\mathbb{R}^2$. As discussed in \cite{Reinhart2018review}, the triggering function $g(t,\bm{s})$ is often taken to be separable as the product of time and spatial components: $g(t,\bm{s}) = g_1(t)g_2(\bm s)$. 
Common choices for $g_2(\sss)$ include the bivariate Gaussian and the power law distribution \citep{ogata1998space} (see \cref{tab:specification}).

\begin{table}[!t]
\centering
\caption{Common choices for $g_1(t)$ and $g_2(\sss)$. The ones considered in this manuscript are marked with an asterisk. For the temporal models, $g(t) = g_1(t).$} \label{tab:specification}

\scalebox{0.9}{
\begin{tabular}{ c c c  c c}
\\[-5pt]
\toprule
  & \multicolumn{2}{c}{$g_1(t)$} &  \multicolumn{2}{c}{$g_2(\sss)$}\\
\cmidrule(lr){2-3}
\cmidrule(lr){4-5}
& Exponential$^*$ & Lomax$^*$    & Gaussian$^*$ & Power Law\\   
\cmidrule(lr){2-2}
\cmidrule(lr){3-3}
\cmidrule(lr){4-4}
\cmidrule(lr){5-5}
 
Functional  form  & $\beta\exp(-\beta t)$  & $\displaystyle \frac{(p-1)c^{p-1}}{(t+c)^p}$    & $ \displaystyle \frac{1}{2\pi\gamma^2}\exp(-\frac{||\bm s ||^2}{2\gamma^2})$   &   $\displaystyle \frac{(q-1)d^{q-1}}{\pi (||\bm s ||^2+d)^q}$ \\

\midrule
Parameter(s) & $\beta$ & $c,p$ & $\gamma$ & $q,d$\\
\bottomrule

\end{tabular}
}
\end{table}

As in the temporal case, the spatio-temporal Hawkes process can also be viewed as a Poisson cluster process. Consider a realization of a spatio-temporal Hawkes process with $n$ events, $\bm x = 
\{(t_1,\sss_1),\dots,(t_n,\sss_n)\}$, on some observed window $[0,T)\times W$.
The branching structure is defined as in the temporal case, with $\bm Y_i = \{Y_1, Y_2, \dots, Y_n\}.$ Let $\bm{x}_I = \{(t_i,\sss_i): i\in I\}$ and $\bm{x}_{O_j} = \{(t_i,\sss_i):i\in O_j\}$ for all $1\leq j\leq n.$ Then the set of all immigrant events $\bm{x}_I$ are generated from a non-homogeneous Poisson process with intensity $\lambda_I(t,\bm s) = u(t,s)$, and each event $(t_i,\bm{s}_i)\in\bm x$ produces offspring according to a non-homogeneous Poisson process with intensity $\lambda_{O_i}(t,\bm s) = \alpha g(t-t_i,\bm{s}-\bm{s}_i).$ 
The likelihood of $(\bm x, \bm Y)$ can be written as
\begin{equation}\label{eq: exact_likelihood}
   \begin{aligned}
       p(\bm{x},\bm{Y}|\bm \theta) &= p(\bm x_{I} |\bm \theta)\prod_{j=1}^np(\bm x_{O_j} |\bm \theta).
\end{aligned} 
\end{equation}
where 
\begin{equation}\label{eq: exact_likelihood_immi}
  p(\bm x_I |\bm \theta) =     \exp\Big(-\int_0^T \!\!\! \int_W \lambda_I(t,\bm s))\dd \bm s  \dd t\Big)\prod_{i\in I} \lambda_I(t_{i},\bm{s}_{i}),  
\end{equation}
and
\begin{equation}\label{eq: exact_likelihood_offspring}
  p(\bm x_{O_j} |\bm \theta) = \exp\Big(-\int_0^T \!\!\! \int_W \lambda_{O_i}(t,\bm{s}))\dd \bm s \dd t\Big)\prod_{i\in O_j} \lambda_{O_j}(t_i,\bm{s}_i) .
\end{equation}

\subsection{The multivariate Hawkes process}
\label{subsec:continuous_hp_multi}

The Hawkes process has been extended to accommodate multiple processes with mutually exciting behavior. An $L$-dimensional spatio-temporal process gives rise to point patterns of the form $\{(t_{l,i},\bm{s}_{l,i}): l\in\{1,2,\dots L\}, i\in\mathbb{Z}^+\},$ where $t_{l,i}$ and $s_{l,i}$ are the time and location of the $i^{th}$ event from process $l$.  In this context, $L$ is a known value, and we  also know the specific point process to which an event belongs. For example, in \cref{sec:application}, we study the dynamics between $L = 2$ processes corresponding to airstrikes and insurgent violence events. Therefore, our interest is in studying a specific number $(L=2)$ of processes, and this is not a parameter that has to be learnt.
The multivariate Hawkes process can be defined through the conditional intensity of each of its $L$ components as
\begin{equation}
  \lambda_l^*(t,\bm{s}) = u_l(t,\sss)+\sum_{m = 1}^L\sum_{i: t_{l,i}<t} \alpha_{m,l}g_{m,l}(t-t_{l,i},\sss-\sss_{l,i}), 
\end{equation}
where $l = 1, 2, \dots, L$, and (implicitly) the history is extended to include times and locations for the events occurring before time $t$ from {\it all} $L$ processes.
A multivariate Hawkes process can also be viewed as a Poisson cluster process generated by a latent branching structure, $\boldsymbol Y = \{Y_{l,i}: l\in\{1,2,\dots L\}, i\in\mathbb{Z}^+\}$. Here, $Y_{l,i} = (0,0)$ if $t_{l,i}$ is an immigrant and $Y_{l,i}=(m,j)$ if $t_{l,i}$ is an offspring of $t_{m,j}$ for $i,j\in \mathbb{Z}^+$ (see \cref{supp_subsec:multi_notation}). 
Through this equivalence, $u_l(t,\sss)$ represents the background intensity for process $l$ and $\alpha_{m,l}g_{m,l}(t-t_{l,i},\sss-\sss_{l,i})$ the offspring intensity corresponding to either self-excitation ($m=l$) or cross-excitation ($m\neq l$). For $\alpha$ representing the $L \times L$ matrix of all coefficient of this type, $\alpha = (\alpha_{m,l})$, a sufficient condition for the stationarity of the process is that its spectral radius is smaller than 1 \citep{hawkes1971spectra,bremaud1996stability}.

\section{Aggregated Hawkes process}
\label{sec:aggregated}

Standard estimation techniques for the parameters of the Hawkes process require that the available data include exact temporal and spatial information. However, researchers often only have access to limited event information, such as the date or region an event occurred, without its precise time stamp or location. We refer to such data as ``aggregated'' point pattern data, and we consider estimation of parameters of the underlying process when the underlying exact event occurrence follows a Hawkes process.

In this section, we introduce the proposed Bayesian approach for estimating the parameters of Hawkes processes when aggregation occurs over time, space, or both. 
We introduce aggregated point patterns in \cref{subsec: aggregated event}, and we link them to the exact, latent point pattern data in \cref{framework}, which is the basis of our approach.
In \cref{subsec:identifiability}, we show that the true underlying temporal or spatio-temporal Hawkes process model is identifiable based on aggregated data. We introduce specific functional forms in \cref{subsec:model} and give details about the MCMC steps in \cref{subsec:MCMC}.

\subsection{Aggregated event information}
\label{subsec: aggregated event}

We consider three scenarios of aggregated data from Hawkes processes: aggregated temporal event data, aggregated spatio-temporal event data, and aggregated event data on multiple processes. We assume that the process is observed during the $[0,T)$ time window, and, if spatial information is available, over a spatial window $W$.

\begin{figure}[!t]
\centering
\includegraphics[width = 0.95\textwidth,trim =0 10 0 10, clip]{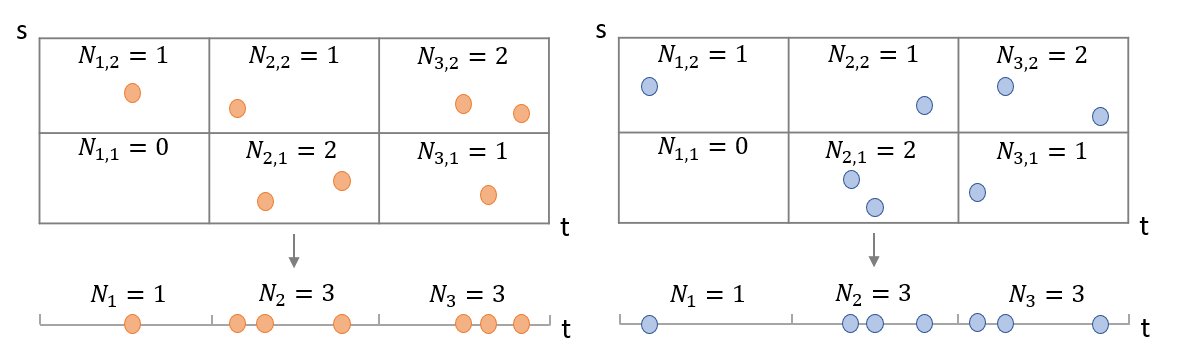}
\caption{Exact and Aggregated Point Pattern Data.
The orange points (left panel) represent the exact spatio-temporal (top) or temporal (bottom) data.
The continuous time (horizontal axis) is divided in three time bins, and the continuous space (one-dimensional, vertical axis) is divided in two bins.
The aggregated data correspond to the count of events occurring within each bin, which might be available with spatial information on the $2 \times 3$ spatio-temporal bins (top), and without spatial information on the 3 temporal bins (bottom). The aggregated data is $\{N_1,N_2,N_3\} = \{1,3,3\}$ when spatial information is not available, and is $\{N_{1,1},N_{2,1},N_{3,1},N_{1,2},N_{2,2},N_{3,2}\}=\{0,2,1,1,1,2\}$ when spatial information is available.
The blue points (right panel) represent an alternative point pattern that would produce the same aggregated data.
}
\label{fig:visulization_data}
\end{figure}

For aggregated temporal data, the time interval $[0,T)$ is partitioned into $K$ bins $B^t_1,B^t_2,\dots,B^t_{K}$.
The observed data correspond to the number of events that occurred within each temporal bin, denoted by $\bm N = \{N_1, N_2, \dots, N_{K}\},$ where $N_k \in \{0\} \cup \mathbb{Z}^+$.
For aggregated spatio-temporal data, the spatial window is also partitioned in $R$ bins, $B^{\sss}_1,B^{\sss}_2,\dots,B^{\sss}_{R}$, and the observed data are of the form $\bm N = \{N_{k,r}: k=1,2,\dots,K,r = 1,2,\dots,R\}$, where $N_{k,r}$ denotes the number of events in $B^t_{k}\times B^{\sss}_{r}.$ A visualization of aggregated data with equal bin sizes is shown in \cref{fig:visulization_data}, though bin sizes need not be equal in general.
When data are available for $L$ processes, we denote the spatio-temporal bins by $B^t_{l,k} \times B^{\sss}_{l,r}: l = 1,2,\dots,L, k=1,2,\dots, K_{l}, r = 1,2,\dots, R_{l},$ which illustrates that the aggregation size in time and space may be different for each process. Then, the aggregated data for multiple processes are in the form $\bm N = \{N_{l,k,r}: \text{all } l,k,r \}$ where $N_{l,k,r}$ is the number of events from process $l$ in $B^t_{l,k}\times B^{\sss}_{l,r}.$

\subsection{Latent variable formulation}
\label{framework}

Even though available data are in aggregated form, the events truly occur in continuous time and space, and ignoring dependencies among events in the same bin can lead to misleading conclusions. Our approach models the aggregated event data at the scale they actually occur, in continuous time and space.
The main idea is to consider the exact time and location of an event as latent variables. 

For aggregated data $\bm{N}$, we use $\bm{x}$ to represent the underlying exact point pattern which is unobserved. If $\bm \theta$ represents the parameters of the Hawkes process driving $\bm x$, we write the likelihood of the observed data given the parameters of the Hawkes process as
\begin{equation}
p(\bm N \mid \bm \theta)
= \int p(\bm N, \bm x, \bm Y \mid \bm \theta) \  \mathrm{d}\bm x \  \mathrm{d} \bm Y
= \int p(\bm N \mid \bm x, \bm Y, \bm \theta) \
p(\bm x, \bm Y \mid \bm \theta) \  \mathrm{d}\bm x \  \mathrm{d} \bm Y,
\label{eq:aggregated_integral}
\end{equation}
integrating over the distribution of the latent exact point pattern $\bm{x}$ and its latent branching structure $\bm Y$ (discussed in \cref{sec:methods}).

The likelihood in \cref{eq:aggregated_integral} can be simplified. Each exact point pattern $\bm x$ maps to a single aggregated point pattern $\bm N$. Thus, it holds that
$p(\bm N \mid \bm x, \bm Y, \bm \theta) = p(\bm N \mid \bm x)$, which is equal to 1 only when the aggregated data match the exact point pattern in terms of the total number of events and the number of events in each bin. For example, for $\bm N$ being the aggregated data from a univariate, spatio-temporal pattern $\bm x$, we have that
\begin{equation}
\label{eq:N_given_x}
p(\bm N \mid \bm x, \bm Y, \bm \theta) =
\begin{cases}
1, & \text{if } 
\sum_{i = 1}^n 1{\{(t_i,\bm{s}_i)\in B^t_{k}\times B^{\sss}_r\}} = N_{k,r} \text{ for all } k,r \\[5pt]
0, & \text{otherwise.}
\end{cases}
\end{equation}
Similar formulas to that in \cref{eq:N_given_x} for a temporal or a multivariate spatio-temporal process are given in \cref{supp_subsec:multi_notation}.

Therefore, the likelihood of an aggregated point pattern in \cref{eq:aggregated_integral} can be written as
\(
\int p(\bm x, \bm Y \mid \bm \theta) \  \mathrm{d}\bm x \  \mathrm{d} \bm Y
\),
where the integral is over exact point patterns that agree with the aggregated one. As an example, \cref{fig:visulization_data} shows two exact point patterns that lead to the same aggregated data.  The form of $p(\bm x, \bm Y \mid \bm \theta)$ is given in \cref{sec:methods}.


\subsection{Identifiability of the Hawkes process based on aggregated data}
\label{subsec:identifiability}

One question we need to address is whether the model parameters of the Hawkes process are identifiable when available data do not include the exact event information, but are instead aggregated. Here, we provide identifiability results under general functional specifications of the Hawkes process, with specific examples in \cref{subsec:model}.
We focus first on temporal Hawkes processes.

\begin{theorem}[Identifiability for a general temporal model]\label{thm: general temporal}
Let $\{t_1, t_2, \dots,t_n\}$ be a realization of a temporal Hawkes process on $[0,T)$ with conditional intensity $\lambda^*(t; \ttheta) = \mu(t; \bm \theta^\mu)+\sum_{i:t_i<t}\alpha g(t-t_i; \bm \theta^g),$ where $\mu(t; \bm \theta^\mu)$ is a positive function on $[0,\infty)$ with parameters $\bm \theta^\mu$, $\alpha$ is a positive constant, $g(t; \bm \theta^g)$ is a density function on $[0,\infty)$ with parameters $\bm \theta^g$, and $\ttheta = (\bm \theta^\mu, \bm \theta^g, \alpha)$ is the vector of all parameters. 
Let $ B_1^t,B_2^t,\dots,B_K^t$ be the temporal bins that partition $[0,T)$ and $\bm N = \{N_1,N_2,\dots,N_K\}$ be the corresponding aggregated temporal data. Suppose that 
\begin{enumerate}[itemsep=0pt,topsep=0pt,leftmargin=*,label=(\alph*)]
    \item For all $\ttheta^\mu_1\neq \ttheta^\mu_2,$ there exists $1\leq k \leq K$ such that 
    $ 
    \int_{B^t_k} \mu(t;\ttheta^{\mu}_1)\dd t\neq \int_{B^t_k} \mu(t;\ttheta^{\mu}_2)\dd t,
    $
    
    \item For $h(t; \alpha, \ttheta^g) = \int_0^t\alpha g(u; \ttheta^g) \dd u$,  
    the function
    $h(t; \alpha_1, \ttheta_1^g)-h(t; \alpha_2, \ttheta_2^g)$ has at most $K-1$ roots on $(0,T)$ whenever $(\alpha_1, \ttheta_1^g) \neq (\alpha_2, \ttheta_2^g)$.
\end{enumerate}
Then the parameters $\ttheta$ are identifiable based on the aggregated data $\bm N.$
\end{theorem}

\cref{thm: general temporal} establishes that, for generic temporal Hawkes processes, and under reasonable conditions on the process' functional specification, model parameters are identifiable based on aggregated data. Intuitively, these conditions imply that different parameter sets lead to a different distribution for  
\begin{enumerate*}[label=(\alph*)]
    \item immigrant events across bins, and
    \item an event's offspring in the same and subsequent bins. 
\end{enumerate*}


The previous result establishes identifiability of the temporal Hawkes process model parameters based on aggregated data. We extend these results to general spatio-temporal Hawkes processes with intensity as in \cref{eq: general_intensity_st} with spatio-temporal background and offspring intensities that are not necessarily separable. It is useful to consider an equivalent form of the conditional intensity for the spatio-temporal Hawkes process, as 
\begin{equation} \label{eq: general_intensity_st2}
    \lambda^*(t) = \mu(t) \ f(\sss \mid t)+\sum_{i:t_i<t}\alpha \ g_1(t-t_i)\ g_2(\sss-\sss_i \mid t-t_i)
\end{equation}
{where}
$\mu(t) = \int_{\mathbb{R}^2}u(t,\sss)\dd \sss,$
$f(\sss|t) = u(t,\sss)/\mu(t),$
$g_1(t) = \int_{\mathbb{R}^2}g(t,\sss)\dd \sss$,
and
$g_2(\sss|t) = g(t,\sss)/g_1(t).$

\begin{theorem}[Identifiability for a general spatio-temporal model]\label{thm: general st}
    Let $\{(t_i,\bm{s}_i):i=1,2,\dots,n\}$ be a realization of a spatio-temporal Hawkes process on $[0,T)\times\mathbb{R}^2$ with conditional intensity 
    $\lambda^*(t,\bm{s};\ttheta) = \mu(t;\ttheta^\mu)f(s\mid t;\ttheta^f)+\sum_{i:t_i<t}\alpha g_1(t-t_i;\ttheta^{g_1})g_2(\sss-\sss_i\mid t-t_i;\ttheta^{g_2}),$ 
    where 
    $\mu(t;\ttheta^\mu)$ is a positive function on $[0,\infty)$ with parameters $\ttheta^\mu$, 
    $\alpha$ is a positive constant, 
    $g_1(t;\ttheta^{g_1})$ is a density function on $[0,\infty)$ with parameters $\ttheta^{g_1}$, and 
    $f(\sss\mid t;\ttheta^f)$ and $g_2(\sss\mid t;\ttheta^{g_2})$ are density functions on $\mathbb{R}^2$ with parameters $\ttheta^f$ and $\ttheta^{g_2}$ respectively.
    Let $\{B_k^t\times B_r^{\sss}: 1\leq k\leq K,1\leq r\leq R\}$ be the spatio-temporal partition of $[0,T) \times \mathbb{R}^2$ for some $K,R\in\mathbb{Z}^+\cup\{\infty\}$, and $\bm N = \{N_{k,r}: 1\leq k\leq K,1\leq r\leq R\}$ be the corresponding aggregated spatio-temporal Hawkes process. Suppose the following conditions hold.
    \begin{enumerate}[itemsep=0pt,topsep=0pt,leftmargin=*,label=(\alph*)]
    \item  For all $\ttheta^{\mu}_1\neq\ttheta^{\mu}_2$, there exists $1\leq k \leq K$ such that 
    $ \displaystyle
    \int_{B^t_k} \mu(t;{\ttheta^\mu_1})\dd t\neq \int_{B^t_k} \mu(t;{\ttheta^\mu_2}) \dd t.$ 
    \item For $h(t;\alpha, \ttheta^{g_1}) = \int_0^t \alpha g_1(u; \ttheta^{g_1})\dd u$, and for all $(\alpha_1,\ttheta^{g_1}_1)\neq (\alpha_2,\ttheta^{g_1}_2)$, $h(t;\alpha_1,\ttheta^{g_1}_1)-h(t;{\alpha_2,\ttheta^{g_1}_2})$ has at most $K-1$ roots on $(0,\infty).$
    \item $\ttheta^f = \varnothing$, or if $\ttheta^f \neq \varnothing$,  for $\ttheta_1^f\neq\ttheta_2^f$ there exist $1\leq k\leq K$ and $1\leq r \leq R$ such that 
    $ \displaystyle
    \int_{B_r^{\sss}} f(\sss \mid t_1;\ttheta^{f}_1)\dd \sss \neq \int_{B_r^{\sss}} f(\sss \mid t_1;\ttheta^{f}_2)\dd \sss,
    $
    for all $t_1\in B^t_{k}.$
    \item  For all $\ttheta^{g_2}_1\neq \ttheta^{g_2}_2 $,
    there exist $1\leq k\leq k'\leq K$ and $1\leq r,r' \leq R$ such that $$\int_{B_r^{\sss}}\int_{B_{r'}^{\sss}} {g_2}(\sss'-\sss \mid t_2-t_1;{\ttheta^{g_2}_1})\dd \sss\dd \sss' \neq \int_{B_r^{\sss}}\int_{B_{r'}^{\sss}} {g_2}(\sss'-\sss \mid t_2-t_1;{\ttheta^{g_2}_2})\dd \sss\dd \sss',$$ for all $t_1\in B^t_{k}$, $t_2\in B^t_{k'}$ with $t_1<t_2.$
\end{enumerate}
Then, all parameters are identifiable based on the aggregated data $\bm N$.
\end{theorem}

Conditions (a) and (b) of \cref{thm: general st} are equivalent to the conditions of \cref{thm: general temporal} and they pertain to the temporal aspect of the immigrant and offspring processes. Conditions (c) and (d) enforce that different parameter values lead to different distributions on the locations of immigrant events, and of offspring events in the current and subsequent bins, respectively. Under these conditions, the parameters of the corresponding spatio-temporal Hawkes process are identifiable based on aggregated data.

The proofs are included in \cref{appendix: proofs}, and they provide an interesting insight.
The identifiability of the parameters in the excitation function is based on the triggering effect of one event in different time bins. Therefore, for Hawkes processes for which most parent-offspring pairs fall in the same aggregation bin, the aggregated data may not provide sufficient information to accurately estimate the parameters of the triggering kernel. This limitation can hinder the practical uncovering of true parameters, despite theoretical identifiability being achievable. Even though our theoretical results for the spatio-temporal process are stated for $W = \mathbb{R}^2$, the statements and proofs readily extend to the scenario where $W \subsetneq \mathbb{R}^2$, as long as there is a zero probability of event occurrence outside of the window $W$.

\subsection{Parametric specification of the Hawkes model}
\label{subsec:model}

Hawkes process models are flexible and there are many parametric choices for the functions in the background and offspring intensities. The theorems in \cref{subsec:identifiability} can be used to established identifiability of model parameters for any adopted functional form. Here, we establish identifiability for the functional form choices in this manuscript.

In simulation studies, we specify that the occurrence of immigrants is homogeneous in time and space by specifying $\mu(t)=\mu$ and
\(
f(\sss) = |W|^{-1}{1{\{s\in W\}}}
\), where $|W|$ is the area of the spatial window. In our study of \cref{sec:application}, we set $\mu(t) = \mu\exp(-\eta t)$ to allow for an overall temporal trend in the occurrence of immigrant events, where $\mu > 0$ ensures that the background intensity is positive, and $\eta > 0$ ensures that it does not diverge with $t$. We also set $f(\sss \mid t) = f(\sss)$ where
\begin{equation}\label{eq: f_form}
 f(\sss) = \frac{\rho}{C_1}\exp\Big\{-\frac{d_1^2(\sss)}{2\gamma^2_{1}}-\frac{d_3^2(\sss)}{2\gamma^2_{3}}\Big\}+\frac{1 - \rho}{C_2}\exp\Big\{-\frac{d_2^2(\sss)}{2\gamma^2_{2}}-\frac{d_3^2(\sss)}{2\gamma^2_{3}}\Big\},   
\end{equation}
for functions $d_1,d_2,d_3$ on $\mathbb{R}^2$ (see \cref{sec:application}) and $C_1,C_2$ normalizing constants. 
Throughout we adopt a separable excitation function $g(t,\sss) = g_1(t)g_2(\sss)$  with an exponential or Lomax density for $g_1(t)$ and a Gaussian density for $g_2(\sss)$.
The following propositions show the identifiability results for these special cases of Hawkes processes.

\begin{proposition}\label{thm: temporal}
For the setting of \cref{thm: general temporal} with constant background intensity
\(\mu(\cdot; \bm \theta^\mu) = \mu > 0\), exponential or Lomax excitation density function 
$g(t; \ttheta^g)$ with parameters $\ttheta^g$ as defined in \cref{tab:specification}, and number of temporal bins $K \geq 3$, the parameters $\ttheta = (\mu, \alpha, \ttheta^g)$ are identifiable based on the aggregated data $\bm N.$
\end{proposition}

\begin{proposition}\label{thm: st}
For the setting of \cref{thm: general st} with intensity of the form $\lambda^*(t,\bm{s};\ttheta) = \mu(t;\ttheta^{\mu}) f(s;\ttheta^f)+\sum_{i:t_i<t}\alpha g_1(t-t_i;\ttheta^{g_1})g_2(s-\bm{s}_i;\ttheta^{g_2}),$ suppose that the following conditions hold:
\begin{enumerate}[itemsep=0pt,topsep=0pt,leftmargin=*,label=(\alph*)]
    \item $\mu(\cdot;\ttheta^{\mu}) = \mu$ with $\ttheta^{\mu} = \mu$, or $\mu(t;\ttheta^\mu) = \mu\exp(-\eta t)$ for $t > 0$ with $\ttheta^{\mu} = (\mu, \eta)$.
    \item For the background spatial density $f(\sss;\ttheta^f)$ it holds that either
    \begin{itemize}[leftmargin=*,topsep=0pt,itemsep=3pt]
        \item 
        there are no unknown parameters in $f(\sss;\ttheta^f)$, i.e. $\ttheta^f = \varnothing$, or
        \item $f(\sss|\ttheta^f)$ is of the form specified in \cref{eq: f_form} with $\ttheta^f = (\rho, \gamma_1, \gamma_2, \gamma_3)$,
        where for each pair $i \neq j\in\{1,2,3\}$, there exist $r_1,r_2,r_3,r_4$ such that $d_i(\sss_1)<d_j(\sss_2)$ and $d_i(\sss_3)>d_j(\sss_4)$ for all $\sss_1\in B_{r_1},
        \sss_2\in B_{r_2}, \sss_3\in B_{r_3},$ and $\sss_4\in B_{r_4}.$
    \end{itemize}
    \item $g_1(t;\ttheta^{g_1})$ is the exponential or Lomax density function, and $g_2(\sss;\ttheta^{g_2})$ is the bivariate Gaussian density defined in \cref{tab:specification}.

    \item For the number of temporal and spatial bins it holds that $K,R\geq 3.$
\end{enumerate}
Then, the parameters $\ttheta = (\ttheta^\mu,\ttheta^f,\ttheta^{g_1},\ttheta^{g_2})$ of the Hawkes process are identifiable based on the aggregated data $\bm N$.
\end{proposition}

The number of temporal and spatial bins required for identifiability is determined by the complexity of the background intensity and triggering kernel, with a higher number of bins required for more flexible background intensity and triggering functions. 
For the multivariate Hawkes process, we maintain these specifications while allowing for different parameters for each component of the process.

\subsection{Estimation and inference within the Bayesian framework}
\label{subsec:MCMC}

We propose a Bayesian approach to estimate the parameters of the Hawkes process based on aggregated point pattern data.
We use MCMC methods to acquire the posterior distribution of model parameters $\ttheta$ based on the likelihood in \cref{eq:aggregated_integral}. Specifically, we iteratively impute the exact point pattern and branching structure, and update the model parameters given the imputed exact data to acquire samples from the joint distribution $p(\bm x, \bm Y, \bm \theta \mid \bm N)$. We then discard samples of $(\bm x, \bm Y)$, and the resulting samples of $\ttheta$ are samples from the desired posterior distribution $p(\ttheta \mid \bm N)$.
Bayesian approaches that augment their sampling scheme using latent variables in an iterative manner have been previously used in modeling event occurrence with missing or aggregate data \citep{tucker2019handling, bu2022likelihood}.

We present the details for the case of spatio-temporal aggregated data on $[0,T)\times W$, discussed in \cref{framework}, under the exponential-Gaussian specification with constant background intensity, $\mu(t) = \mu$ and $f(s) = |W|^{-1}1\{s \in W\}$. MCMC methods for alternative specifications are discussed in \cref{a: sec: MCMC}.

The spatial window $W$ over which the process is observed is often a bounded subset of $\mathbb{R}^2$ and the integrals in \cref{eq: exact_likelihood_immi} and \cref{eq: exact_likelihood_offspring} over $W$ may be difficult to evaluate. When $W$ is sufficiently large, \cite{Reinhart2018review} and \cite{schoenberg2013facilitated} advocate to approximate these integrals by integrating over $\mathbb{R}^2$.
The exact data likelihood in \cref{eq: exact_likelihood} then becomes 
\vspace{-5pt}
\begin{align*}
    p(\bm{x},\bm{Y}|\bm \theta) 
    &=\exp(-\mu T)(\mu|W|^{-1})^{|I|} \times\prod_{j=1}^n\Bigg\{\exp\big \{ -\alpha \{ 1-\exp[-\beta(T-t_j)]\} \big\}\\[-10pt]
    &\hspace{125pt} \times\prod_{i\in O_j}\frac{\alpha\beta}{2\pi\gamma^2}\exp\Big[-\beta(t_i-t_j)-\frac{\|\bm{s}_i-\bm{s}_j\|^2}{2\gamma^2}\Big] \Bigg\}
\end{align*}
where $|I|$ denotes the number of immigrants.

We adopt (independent) priors,
$\mu\sim\ga (a_1,b_1),$ $\alpha\sim\text{Trunc}\ga (a_2,b_2),$ and $\beta\sim\ga (a_3,b_3)$, where the second argument is the rate parameter and $\text{Trunc}\ga$ denotes the truncated gamma distribution on $(0,1)$. We adopt an inverse gamma prior $\gamma^2\sim \ig(a_4,b_4)$.
We choose $a_1 = a_2 = 1$ and $b_1 = b_2 = 0.1$ for the priors on $\mu$ and $\alpha$, and $a_4 = b_4 = 0.001$ for the prior on $\gamma^2$. We set $a_3 = 1$ and take $b_3 = 0.1$ for the simulations and $b_3 = 1$ for our study (see \cref{sec:application} for this choice). The priors we choose are relatively flat to depict that we do not have prior knowledge about the parameters as also done in other work \citep{tucker2019handling,ross2021bayesian,Rasmussen2013bayesian}, and the prior on $\alpha$ specifies that the process is stationary.
If there exists prior knowledge about the parameter, hyper-parameters can be specified accordingly, as in \cite{darzi2023calibration}. 

We use Gibbs sampling for parameters $\mu,\alpha,$ and $\gamma$, and for the latent branching structure $\bm Y = (Y_1,Y_2, \dots,Y_n)$.
Specifically, we iteratively sample from
\begin{align*}
 \mu|\bm{x},\bm \theta_{-\mu},\bm{Y}, \bm N &\sim \ga \big(a_1+|I|,b_1+T \big) \\
 \alpha|\bm{x},\bm \theta_{-\alpha},\bm{Y}, \bm N &\sim \text{Trunc} \ga \big(a_2+\sum_{j=1}^n|O_j|,b_2+n-\sum_{j=1}^n \exp(-\beta(T-t_j))\big)\\
 \gamma^2|\bm{x},\bm \theta_{-\gamma},\bm{Y}, \bm N &\sim \ig\big(a_4+\sum_{j=1}^n |O_j|,b_4+\frac{1}{2}\sum_{j=1}^n\sum_{i\in O_j}\|\bm{s}_i-\bm{s}_j\|^2\big),
\end{align*}
where $\bm \theta = (\mu, \alpha, \beta, \gamma)$ and $\bm \theta_{-(\ )}$ excludes the one in the subscript.
The full conditional distribution of $Y_i$ is a categorical distribution with
\begin{align*}
      &P( Y_i = j \mid \bm \theta,\bm{x},\bm{N}) \propto 
      \begin{cases} 
      \mu|W|^{-1}, & j= 0, \\
      \dfrac{\alpha\beta}{2\pi\gamma^2} \exp \left( - \beta(t_i-t_j) -\dfrac{\|\bm{s}_i-\sss_j\|^2}{2\gamma^2} \right)
      , & j=1,2, \dots, i-1, \\
      0, & j = i, i +1, \dots, n.
   \end{cases}  
\end{align*}
In principle, this conditional probability would have to be evaluated for all $j$ for which $j < i$, which can be computationally burdensome.
We improve the computational efficiency of the algorithm by avoiding evaluating this conditional probability for potential parent events $j$ that precede event $i$ by an unrealistically long amount of time, which are deemed too distant in time to serve as a parent event. We do so by setting this conditionally probability to 0 for $j \in \{1, 2, \dots, i-1\}$ if $(t_i - t_j)$ is above the $99+^{th}$ quantile of the Exponential($\beta$) distribution in the excitation function. 
The details of the truncation and the analysis on its performance are presented in \cref{a: subsec: truncation}.

We use a Metropolis-Hasting step to sample $\beta$ from its posterior conditional distribution. Specifically, we propose value $\beta'$ from a normal distribution with mean $\beta^c$ and standard deviation $\sigma_\beta$, where superscripts $c$ denotes the current value and $'$ denotes the proposed value. We accept the move with probability
\begin{align*}
H_{\beta} &= 
\min \Big(1,\ (\beta'/\beta^c)^{a_3-1}\exp(b_3(\beta^c-\beta'))\\
& \hspace{60pt} \times\prod_{j = 1}^n\big\{\exp\Big(\alpha\exp(-\beta'(T-t_j))-\alpha\exp(-\beta^c(T-t_j))\Big) \\
& \hspace{90pt} \times \prod_{i\in O_j}(\beta'/\beta^c)\exp\Big(-(\beta'-\beta^c)(t_i-t_j)\Big)\big\}\Big)
\end{align*}
The value of $\sigma_\beta$ controls the acceptance rate of $\beta$. For all the simulation and application analysis, we choose $\sigma_\beta$ such that the acceptance rate is between 20\% and 40\%.

Lastly, we sample the latent exact data $\bm x$ from its full conditional distribution $p(\bm x \mid 
\bm \theta, \bm Y, \bm N)$.
From \cref{eq:N_given_x}, we know that $\bm x$ has to be imputed in a way that the exact data agree with the aggregated data $\bm{N}$. We decompose $\bm x$ to $(\bm t, \bm s)$ for $\bm t = (t_1,t_2,\dots,t_n)$ and $\bm s = (\bm{s}_1,\bm{s}_2,\dots,\bm{s}_n)$, and perform element-wise Metropolis-Hastings updates.
For $i = 1,2\dots, n$, the proposed time $t_i'$ is drawn from a continuous uniform distribution that satisfies the following restrictions: 
\begin{enumerate*}[label=(\alph*)]
\item $t_i'$ is within the time bin of event $i$,
\item if event $i$ is an offspring of event $j$ ($Y_i= j$), then it occurs after its parent, $t_i' > t_j$, and similarly
\item it occurs before all of its offspring (if any).
\end{enumerate*}
By imposing these restrictions, we ensure that the proposed latent time agrees with the observed counts in the temporal bins. The proposed value of $t_i$ is accepted with probability
\begin{align*}
&H_{t_i} = \min\Big(1,\ \exp\Big(\alpha\exp(-\beta(T-t_i'))-\alpha\exp(-\beta(T-t_i^c)))\\
& \hspace{85pt}+(|O_i|-1{\{Y_i>0\}})\beta(t_i'-t_i^c)\Big)\Big).
\end{align*}
Similarly, the proposed location $\bm{s}_i'$ is drawn from a uniform distribution on the spatial bin containing event $i$. The proposed move is accepted with probability
\begin{align*}
&H_{\bm{s}_i} = \min\Big(1, \ \exp\Big(\frac{-\|\bm{s}_i'-\bm{s}_{\textrm{pa}}\|^2+\|\bm{s}_i^c-\bm{s}_{\textrm{pa}}\|^2}{2\gamma^2}1{\{Y_i>0\}}\\
&\hspace{85pt}+\sum_{j\in O_i}\frac{-\|\bm{s}_i'-\bm{s}_{j}\|^2+\|\bm{s}_i^c-\bm{s}_{j}\|^2}{2\gamma^2}\Big)\Big),
\end{align*} where $\bm{s}_{\textrm{pa}}$ is the location of the parent event of $i$, if it exists. The derivation of all acceptance ratios are shown in \cref{a: sec: acceptance}.

\begin{figure}[!b]
\centering
\includegraphics[width = \textwidth,trim=10 10 14 10, clip]{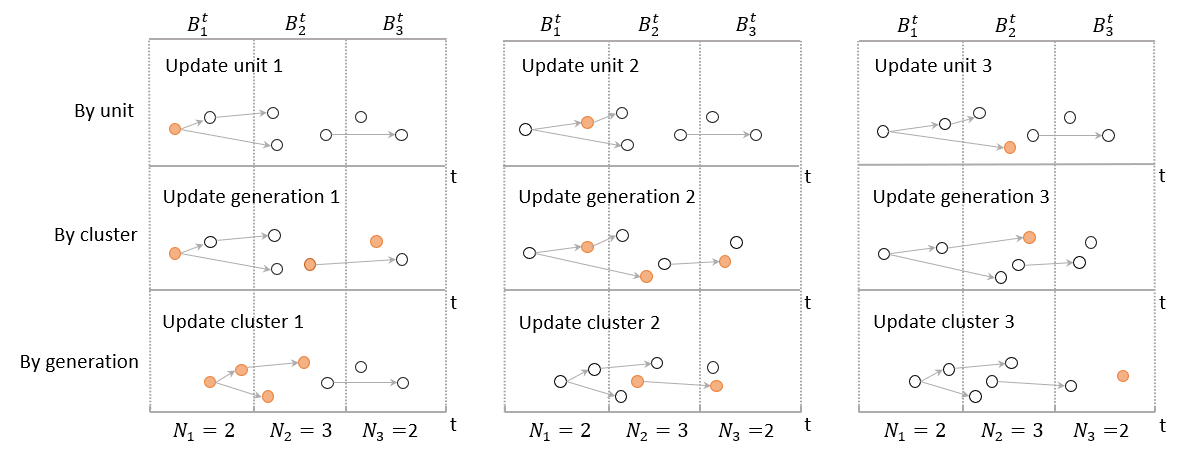}
\caption{Illustration of different ways of updating the latent event times on an aggregated temporal Hawkes process with 3 temporal bins and observed data $(N_1 = 2, N_2 = 3, N_3 = 2)$. We show the first three updates of three methods: separate update by unit (top row), block update by generation (middle row), and block update by cluster (bottom row). The points represent the latent times of events, with the orange color indicating the times currently under consideration for updating. The arrows show the branching structure of the events. For the update by unit, the procedure continues for seven iterations (one for each event).}
\label{fig:updating_time}
\end{figure}

Even though we propose updating the exact data for time and location for each event one-at-a-time, it is possible to design simultaneous MCMC updates of multiple latent parameters, though designing proposal distributions that lead to admissible acceptance rates for simultaneous updates is challenging. 
Simultaneous updates of events' exact time is restricted by the events' branching structure, which in turn imposes no restrictions on their exact locations.
We consider two possible alternative strategies: updating all events in the same generation, or in the same cluster simultaneously. A cluster $C_i$ is formed by an immigrant event $i$, and its descendants of all generations. Let $T_{C_i} = \{t_j: j\in C_i\}$ denote the set of event times in that cluster. We depict these approaches in \cref{fig:updating_time}.
In the former, timestamps for all events in the same generation are proposed simultaneously. Since the timestamps and locations of events in the same generation are independent, the posterior distribution under this block update and the one-at-a-time updates is essentially equivalent.
In the latter,  
we propose new values for $T_{C_i}$ using $p(T_{C_i}|\bm{Y},\bm \theta)$ as the proposal distribution, with the restriction that for all $t_j \in T_{C_i}$, the proposed value $t_j'$ is within the time bin as the current value $t_j^c$. Event locations are updated in a similar manner. Algorithmic details are included in \cref*{appendix: updates}. Block updates of the latent point pattern by cluster can reduce the autocorrelation among posterior samples, but this approach is computationally demanding and generally leads to lower acceptance rates than the other methods two methods. In \cref{appendix: updates}, we investigate the relative performance of these updating methods for aggregated temporal data by comparing the Gelman-Rubin statistic \citep{gelman1992inference} (Rhat), effective sample size (ESS), and computing time. We find that updating the latent, exact point pattern one-at-a-time is computationally efficient and leads to good MCMC mixing.

Prior choices and MCMC updates for the temporal Hawkes process or the mutually-exciting spatio-temporal Hawkes process under exponential, Gaussian, or other specifications are discussed in \cref{a: sec: MCMC}, along with a prior specification that ensures stationarity of the multivariate Hawkes process.

\section{Simulations}
\label{sec:simulations}

We investigate the performance of our method on simulated data in different scenarios.
Particularly, we explore the effect that fine or coarse aggregations of time and space have on the method's performance, and in comparison with the gold standard of fitting the Hawkes model on the exact data.
We focus on the univariate temporal case in \cref{subsec: temporal sim}, the spatio-temporal case in \cref{subsec: spatio-temporal sim}, and the multivariate spatio-temporal case in \cref{subsec: multivariate}.

For our simulations, we consider the time window $[0, 500)$, a rectangular spatial window $W = [0, 100]\times[0,100]$, and bins of equal size. 
Specifically, for $\Delta^t$ and $\Delta^{\sss}$ being the size of aggregation in time and space, respectively,
we set $B^t_k = [(k-1)\Delta^t,k\Delta^t)$ and $B^{\sss}_{r}$ is of the form $[(c-1)\Delta^{\sss},c\Delta^{\sss})\times[(d-1)\Delta^{\sss},d\Delta^{\sss})$ for some $c,d\in\mathbb{Z}^+$.  We present a subset of our simulations here, and we include additional simulations in \cref{appendix:add_sim}, which we summarize where applicable and in \cref{subsec:additional_sims}.
 Prior choices are discussed in \cref{subsec:MCMC} and in \cref{a: sec: MCMC}.
MCMC convergence was investigated using the Gelman-Rubin Rhat statistic, and by traceplots (shown in \cref{appendix:add_sim}).

\subsection{Temporal simulations}
\label{subsec: temporal sim}

Data for the temporal and spatio-temporal simulations were generated simultaneously, and they were aggregated in pre-defined bins.
We simulate spatio-temporal data sets from a Hawkes process 
with the exponential-Gaussian specification of the excitation function and parameters $\beta = \gamma = 1$, and $\mu = 1 - \alpha$, where $\alpha \in \{0.1, 0.3,  0.5, 0.7, 0.9\}$.
Data sets generated by these parameter sets have approximately 500 events on average, but the process produces fewer immigrants and more offspring for larger values of $\alpha$.
Since the branching ratio $\alpha$ does not depend on an event's location, simulated data sets from the corresponding {\it temporal} Hawkes process are acquired simply by removing the spatial information.
For each parameter set, we simulate 400 data sets. 

In this section, we consider estimation of the parameters of the Hawkes process based on aggregated temporal data. We consider time aggregations in terms of the true value of $1/\beta$, which is the expected length of time between an offspring and its parent. We set the length of aggregation to $\Delta^t \in \{ 0,0.5,1,2,3,4,5 \}$ (times 1/$\beta$), where $\Delta^t = 0$ refers to no aggregation. We consider the performance of the proposed Bayesian method, and the frequentist MC-EM approach proposed by \cite{shlomovich2022parameter} on aggregated data with $\Delta^t > 0$. We use $\Delta^t = 0$ to denote results from a gold-standard Bayesian approach that uses the {\it exact} data which, in reality, would not be applicable when the available data are aggregated. 

\begin{table}[!htb]
\centering
\caption{Results of the temporal simulation. The table shows the average posterior mean (``Estimate''), average 95\% credible interval length (``CI length''), and coverage rate of the 95\% credible interval (``Coverage'') for parameters $\mu, \alpha$ and $\beta$ of the Hawkes process, across 400 simulated data sets, for the exact data ($\Delta^t = 0$), temporal aggregations of different sizes, and two sets of parameters. 95\% credible interval is determined by the 2.5 and 97.5 quantiles of the posterior samples.
}

\label{tab:sim_temporal}
\scalebox{0.9}{
\begin{tabular}{ c c c c c c c c c } 
\toprule 
& \multicolumn{4}{c}{Parameter set 1: $(\mu, \alpha, \beta) = (0.3, 0.7, 1)$} & \multicolumn{4}{c}{Parameter set 2: $(\mu, \alpha, \beta) = (0.5, 0.5, 1)$}  \\
\cmidrule(lr){2-5} \cmidrule(lr){6-9}
&$\Delta^t$  &  Estimate & CI length & Coverage &$\Delta^t$  &  Estimate & CI length & Coverage\\   
\midrule
\multirow{7}*{$\mu$} 
& 0  & 0.3115  & 0.1666  & 0.945\phantom{0}  &  0 & 0.5141	& \phantom{0}0.2421	&0.9575\\
& 0.5  & 0.3116  & 0.1669  & 0.9425 &  0.5 & 0.5149	& \phantom{0}0.2436	&0.9625\\
& 1  & 0.3122  & 0.1679  & 0.945\phantom{0} &  1 & 0.5166	& \phantom{0}0.2456	&0.9475\\
& 2  & 0.3139  & 0.171\phantom{0}  & 0.95\phantom{00} & 2  & 0.5267  & \phantom{0}0.2571  & 0.94\phantom{00} \\
& 3  & 0.3169  & 0.1756  & 0.9496 & 3  & 0.5388  & \phantom{0}0.2641  & 0.9125 \\
& 4  & 0.3225  & 0.1822  & 0.915\phantom{0} & 4  & 0.544\phantom{0}  & \phantom{0}0.2621  & 0.885\phantom{0}\\
& 5  & 0.3283  & 0.1866  & 0.8975 & 5  & 0.5492  & \phantom{0}0.2589  & 0.8675\\
\hline
\multirow{7}*{$\alpha$} 
& 0  & 0.6854  & 0.2024  & 0.9375  &  0 & 0.4829	& \phantom{0}0.2442	&0.9375\\
& 0.5  & 0.6853  & 0.2029  & 0.935\phantom{0} &  0.5 & 0.482\phantom{0}	& \phantom{0}0.2457	&0.95\phantom{00} \\
& 1  & 0.6847  & 0.2036  & 0.945\phantom{0} &  1 & 0.4802	& \phantom{0}0.2475	&0.935\phantom{0}\\
& 2  & 0.6829  & 0.206\phantom{0}  & 0.9375 & 2  & 0.471\phantom{0}  & \phantom{0}0.2575  & 0.9125\\
& 3  & 0.6794  & 0.2101  & 0.927\phantom{0} & 3  & 0.4586  & \phantom{0}0.2628  & 0.875\phantom{0}\\
& 4  & 0.6734  & 0.2152  & 0.8975 & 4  & 0.4532  & \phantom{0}0.2595  & 0.855\phantom{0}\\
& 5  & 0.6672  & 0.2179  & 0.8675 & 5  & 0.4477  & \phantom{0}0.2557  & 0.855\phantom{0}\\
\hline
\multirow{7}*{$\beta$} 
& 0  & 1.0567  & 0.5995  & 0.95\phantom{00} &  0 & 1.0909	& \phantom{0}0.9839	&0.9575\\
& 0.5  & 1.0599  & 0.6163  & 0.955\phantom{0} &  0.5 & 1.1072	& \phantom{0}1.063\phantom{0}	&0.95\phantom{00} \\
& 1  & 1.07\phantom{00}  & 0.6587  & 0.955\phantom{0} &  1 & 1.1667	& \phantom{0}1.3206	&0.9275\\
& 2  & 1.0991  & 0.7968  & 0.965\phantom{0} & 2  & 1.5464  & \phantom{0}3.4335  & 0.925\phantom{0} \\
& 3  & 1.1935  & 1.284\phantom{0}  & 0.9345 & 3  & 3.0082  & \phantom{0}9.7799  & 0.8725\\
& 4  & 1.6732  & 3.2711  & 0.9025 & 4  & 4.6619  & 15.0116  & 0.78\phantom{00}\\
& 5  & 2.4718  & 6.3484  & 0.85\phantom{00} & 5  & 6.4988  & 20.6779  & 0.7425\\
\bottomrule
\end{tabular}
}
\end{table}

First, we investigate how results from our model that uses aggregated data compare to the results from the gold standard approach ($\Delta^t = 0$).
\cref{tab:sim_temporal} shows the average estimate (posterior mean), average length of 95\% credible interval, and coverage rate for each parameter under the different aggregations.
For space considerations, we include results for $\mu = 0.3, \alpha = 0.7$ (referred to as {\it parameter set 1}), and $\mu = \alpha = 0.5$ (referred to as {\it parameter set 2}) here, and results for the other parameter choices are shown in \cref{sec: sim_t}. 
For all parameter sets, the estimates and credible interval lengths for $\mu$ and $\alpha$ are generally unaffected by the data's temporal aggregation, especially for aggregation size less than or equal to 2. On the other hand, the credible interval length for $\beta$ is on average wider, and the estimate of $\beta$ is on average larger for coarser temporal aggregations. 
The efficiency loss in estimating $\beta$ is expected for larger aggregations, since coarser temporal aggregations would result in fewer parent-offspring pairs that fall at different temporal bins.
The coarseness of the temporal aggregation has a larger effect for $\beta$ under smaller values of $\alpha$. 
This is also expected as data sets with a smaller value of $\alpha$ contain fewer offspring, and as a result, even fewer parent-offspring pairs in separate bins. 
Moreover, we observe an increase in bias and a drop in coverage rate for parameter $\beta$ in  scenarios with a small $\alpha$ and a large aggregation size. This could be attributed to insufficient information in the data, leading to a posterior distribution that is heavily influenced by the prior.
These simulation results are in agreement with our theoretical identifiability results where we found that the information to estimate the parameters of the excitation function lies in the parent-offspring pairs that are in different bins. Therefore, researchers can investigate whether there is indication that the parameters of the excitation function are not well-estimable based on the observed aggregated data by investigating the number of parent-offspring pairs in different bins across MCMC samples, as we do in \cref{sec:application}. 
The efficiency loss and bias with larger temporal aggregations is alleviated in the presence of spatial information (see \cref{subsec: spatio-temporal sim}).  

We compare the performance of the proposed Bayesian method and the MC-EM method on the same data sets in terms of estimation only since the MC-EM approach does not offer a way for inference. The results are shown in \cref{appendix:MCEM}.
We observe that the Monte Carlo variance for estimating $\mu$ and $\beta$ is similar for the two methods, but the Monte Carlo variability for $\alpha$ is lower based on our Bayesian method. 
The performance of the MC-EM algorithm {\it deteriorates significantly} under parameter set 2 and large aggregations. Results from MC-EM are unstable, yielding unreasonably large estimates when $\Delta^t \geq 4$.
In contrast, the proposed Bayesian approach leads to accuracy gains in terms of estimation of $\alpha$, is stable across different aggregation sizes, and   it offers a straightforward inference procedure. 
Since the MC-EM approach is only applicable for the temporal Hawkes process, we do not use it for the spatio-temporal simulations.

%

%

\subsection{Spatio-temporal simulations}
\label{subsec: spatio-temporal sim}

We aggregate the exact data using the true values of $\beta$ and $\gamma$ by $\Delta^t \in \{ 0,1,3,5 \} $ (times 1/$\beta$) and $\Delta^{\sss} \in \{ 0,1,3,5 \}$ (times $\gamma$), where $\Delta^t = 0$ and $\Delta^{\sss} = 0$ denotes no aggregation in time and space, respectively. We summarize the findings here, and we present the full results of the spatio-temporal simulation in \cref{sec: sim_st}.

Across all aggregations over time and space and all parameter sets, the bias for parameters $\mu$ and $\alpha$ is negligible. Estimation of $\gamma$ is essentially unbiased, except under very large spatial aggregations, when the spatial aggregation size is 5 times the standard deviation of the spatial triggering function. Estimation of $\beta$ is essentially unbiased across all scenarios; the only exception being the case under $\alpha = 0.1$, where even the gold standard that uses the exact data exhibits bias due to the fact that the number of parent-offspring pairs is small. 
Once more, researchers can investigate whether there appears to be sufficient information in their aggregated data to estimate the parameters of the excitation function by checking the number of parent-offspring pairs in different temporal and spatial bins across MCMC iterations. 
Credible intervals for the excitation parameters are wider under coarser aggregations, while coverage is above 90\% throughout, with some exceptions in the challenging cases where $\alpha$ is small and $\Delta^t$ is large.
Similar to the temporal case, the credible interval length for $\mu$ and $\alpha$ remains mostly unaltered as the aggregation in time and space becomes coarser.
\cref{fig:heatmap_st} shows the mean length of 95\% credible intervals for $\beta$ and $\gamma$ under different aggregation sizes. 
Even though aggregation in either time or space affects the credible interval lengths for both $\beta$ and $\gamma$, the credible interval lengths of $\beta$ are mainly influenced by time aggregation, while the credible interval lengths of $\gamma$ are mainly influenced by space aggregation.

Lastly, we compared the results from the temporal model (\cref{subsec: temporal sim}) and the spatio-temporal model (see \cref{sec: sim_st}).
Incorporating the spatial locations leads to a reduction in bias and coverage closer to the nominal level, irrespective of the size of the time and space aggregation. In fact, the bias and efficiency loss observed for $\beta$ under large temporal aggregations in parameter set 2 is almost entirely mitigated when spatial information is available, even at coarse spatial bins.
In terms of estimation efficiency, 
even with coarsely aggregated spatial locations, the CI lengths of all parameters are much smaller based on the spatio-temporal model than based on the temporal model (\cref*{fig:compare_t_st}). This is true even when the event times are known exactly $(\Delta^t = 0)$. These results indicate that, if spatial information is available, incorporating it leads to more accurate estimation of the model parameters for the underlying Hawkes process, even when the location of the events is known at coarse levels. Our approach provides a framework to incorporate such coarse spatial information in the estimation procedure, and improve estimation efficiency.

\subsection{Simulations on the multivariate spatio-temporal process}
\label{subsec: multivariate}
We simulate 400 data sets from a bivariate spatio-temporal Hawkes process on the time window $[0,500]$ and the spatial window $[0,100]\times[0,100]$, using parameters
$$
\mu = \begin{pmatrix}
0.3 \\
0.5
\end{pmatrix}, \quad
\alpha = \begin{pmatrix}
0.7 & 0.15 \\
0.3 & 0.5 
\end{pmatrix}, \quad
\beta = \begin{pmatrix}
1 & 1\\
1 & 1 
\end{pmatrix}, \quad \text{and} \quad
\gamma = \begin{pmatrix}
1 & 1\\
1 & 1 
\end{pmatrix}.
$$
The spectral radius of the specified $\alpha$ matrix is less than 1, ensuring the stationarity of the generated process. The parameters in the $\beta$ and $\gamma$ matrices control the expected time and distance in self- and cross-excitation, and their diagonal and off-diagonal elements are not inherently related. We consider aggregations over both time and space for both processes. For ease of visualization, we aggregate data using the same coarseness for time and space, and separately for each process, i.e $\Delta_1^t = \Delta_1^{\sss} = \Delta_1$ and $\Delta_2^t = \Delta_2^{\sss} = \Delta_2,$ where the subscript is used to indicate process 1 and 2. 
We consider 9 different data aggregations by varying $\Delta_1$ and $\Delta_2$ over the values 0, 1, 3. 

Applying our method on the aggregated data sets results in good estimates on average (\cref*{tab:sim_st2}). The bias for the estimated parameters is consistently low, ranging from -0.008 to 0.033, and coverage is close to nominal, varying from 92.25 to 96.5\%, across all parameters and aggregations. Similar to the univariate case, the estimates of $\mu$ and $\alpha$ are very stable over different aggregations, with credible interval lengths that increase very slightly as the aggregations get coarser.
Compared to these parameters, $\beta$ and $\gamma$ are more sensitive to coarser aggregations. 
\cref{fig:heatmap_st2} shows the mean credible interval lengths for $\beta$ and $\gamma$ over different aggregation sizes.
We find that the credible interval length for parameters that correspond to self-excitation ($\beta_{11},\beta_{22},\gamma_{11},\gamma_{22}$) are affected mostly by aggregation in the process they represent; coarser aggregation in process 1 leads to higher credible interval length for $\beta_{11}, \gamma_{11}$ and little impact on the credible interval lengths for $\beta_{22}, \gamma_{22}$, while the reverse is true for coarser aggregation in process 2.
In contrast, parameters which correspond to cross-excitation ($\beta_{12},\beta_{21},\gamma_{12},\gamma_{21}$) are similarly affected by aggregation in either process.

\subsection{Additional simulation results}
\label{subsec:additional_sims}

We analyze the performance of the proposed method for data generated based on different values of the parameter $\beta$ of the exponential excitation function. This parameter controls the expected time for an event to trigger a new event.  Since this is a parameter corresponding to the underlying exact Hawkes process, its interpretation remains consistent regardless of the aggregation size. For given values of the remaining parameters, increasing the value of $\beta$ results in a shorter average time between parent and offspring events, which in turn leads to a larger number of empty bins. This effect is illustrated in \cref{tab:temporal_data_summary}. For these simulations, we generate datasets based on $\beta = 0.5,1,3,5,7$ and fix $\mu = 0.3$, $\alpha = 0.7$, $\Delta^t=3$ throughout. The results are shown in \cref{sec: sim_t}.
Estimates for $\mu$ and $\alpha$ are relatively accurate across all true values of $\beta$, though wider credible intervals are observed under larger $\beta$ values. The parameter $\beta$ itself is well-estimated when its true value is small, and it is overestimated with large credible interval lengths as the true $\beta$ values increases, though coverage is close to nominal for all parameters. We suspect that the deterioration in estimation performance is because large values of $\beta$ lead to a small expected time from parent to offspring ($1 / \beta$), and therefore few parent-offspring pairs in different bins, which as we have previously noted, can lead to practical lack of estimability.

We investigate the performance of our method under alternative excitation functions under the Lomax offspring kernel for the temporal component, in both temporal and spatio-temporal simulations (\cref{appendix: Lomax_sim}). 
Average estimates and coverage for the parameters $\mu, \alpha$ and $\gamma$ are similar between our approach that uses aggregated data and the gold-standard that uses exact data, and the coverage rates in all simulations are above 90\% throughout.
The estimation performance of our method under the Lomax kernel is comparable to that under the exponential kernel when the mean and variance of the excitation times are similar for both kernels.

In \cref{supp_sec:sim_temporal_computing} we investigate the computational time required to fit the proposed approach on temporal and spatio-temporal data. Aggregation over space or time increases computational time by 30-40\% each due to the imputation of the underlying exact point pattern, in comparison to the model that uses the exact data. We find that the computational time is higher when the proportion of offspring events is higher, and that it grows linearly with the total number of events.

\section{Application}
\label{sec:application}

We apply our method to real data on insurgent violence incidents and airstrikes in Iraq. Insurgent violence events and airstrikes are expected to exhibit self-exciting behaviors and potentially cross-excitation \citep{lewis2011nonparametric}.

We focus on airstrikes and improvised explosive device (IED) attacks occurring between February 23, 2007, and June 30, 2008. 
Our data record the time and location (latitude and longitude) of each event.
Airstrikes conducted by the same airplane at the same location on the same day are considered as a single event. During the study period, there were 15,424 IED attacks, and 600 airstrikes. 
The spatial resolution for IED attacks is at the $10^{-5}$ scale in terms of longitude and latitude, corresponding to a precision of approximately 1 meter. The airstrike data also contain high-resolution spatial information, so the events are not considered aggregated in space. In contrast, the times of all airstrikes and 214 IED attacks are recorded only at the day level, with exact times unknown, which results in these events being temporally aggregated. To address this temporal aggregation, we apply our methodology. The remaining IED attacks are recorded with minute-level precision and are treated as not aggregated. 

\begin{figure}[!t]
\centering
\includegraphics[width = 0.55\textwidth,trim=0 0 0 50, clip]{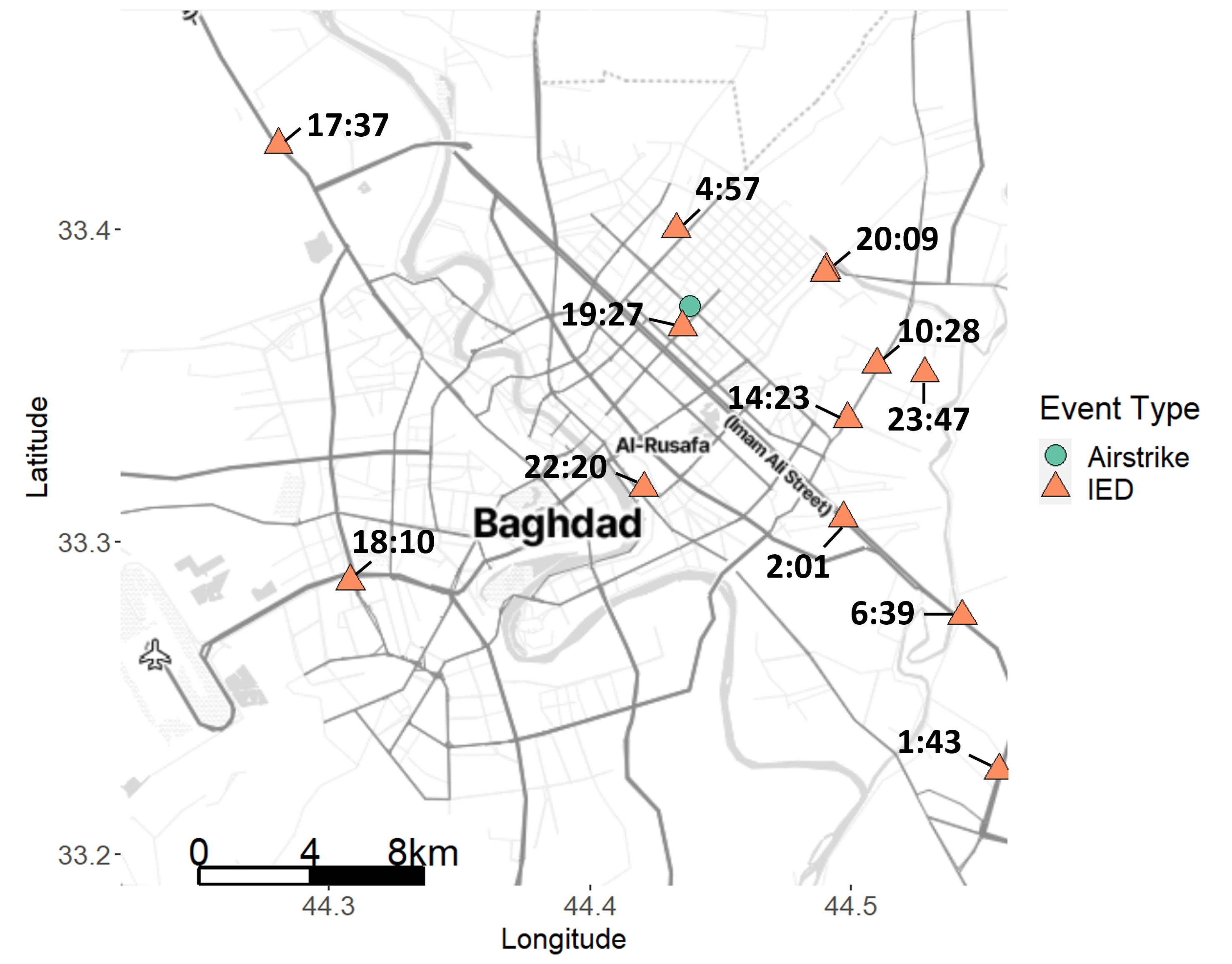}
\caption{Airstrikes and IED attacks in a bounded region of Iraq around Baghdad on May 1st, 2008. The green circle indicates the location of the one airstrike, and the orange triangles show the locations of the IED attacks. The exact time of the airstrike is unknown, and is not reported on the figure. Even though the exact time of these IED attacks are available in our data, the times shown in the figure are generated from a uniform distribution due to data privacy constraints.}
\label{fig:event_oneday}
\end{figure}

\cref{fig:event_oneday}
shows IED attacks and airstrikes on May 1, 2008 in an area around Baghdad in Iraq.

We consider both separate spatio-temporal aggregated Hawkes process models for IED attacks and airstrikes, as well as a joint model that accounts for cross-excitation between these two processes. We fit the models using events from February 23rd, 2007 to June 25th, 2008, with the last five days of data reserved as a holdout set.

\subsection{Data-driven specification of the background intensity}\label{subsec: app_specification}

\begin{figure}[!t]
\centering
\subfloat[Airstrikes]{
\includegraphics[width = 0.33\textwidth,trim = 0 0 0 0, clip]{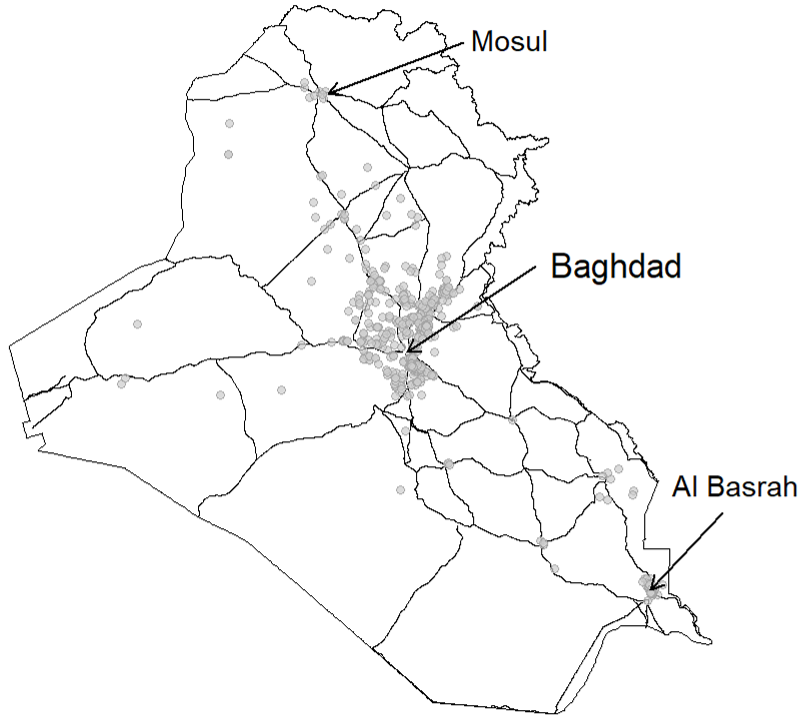}
\label{fig:airstrikes_spatial}
} 
\subfloat[IED attacks]{
\includegraphics[width = 0.33\textwidth,trim = 0 0 0 0, clip]{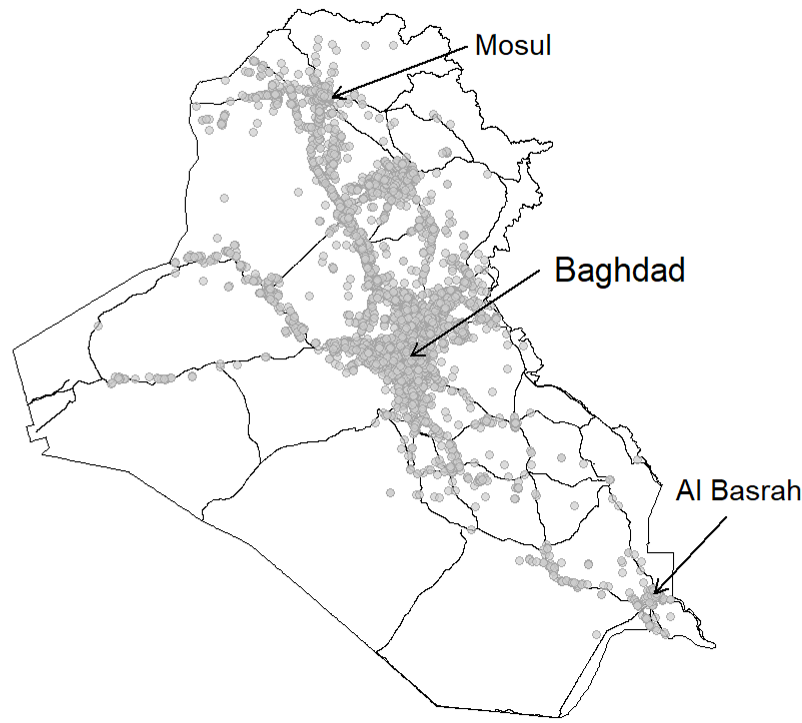}
\label{fig:IED_spatial}
}
\caption{Distribution of airstrikes and insurgent violence. Panel (a) and panel (b) show spatial locations of all airstrikes and IED attacks during February 23, 2007 to June 30, 2008, respectively.}
\label{fig:event_spatial}
%
%
%
\includegraphics[width = 0.98\textwidth,trim=0 15 0 20, clip]{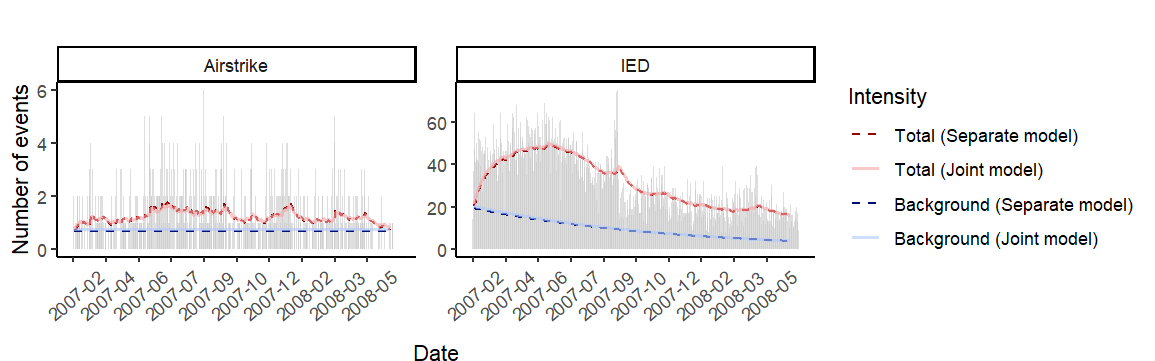}
\caption{Histogram of airstrikes and IED attacks. This histogram displays the frequency of airstrikes and IED attacks over time, grouped by days, across the study period. The dashed blue and pink lines represent the fitted conditional intensity and background intensity, respectively, estimated using the separate model. The solid lines indicate the fitted conditional intensity and background intensity as estimated by the joint model.}
\label{fig:fitted_intensity}
\end{figure}

\cref{fig:event_spatial} shows that most airstrikes and insurgent violence events concentrate close to the road network, and around the cities of Baghdad, Mosul and Al Basrah. Furthermore, \cref{fig:fitted_intensity} shows that the number of insurgent violence IED attacks might have an overall decreasing trend over time, whereas the pattern for the number of airstrikes remains relatively constant. We incorporate this information to design a realistic specification for the data's background intensity. 

First of all, we specify that the location of immigrant events of either type arise from a mixture distribution as
\begin{equation*}
    f(\sss) = \frac{\rho}{C_1}\exp\Big\{-\frac{d^2_{c_1}(\sss)}{2\gamma^2_{c_1}}-\frac{d^2_{r}(\sss)}{2\gamma^2_{r}}\Big\}+\frac{1 - \rho}{C_2}\exp\Big\{-\frac{d^2_{c_2}(\sss)}{2\gamma^2_{c_2}}-\frac{d^2_{r}(\sss)}{2\gamma^2_{r}}\Big\},
\end{equation*}
where $d_r(\sss)$ is the closest distance of point $\sss$ from the road network, $d_{c_1}(\sss)$ is its distance from Baghdad, and $d_{c_2}(\sss)$ is the closest distance of point $\sss$ from Mosul and Al Basrah. $C_1$ and $C_2$ are normalizing constants. One could consider a 3-way mixture distribution by allowing a separate distance from Mosul and Al Basrah. However, we find that the patterns of events near those two cities are similar, and therefore we consider them simultaneously. The parameter $\rho \in (0, 1)$ denotes the mixture parameter, specifying the relative prevalence of immigrants arising from one versus the other distribution. 
The prior distributions for the additional parameters are specified to be independent conjugate priors:
$
\gamma_r^2\sim \ig(0.001,0.001), 
\gamma_{c_1}^2\sim \ig(0.001,0.001),$ and $\gamma_{c_2}^2\sim \ig(0.001,0.001),$
and the prior for $\rho$ is assumed to be $\rho\sim \textrm{Beta}(1,1).$
To improve the efficiency of the MCMC algorithm, we introduce a latent variable $z \sim \textrm{Bernoulli}(\rho)$ conditional on which
%
$ f(\sss\mid z=1) = \frac{1}{C_1} \exp\Big\{-\frac{d^2_{c_1}(\sss)}{2\gamma^2_{c_1}}-\frac{d^2_{r}(\sss)}{2\gamma^2_{r}}\Big\} $ and 
$f(\sss\mid z=0) = \frac{1}{C_2}\exp\Big\{-\frac{d^2_{c_2}(\sss)}{2\gamma^2_{c_2}}-\frac{d^2_{r}(\sss)}{2\gamma^2_{r}}\Big\}.$   
%
Note that we only need the latent variable $z$ for immigrant events.

To accommodate the decreasing temporal trend in the number of IED attacks, we adopt the time-dependent function $\mu(t) = \mu\exp(-\eta t )$ where $\mu,\eta > 0$ for the IED attacks. 
We adopt a $\ga$ prior for the additional parameter $\eta$, as $\eta \sim \ga(1,0.01)$.
Since there is no obvious temporal trend for airstrikes,  we assume a constant background intensity for this process.
We adopt the exponential-Gaussian specification for the triggering kernel. The prior for the other parameters is specified in \cref{subsec:MCMC} and in \cref{supp_sec:multi_HP}.

We modify the sampling procedure described in \cref{subsec:MCMC} and in \cref{supp_sec:multi_HP} to sample from posterior distribution of all parameters under this model specification (see \cref{a: subsec: MCMC} for details). Our MCMC imputes of the exact time of all airstrikes and the 214 IED attacks with incomplete temporal information.

\subsection{Analyzing self and cross excitation of airstrikes and IED attacks}\label{subsec: app_result}

Estimated parameters for IED attacks (Process 1) and airstrikes (Process 2) from separate and joint models are shown in \cref{tab:estimates}. The estimates for the parameters representing background intensities and self-excitement are overall consistent between the separate and joint models. 
We observe that the estimated $\alpha$ for IED attacks is about 0.7, and $1/\beta$ is about 24.5 in both models. These results indicate that, on average, an IED attack triggers 0.7 IED attacks, and the average time to produce an event of the same type is 24.5 days. For airstrikes, the estimated value for $\alpha$ is 0.46, and for $1/\beta$ is 8.05 in the separate model, showing that, on average, an airstrike triggers 0.46 airstrikes, and the average time from one airstrike to an offspring airstrike is 8.05 days. The estimate for $\alpha$ for the airstrike process is slightly smaller, and $1/\beta$ is larger in the joint model, indicating a smaller number of offspring events and a longer time to produce these offspring events under this model. This change can be partially explained by the reclassification of events from immigrants to offspring and vice versa in the separate and joint models (see \cref{a:subsec:event_label}). The similarities in the estimates across the separate and joint models are also evident in \cref{fig:fitted_intensity}, which shows that the fitted total and background intensities are very similar.

The location parameter $\gamma$ controls the distance between parent and offspring events. We observe that the estimated value of $\gamma$ for IED attacks is about 0.003 in both models, and the estimated value of $\gamma$ for airstrikes is about 0.02 in both models. The unit for $\gamma$ is degrees, where 0.003 and 0.018 degrees correspond to approximately 0.33 and 2 kilometers, respectively. 
Interpreting the values of $\gamma_{c_1}, \gamma_{c_2}$ and $\gamma_r$ directly is not possible since they influence
the distribution of the location for immigrant events jointly. To have a better understanding of the magnitude of these values, we use the estimated background intensity from the joint model to generate points from the two distributions in the mixture $f(\sss)$ (see \cref{a:sec:scale_parameters} for details).

We observe that 90\% of airstrikes from the first distribution are within 194 kilometers of Baghdad and 13.4 kilometers of the road network. Additionally, 90\% of airstrikes from the second distribution are within 26.1 kilometers of Mosul or Al Basrah and 9.12 kilometers of the road network. For IED attacks from the first distribution, 90\% are within 304 kilometers of Baghdad and 10.9 kilometers of the road network, whereas from the second distribution, 90\% are within 19.3 kilometers of Mosul or Al Basrah and 7.39 kilometers of the road network. Since $\rho \approx 0.95$ for IED and $0.91$ for airstrikes, these results imply that immigrant IED attacks are generally closer to the road network and farther from the cities, compared to immigrant airstrikes.

\begin{table}[!t]
\centering
\caption{Results of the fitted model. The table shows the posterior mean (``Estimate''), lower bound (``Lower'') and upper bound  (``Upper'') of the 95\% credible interval for all parameters in the separate model and joint model on IED attacks and airstrikes. The subscript $_{ij}$ is used for cross-excitation parameters of Process $i$ towards Process $j$ where the index 1 is used for IED attacks and the index 2 is used for airstrikes. \\
}

\label{tab:estimates}
\scalebox{0.85}{
\begin{tabular}{ c c c c c c c c c} 
\toprule 
 & & \multicolumn{3}{c}{Separate} & \multicolumn{3}{c}{Joint}  \\
\cmidrule(lr){3-5} \cmidrule(lr){6-8}
&  &  Estimate & Lower & Upper  &  Estimate & Lower & Upper\\   
\midrule
\multirow{9}*{{IED}} 
& $\mu$ & 19.3113 & 18.2728 & 20.3480 & 20.4092 & 19.3353 & 21.5181 \\ 
& $\eta$ & 0.0033 & 0.0031 & 0.0035 & 0.0033 & 0.0031 & 0.0036 \\ 
& $\alpha$ & 0.7166 & 0.7022 & 0.7315 & 0.6957 & 0.6812 & 0.7101 \\ 
& $1/\beta$ & 24.7783 & 24.1420 & 25.4050 & 24.5517 & 23.8945 & 25.2318 \\
& $\gamma$ & 0.0033 & 0.0032 & 0.0034 & 0.0032 & 0.0031 & 0.0032 \\ 
& $\gamma_r$ & 0.0774 & 0.0757 & 0.0791 & 0.0668 & 0.0649 & 0.0689 \\ 
& $\gamma_{c_1}$ & 1.6522 & 1.6179 & 1.6878 & 1.5154 & 1.4776 & 1.5540 \\ 
& $\gamma_{c_2}$ & 0.6574 & 0.0866 & 1.8728 & 0.0853 & 0.0765 & 0.0953 \\ 
& $\rho$ & 0.9998 & 0.9991 & 1.0000 & 0.9526 & 0.9422 & 0.9633 \\ 
\hline
\multirow{8}*{{Airstrikes}} 
& $\mu$ & 0.6709 & 0.5833 & 0.7631 & 0.7433 & 0.6563 & 0.8353 \\ 
& $\alpha$ & 0.4611 & 0.3968 & 0.5310 & 0.3907 & 0.3288 & 0.4555 \\ 
& $1/\beta$ & 8.0530 & 5.8262 & 10.6667 & 9.8231 & 7.2627 & 13.5827 \\
& $\gamma$ & 0.0205 & 0.0174 & 0.0235 & 0.0180 & 0.0155 & 0.0212 \\ 
& $\gamma_{r}$ & 0.1075 & 0.0993 & 0.1168  & 0.0852 & 0.0785 & 0.0923 \\ 
& $\gamma_{c_1}$ & 1.0773 & 0.9838 & 1.1820 & 0.8902 & 0.7984 & 0.9864 \\ 
& $\gamma_{c_2}$ & 0.1188 & 0.0873 & 0.1620 & 0.1162 & 0.0894 & 0.1528 \\ 
& $\rho$ & 0.9239 & 0.8910 & 0.9523 & 0.9055 & 0.8713 & 0.9366 \\ 
\hline
\multirow{6}*{{Cross}} 
& $\alpha_{21}$ &  &  &  & 0.1235 & 0.0838 & 0.1679 \\ 
& $\alpha_{12}$ &  &  &  & 0.0005 & 0.0001 & 0.0010 \\ 
& $1/\beta_{21}$ &  &  &  & 13.2527 & 5.8188 & 27.7880 \\ 
& $1/\beta_{12}$ &  &  &  & 1.8551 & 0.4228 & 41.0310 \\
& $\gamma_{21}$ &  &  &  & 1.4697 & 1.2036 & 1.7674 \\ 
& $\gamma_{12}$ &  &  &  & 1.6742 & 1.0483 & 1.9883 \\ 
\bottomrule
\end{tabular}
}
\end{table}

We observe relatively weak cross-excitation between IED attacks and airstrikes. The estimated value for $\alpha_{12}$ is close to 0, indicating that nearly no airstrikes are triggered by IED attacks. The estimated values for $\alpha_{21}$ and $1/\beta_{21}$ show that, on average, one airstrike triggers approximately 0.12 IED attacks, and it takes 13 days for this to occur. When comparing the estimates for the location parameter in the self- and cross-triggering kernel for airstrikes in the joint model, we see that the cross-excitation location parameter $\gamma_{21}$ is much larger than the self excitation parameter $\gamma_{22}$ (the former is 1.4697 while the latter is 0.018 degrees, corresponding to 163.58 and 2 kilometers). This result indicates that while an airstrike is often followed by an offspring airstrike nearby, it triggers IED attacks that can be relatively far away. 
The matrix of the parameters $\alpha$ had spectral radius below 1 across all posterior samples, indicating stationarity of the multivariate Hawkes process.

We analyze the number of parent-offspring pairs that occurred in the same or in a different day to check estimability of the corresponding parameters. Since the time of IED attacks is known exactly for most events, we focus on parent-offspring pairs that involve airstrikes  and parameters $\beta_{12},\beta_{21},\beta_{22}.$ We observe that the number of parent-offspring pairs in different days is relatively large when the parent event is an airstrike, suggesting no estimability concerns with $\beta_{21}$ and $\beta_{22}.$ However, the number of parent-offspring pairs is very small when the parent is an IED attack and offspring is an airstrike (since $\alpha_{12} \approx 0$), indicating that the estimate for $\beta_{12}$ might not be interpretable (See \cref{a:subsec: parent_offspring} for details).

\subsection{Forecasting the number of airstrikes and IED attacks}\label{subsec: app_forecast}

We examine the separate and joint models' forecasting ability during the following five days (June 26-30, 2008) and over five regions around cities in Iraq. 
Model assessment for point process models was introduced in \cite{leininger2017bayesian} employing the posterior predictive distribution. In our case, the posterior predictive distribution is defined as
$p(\tilde{\bm x}|\bm N)  = \int p(\tilde{\bm x}|\bm \theta,\bm x)p(\bm x,\bm \theta|\bm N) \dd \bm{x}\dd\bm \theta,$ where $\tilde{\bm x}$ is the predicted data, $\bm N$ is the observed data and $\bm x$ is the corresponding latent exact point pattern. We draw forecasts $\tilde{\bm x}$ by iteratively sampling the parameters $\bm \theta$ and the latent point pattern $\bm x$ from their posterior distribution, and imputing $\tilde{\bm x}$ from $p(\tilde{\bm x}|\bm \theta,\bm x)$ using Algorithm 1 in \cite{tucker2019handling}. Then, we forecast the number of events in a certain time period or region, by counting the number of predicted events that fall in that window. We compare our forecasts to the observed data.

\begin{figure}[!t]
\centering
\vspace{-10pt}
\includegraphics[width = 7cm]{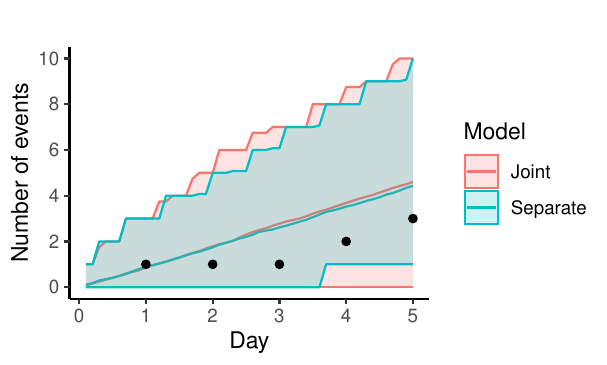}
\caption{Number of airstrikes over time. The x-axis represents the period of time starting at June 26, 2008, while the y-axis indicates the number of airstrikes. The solid line shows the posterior mean of the predicted value, and the shaded region represents the 95\% credible intervals based on the separate (blue) and joint (red) models. The black points indicate the true number of events observed.}
\label{fig: AIR_forecast_time}
%
\includegraphics[width = 12cm]{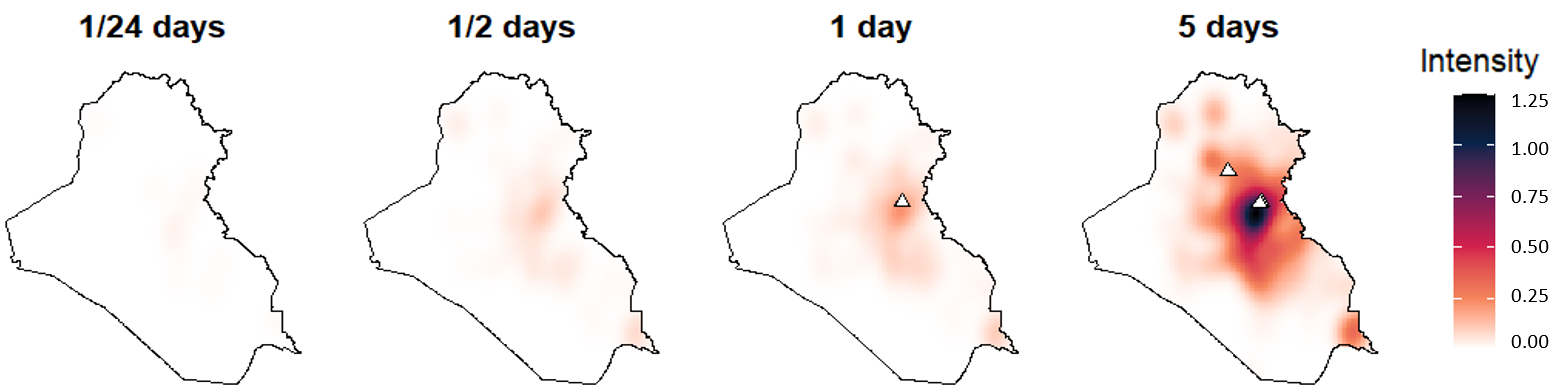}
\caption{Total estimated airstrike intensity over Iraq for during 1/24, 1/2, 1, and 5 days after June 25, 2008 using the separate model. White triangles indicate the locations of observed airstrikes during the same time period. The observed airstrike illustrated for day 1, could have occurred during any part of the day.}
\label{fig: AIR_forecast_location}

\end{figure}

Although airstrike times are aggregated by days, our model can forecast the number of events over any time period. \cref{fig: AIR_forecast_time} shows the estimated event counts starting on June 26, 2008, with 95\% credible intervals from the separate and joint models, alongside observed counts. Both models produce similar predictions, with the realized counts falling within the credible intervals. 
We also examine areas where airstrikes are likely to concentrate. \cref{fig: AIR_forecast_location} shows the estimated total intensity of airstrikes over 1/24, 1/2, 1, and 5 days following June 25, 2008, by applying kernel density estimation to the posterior predictive samples in the separate model. We find that the locations of observed airstrikes coincide with areas of high intensity.

In \cref{appendix: prediction}, we provide an extensive investigation of the models' predictive accuracy as a function of time and space. The models produce similar forecasts, and they predict the total number of airstrikes and IED attacks in the country very accurately. Regionally, they perform overall well in forecasting the number of events in the different regions of Iraq, except they tend to over-estimate the number of IED attacks in the region between Baghdad and Al Basrah.

\section{Discussion}
\label{sec: discussion}


Additional work on aggregated point pattern data can extend in various directions.
Firstly, extensions to accommodate edge effects in time and space can be investigated. Events that happen outside the observed temporal and spatial window
can affect parameter estimates. Edge correction methods for exact data are discussed in \cite{lapham2014hawkes}, \cite{diggle2013statistical} and \cite{cronie2011some}. Extending these to the context of aggregated point processes can be explored in future work.
Secondly, an interesting extension of our work could target marked Hawkes processes where features of events are  available and might drive the process dynamics. For example, in our study of \cref{sec:application}, the number of civilian casualties caused by an event, if available, can be included as a mark in the model. In such cases, one might want to allow an event's number of offspring $\alpha$ to depend on the mark, since the number of casualties it caused may affect the number of events it can incite. Doing so would be of explicit interest and it could complicate estimation.
Thirdly, it would be interesting to investigate  nonparametric Hawkes process models in the presence of aggregated point pattern data, which are used in many applications for extra flexibility. For instance, \cite{lewis2011nonparametric}, \cite{fox2016spatially} and \cite{marsan2008extending} introduce ways to incorporate nonparametric inhomogeneous $\mu(t)$, and \cite{zhou2013learning} and \cite{kirchner2018nonparametric} use
kernel density estimation for the offspring intensity. Moreover, the multivariate Hawkes process has been extended to allow for inhibition effects within a Bayesian estimation procedure \citep{deutsch2022bayesian}, which can be explored within the realm of aggregated data. Lastly, it is of interest to us to investigate the implications of using aggregated point pattern data for learning causal dependencies among processes \citep{papadogeorgou2022causal}, especially when these effects might manifest almost instantaneously in time.

\bibliographystyle{natbib}
\bibliography{Point-processes}

\begin{thebibliography}{}

\bibitem[Aldor-Noiman \emph{et~al.}(2016)Aldor-Noiman, Brown, Fox, and
  Stine]{aldor2016spatio}
Aldor-Noiman, S., Brown, L.~D., Fox, E.~B., and Stine, R.~A. (2016).
\newblock Spatio-temporal low count processes with application to violent crime
  events.
\newblock \emph{Statistica Sinica}  1587--1610.

\bibitem[Br{\'e}maud and Massouli{\'e}(1996)]{bremaud1996stability}
Br{\'e}maud, P. and Massouli{\'e}, L. (1996).
\newblock Stability of nonlinear hawkes processes.
\newblock \emph{The Annals of Probability}  1563--1588.

\bibitem[Bu \emph{et~al.}(2022)Bu, Aiello, Xu, and Volfovsky]{bu2022likelihood}
Bu, F., Aiello, A.~E., Xu, J., and Volfovsky, A. (2022).
\newblock Likelihood-based inference for partially observed epidemics on
  dynamic networks.
\newblock \emph{Journal of the American Statistical Association} \textbf{117},
  537, 510--526.

\bibitem[Chiang \emph{et~al.}(2022)Chiang, Liu, and Mohler]{chiang2022hawkes}
Chiang, W.-H., Liu, X., and Mohler, G. (2022).
\newblock Hawkes process modeling of covid-19 with mobility leading indicators
  and spatial covariates.
\newblock \emph{International journal of forecasting} \textbf{38}, 2, 505--520.

\bibitem[Cronie and S{\"{a}}rkk{\"{a}}(2011)]{cronie2011some}
Cronie, O. and S{\"{a}}rkk{\"{a}}, A. (2011).
\newblock {Some edge correction methods for marked spatio-temporal point
  process models}.
\newblock \emph{Computational Statistics and Data Analysis} \textbf{55}, 7,
  2209--2220.

\bibitem[Daley and Vere-Jones(2003)]{daley2003introduction}
Daley, D.~J. and Vere-Jones, D. (2003).
\newblock \emph{An introduction to the theory of point processes: volume I:
  elementary theory and methods}.
\newblock Springer.

\bibitem[Darolles \emph{et~al.}(2019)Darolles, Le~Fol, Lu, and
  Sun]{darolles2019bivariate}
Darolles, S., Le~Fol, G., Lu, Y., and Sun, R. (2019).
\newblock Bivariate integer-autoregressive process with an application to
  mutual fund flows.
\newblock \emph{Journal of Multivariate Analysis} \textbf{173}, 181--203.

\bibitem[Darzi \emph{et~al.}(2023)Darzi, Halldorsson, Hrafnkelsson, Ebrahimian,
  Jalayer, and Vogfj{\"o}r{\dh}]{darzi2023calibration}
Darzi, A., Halldorsson, B., Hrafnkelsson, B., Ebrahimian, H., Jalayer, F., and
  Vogfj{\"o}r{\dh}, K.~S. (2023).
\newblock Calibration of a bayesian spatio-temporal etas model to the june 2000
  south iceland seismic sequence.
\newblock \emph{Geophysical Journal International} \textbf{232}, 2, 1236--1258.

\bibitem[Deutsch and Ross(2022)]{deutsch2022bayesian}
Deutsch, I. and Ross, G.~J. (2022).
\newblock Bayesian estimation of multivariate hawkes processes with inhibition
  and sparsity.
\newblock \emph{arXiv preprint arXiv:2201.05009} .

\bibitem[Diggle(2013)]{diggle2013statistical}
Diggle, P.~J. (2013).
\newblock \emph{Statistical analysis of spatial and spatio-temporal point
  patterns}.
\newblock CRC press.

\bibitem[Fox \emph{et~al.}(2016)Fox, Schoenberg, and Gordon]{fox2016spatially}
Fox, E.~W., Schoenberg, F.~P., and Gordon, J.~S. (2016).
\newblock Spatially inhomogeneous background rate estimators and uncertainty
  quantification for nonparametric hawkes point process models of earthquake
  occurrences.
\newblock \emph{The Annals of Applied Statistics} \textbf{10}, 3, 1725--1756.

\bibitem[Gelman and Rubin(1992)]{gelman1992inference}
Gelman, A. and Rubin, D.~B. (1992).
\newblock Inference from iterative simulation using multiple sequences.
\newblock \emph{Statistical science} \textbf{7}, 4, 457--472.

\bibitem[Hawkes(1971)]{hawkes1971spectra}
Hawkes, A.~G. (1971).
\newblock Spectra of some self-exciting and mutually exciting point processes.
\newblock \emph{Biometrika} \textbf{58}, 1, 83--90.

\bibitem[Hawkes and Oakes(1974)]{hawkes1974cluster}
Hawkes, A.~G. and Oakes, D. (1974).
\newblock A cluster process representation of a self-exciting process.
\newblock \emph{Journal of Applied Probability} \textbf{11}, 3, 493--503.

\bibitem[Kirchner(2016)]{kirchner2016hawkes}
Kirchner, M. (2016).
\newblock Hawkes and {INAR} ($\infty$) processes.
\newblock \emph{Stochastic Processes and their Applications} \textbf{126}, 8,
  2494--2525.

\bibitem[Kirchner(2017)]{kirchner2017estimation}
Kirchner, M. (2017).
\newblock An estimation procedure for the hawkes process.
\newblock \emph{Quantitative Finance} \textbf{17}, 4, 571--595.

\bibitem[Kirchner and Bercher(2018)]{kirchner2018nonparametric}
Kirchner, M. and Bercher, A. (2018).
\newblock A nonparametric estimation procedure for the hawkes process:
  comparison with maximum likelihood estimation.
\newblock \emph{Journal of Statistical Computation and Simulation} \textbf{88},
  6, 1106--1116.

\bibitem[Lapham(2014)]{lapham2014hawkes}
Lapham, B.~M. (2014).
\newblock \emph{Hawkes processes and some financial applications}.
\newblock Master's thesis, University of Cape Town.

\bibitem[Leininger and Gelfand(2017)]{leininger2017bayesian}
Leininger, T.~J. and Gelfand, A.~E. (2017).
\newblock Bayesian inference and model assessment for spatial point patterns
  using posterior predictive samples .

\bibitem[Lewis and Mohler(2011)]{lewis2011nonparametric}
Lewis, E. and Mohler, G. (2011).
\newblock A nonparametric em algorithm for multiscale hawkes processes.
\newblock \emph{Journal of Nonparametric Statistics} \textbf{1}, 1, 1--20.

\bibitem[Lewis \emph{et~al.}(2012)Lewis, Mohler, Brantingham, and
  Bertozzi]{lewis2012self}
Lewis, E., Mohler, G., Brantingham, P.~J., and Bertozzi, A.~L. (2012).
\newblock Self-exciting point process models of civilian deaths in iraq.
\newblock \emph{Security Journal} \textbf{25}, 3, 244--264.

\bibitem[Marsan and Lengline(2008)]{marsan2008extending}
Marsan, D. and Lengline, O. (2008).
\newblock Extending earthquakes' reach through cascading.
\newblock \emph{Science} \textbf{319}, 5866, 1076--1079.

\bibitem[Ogata(1998)]{ogata1998space}
Ogata, Y. (1998).
\newblock Space-time point-process models for earthquake occurrences.
\newblock \emph{Annals of the Institute of Statistical Mathematics}
  \textbf{50}, 2, 379--402.

\bibitem[Papadogeorgou \emph{et~al.}(2022)Papadogeorgou, Imai, Lyall, and
  Li]{papadogeorgou2022causal}
Papadogeorgou, G., Imai, K., Lyall, J., and Li, F. (2022).
\newblock Causal inference with spatio-temporal data: estimating the effects of
  airstrikes on insurgent violence in iraq.
\newblock \emph{Journal of the Royal Statistical Society Series B: Statistical
  Methodology} \textbf{84}, 5, 1969--1999.

\bibitem[Porter and White(2012)]{porter2012self}
Porter, M.~D. and White, G. (2012).
\newblock Self-exciting hurdle models for terrorist activity.
\newblock \emph{The Annals of Applied Statistics} \textbf{6}, 1, 106--124.

\bibitem[Rasmussen(2013)]{Rasmussen2013bayesian}
Rasmussen, J.~G. (2013).
\newblock {Bayesian Inference for Hawkes Processes}.
\newblock \emph{Methodology and Computing in Applied Probability} \textbf{15},
  623--642.

\bibitem[Reinhart(2018)]{Reinhart2018review}
Reinhart, A. (2018).
\newblock {A Review of Self-Exciting Spatio-Temporal Point Processes and Their
  Applications}.
\newblock \emph{Statistical Science} \textbf{33}, 3, 299----318.

\bibitem[Ross(2021)]{ross2021bayesian}
Ross, G.~J. (2021).
\newblock Bayesian estimation of the etas model for earthquake occurrences.
\newblock \emph{Bulletin of the Seismological Society of America} \textbf{111},
  3, 1473--1480.

\bibitem[Schoenberg(2013)]{schoenberg2013facilitated}
Schoenberg, F.~P. (2013).
\newblock Facilitated estimation of etas.
\newblock \emph{Bulletin of the Seismological Society of America} \textbf{103},
  1, 601--605.

\bibitem[Shlomovich \emph{et~al.}(2022{a})Shlomovich, Cohen, and
  Adams]{Shlomovich2022parameterb}
Shlomovich, L., Cohen, E.~A., and Adams, N. (2022{a}).
\newblock A parameter estimation method for multivariate binned hawkes
  processes.
\newblock \emph{Statistics and Computing} \textbf{32}, 98.

\bibitem[Shlomovich \emph{et~al.}(2022{b})Shlomovich, Cohen, Adams, and
  Patel]{shlomovich2022parameter}
Shlomovich, L., Cohen, E.~A., Adams, N., and Patel, L. (2022{b}).
\newblock Parameter estimation of binned hawkes processes.
\newblock \emph{Journal of Computational and Graphical Statistics}  1--11.

\bibitem[Tucker \emph{et~al.}(2019)Tucker, Shand, and
  Lewis]{tucker2019handling}
Tucker, J.~D., Shand, L., and Lewis, J.~R. (2019).
\newblock Handling missing data in self-exciting point process models.
\newblock \emph{Spatial statistics} \textbf{29}, 160--176.

\bibitem[Veen and Schoenberg(2008)]{veen2008estimation}
Veen, A. and Schoenberg, F.~P. (2008).
\newblock Estimation of space--time branching process models in seismology
  using an em--type algorithm.
\newblock \emph{Journal of the American Statistical Association} \textbf{103},
  482, 614--624.

\bibitem[Zhou \emph{et~al.}(2020)Zhou, Li, Fan, Wang, Sowmya, and
  Chen]{zhou2020efficient}
Zhou, F., Li, Z., Fan, X., Wang, Y., Sowmya, A., and Chen, F. (2020).
\newblock Efficient inference for nonparametric hawkes processes using
  auxiliary latent variables.
\newblock \emph{Journal of Machine Learning Research} \textbf{21}, 241, 1--31.

\bibitem[Zhou \emph{et~al.}(2013)Zhou, Zha, and Song]{zhou2013learning}
Zhou, K., Zha, H., and Song, L. (2013).
\newblock Learning triggering kernels for multi-dimensional hawkes processes.
\newblock In \emph{International conference on machine learning},  1301--1309.
  PMLR.

\end{thebibliography}

\newpage

\doparttoc 
\faketableofcontents 
\part{} 

\setcounter{page}{1}

\vspace{-20pt}
\begin{center}
{\sc \LARGE Supplementary Appendix for ``Bayesian inference for aggregated Hawkes processes''}
\end{center}

\allowdisplaybreaks
\appendix
\setcounter{equation}{0}
\renewcommand{\theequation}{A.\arabic{equation}}
\setcounter{table}{0}
\renewcommand{\thetable}{A.\arabic{table}}
\setcounter{figure}{0}
\renewcommand{\thefigure}{A.\arabic{figure}}

\AtAppendix{\counterwithin{theorem}{section}}

\vspace{-60pt}
\addcontentsline{toc}{section}{Supplement} 
\part{ } 
\parttoc
\clearpage

\setstretch{1.3}

\section{Table of notation}

Definitions for the notation used in the manuscript are given in \cref{supp_tab:notation}.

\begin{table}[H]
\caption{Notation}
\vspace{1mm}
\label{supp_tab:notation}
\begin{tabularx}{\textwidth}{@{}p{5cm}X@{}}
\toprule
  $i,j$ & Indices for events\\
  $k$ & Indices for temporal bins\\
  $r$ & Indices for space bins\\
  $m,l$ & Indices for dimensions of processes\\
  $X = \{(t_i,s_i)\}$&  A spatio-temporal process\\
  $\ttheta$ & Parameter vector for marked Hawkes process\\
  $\bm Y$& Branching structure\\
  $I$ & Collection of all immigrants \\
  $O_j$ & Collection of all offspring indices of events $j$ \\
  $\alpha$   & Total offspring intensity\\
  $g(t,\sss)$   & Normalized offspring intensity\\
  $\Delta^t$    & Aggregation size in time\\
  $\Delta^{\sss}$    & Aggregation size in space\\
  $B^t_k$   & $k^{th}$ Time bins\\
  $B^{\bm s}_r$  & $r^{th}$ Space bins\\
  $N_{k,r}$   & Number of events in $B^t_k\times B^{\sss}_r$ \\
\bottomrule
\end{tabularx}
\end{table}


\section{Latent variable formulation for the Hawkes process with aggregated data}

In all cases, the latent exact point pattern has to lead to aggregated data that match the observed ones. Therefore, the aggregated data need to ``agree'' with the imputed exact point pattern. For spatio-temporal Hawkes processes, the required condition for $p(\bm N \mid \bm x, \bm Y, \bm \theta)$ is given in equation (11). We introduce the equivalent conditions for aggregated temporal and aggregated multivariate spatio-temporal Hawkes processes here.

\subsection{Aggregated temporal Hawkes process}

For temporal data, $\bm N = \{N_1, N_2, \dots, N_K\}$ where $N_k$ represents the number of observed events in bin $B_k^t$. The required condition for the agreement of exact and aggregated data becomes:
\[
p(\bm N \mid \bm x, \bm Y, \bm \theta) =
\begin{cases}
1, & \text{if } \sum_{i = 1}^n 1\{t_i \in B_k^t\} = N_k, \text{ for all } k, \\
0, & \text{otherwise}.
\end{cases}
\]

\subsection{Aggregated multivariate spatio-temporal Hawkes process}
\label{supp_subsec:multi_notation}

For multivariate spatio-temporal data, we need to introduce some of the notation first.
Suppose we observe $L$ aggregated Hawkes processes on $[0, T]\times W$. Assuming the same notations in Section 3.1 of the manuscript, let $x = \{(t_{l,i},\sss_{l,i}) : l=1,\dots,L, i = 1,\dots, n_{l}\}$ represent the underlying exact data which is unobserved, where $n_l$ is the number of events from process $l$, and $\bm Y = \{Y_{l,i}:l = 1,\dots,L, i = 1,\dots,n_l\}$ be the latent branching structure. The temporal bins and spatial bins corresponding to the aggregations are defined similarly as in Section 3.1 in the manuscript. We allow the aggregations to differ for different processes. Specifically, for process $l$, we have temporal bins $B^t_{l,1},\dots,B^t_{l,K_l}$ and spatial bins $B^{\sss}_{l,1},\dots, B^{\sss}_{l,R_l}.$ Then the observed data is $\bm N = \{N_{l,k,r}:l=1,\dots,L,k=1,\dots,K_l,r = 1,\dots,R_l\}$ where $N_{l,k,r}$ is the number of events in spatio-temporal bin $B^t_{l,k}\times B^{\sss}_{l,r}.$

In this case, the distribution of aggregated data $p(\bm N|\bm x, \bm Y,\theta)$ is equal to 
\[
p(\bm N \mid \bm x, \bm Y, \bm \theta) =
\begin{cases}
1, & \text{if } \sum_{k = 1}^{K_l}\sum_{r = 1}^{R_l} N_{l,k,r} = n_l \text{ and } \\
& \hspace{20pt}
\sum_{i = 1}^{n_{l}} = 1{\{(t_{l,i},s_{l,i})\in B_{l,k}^t\times B_{l,r}^{\sss}\}} = N_{l,k,r}, \text{ for all } l,k,r, \\
0, & \text{otherwise}.
\end{cases}
\]
Similar to Section 3.2 of the manuscript, we need to sample from $p(\bm x|\bm Y, \theta)$ and $p(\theta,\bm Y|\bm x)$ iteratively.

\section{Supplementary Estimation Procedures for Hawkes Processes} \label{a: sec: MCMC}

In Supplement \ref{supp_sec:temporal_HP} and \ref{supp_sec:multi_HP}, we provide details on the prior choices and estimation technique for temporal and multivariate HPs under a specific parametric similar to the specification introduced in Section 3.5. In Supplement \ref{a: subsec: MCMC}, we discuss the MCMC steps under other specifications used in the main manuscript.

\subsection{Estimation for temporal Hawkes processes within the Bayesian framework}\label{supp_sec:temporal_HP}

In this section, we introduce the prior choices and MCMC updates for the temporal Hawkes process. The temporal Hawkes process can be viewed as a special case of a spatio-temporal Hawkes process with $g_2(\cdot)\propto 1.$ Thus, the MCMC updates for a temporal Hawkes process are similar to these introduced in Section 3.5.

We adopt (independent) priors,
$\mu\sim\ga (a_1,b_1),$ $\alpha\sim\text{Trunc}\ga (a_2,b_2),$ and $\beta\sim\ga (a_3,b_3)$, where the second argument is the rate parameter and $\text{Trunc}\ga$ denotes the truncated gamma distribution on $(0,1)$. 
In all simulations, we choose $a_1 = a_2 = 1$ and $b_1 = b_2 = 0.1$ and for the prior on $\mu$ and $\alpha$. Moreover, we set $a_3 = 1$ and take $b_3 = 0.1$ for the prior on $\beta.$ 

Let $\bm x = \{t_1, t_2, \dots,t_n\}$ be a realization of a Hawkes process on the temporal window $[0,T)$ with $n\in\mathbb{Z}^+$ number of events. 
We use Gibbs sampling for parameters $\mu$ and  $\alpha$, and for the latent branching structure $\bm Y = (Y_1,Y_2, \dots,Y_n)$.
Specifically, we iteratively sample from
\begin{align*}
 \mu|\bm{x},\bm \theta_{-\mu},\bm{Y}, \bm N &\sim \ga \big(a_1+|I|,b_1+T \big) \\
 \alpha|\bm{x},\bm \theta_{-\alpha},\bm{Y}, \bm N &\sim \text{Trunc} \ga \big(a_2+\sum_{j=1}^n|O_j|,b_2+n-\sum_{j=1}^n \exp(-\beta(T-t_j))\big),
\end{align*}
where $\bm \theta = (\mu, \alpha, \beta)$ and $\bm \theta_{-(\ )}$ excludes the one in the subscript.
The full conditional distribution of $Y_i$ is a categorical distribution with
\begin{align*}
      &P( Y_i = j \mid \bm \theta,\bm{x},\bm{N}) \propto 
      \begin{cases} 
      \mu, & j= 0, \\
      \alpha\beta \exp \left( - \beta(t_i-t_j)\right)
      , & j=1,2, \dots, i-1, \\
      0, & j = i, i +1, \dots, n.
   \end{cases}  
\end{align*}

We use a Metropolis-Hasting step to sample $\beta$ from its posterior conditional distribution. Specifically, we propose value $\beta'$ from a normal distribution with mean $\beta^c$ and standard deviation $\sigma_\beta$, where superscripts $c$ denotes the current value and $'$ denotes the proposed value. We accept the move with probability
\begin{align*}
H_{\beta} &= 
\min \Big(1,\ (\beta'/\beta^c)^{a_3-1}\exp(b_3(\beta^c-\beta'))\\
& \hspace{60pt} \times\prod_{j = 1}^n\big\{\exp\Big(\alpha\exp(-\beta'(T-t_j))-\alpha\exp(-\beta^c(T-t_j))\Big) \\
& \hspace{90pt} \times \prod_{i\in O_j}(\beta'/\beta^c)\exp\Big(-(\beta'-\beta^c)(t_i-t_j)\Big)\big\}\Big)
\end{align*}
The value of $\sigma_\beta$ controls the acceptance rate of $\beta$. For all the simulation and application analysis, we choose $\sigma_\beta$ such that the acceptance rate is between 20\% and 40\%.

We also use Metropolis-Hasting updates for the latent times. For $i = 1,2\dots, n$, the proposed time $t_i'$ is drawn from a continuous uniform distribution that satisfies the following restrictions: 
\begin{enumerate*}[label=(\alph*)]
\item $t_i'$ is within the time bin of event $i$,
\item if event $i$ is an offspring of event $j$ ($Y_i= j$), then it occurs after its parent, $t_i' > t_j$, and similarly
\item it occurs before all of its offspring (if any).
\end{enumerate*}
By imposing these restrictions, we ensure that the proposed latent time agrees with the observed counts in the temporal bins. The proposed value of $t_i$ is accepted with probability
\begin{align*}
&H_{t_i} = \min\Big(1,\ \exp\Big(\alpha\exp(-\beta(T-t_i'))-\alpha\exp(-\beta(T-t_i^c)))\\
& \hspace{85pt}+(|O_i|-1{\{Y_i>0\}})\beta(t_i'-t_i^c)\Big)\Big).
\end{align*}

\subsection{Estimation for multivariate Hawkes processes within the Bayesian framework}
\label{supp_sec:multi_HP}

We discuss prior choice and MCMC for the multivariate HP under the exponential and Gaussian specification for the offspring densities.
The MCMC sampling scheme discussed below corresponds to the formulation where the background intensity is constant over time, the immigrant distribution is uniform over the observed window, and exponential and Gaussian densities are chosen for the temporal and spatial components of the excitation function, respectively, for all components of the multivariate HP.
Specifically, we consider the following parametric form:
\begin{align*}
\mu_{l}(t) &= \mu_l \\
g_{1,m,l}(t) &= \beta_{m,l} \exp \{- \beta_{m,l} t \} \\
g_{2, m,l}(\sss) &= \frac1{\sqrt{2\pi \gamma_{m,l}^2}} \exp\left\{-\frac{ \| \sss\|^2} { 2\gamma_{m,l}^2} \right\}
\end{align*}
Then, the parameters for the continuous Hawkes process are $\ttheta = (\bm \mu,\bm\alpha,\bm\beta,\bm\gamma)$, where $\bm\mu = (\mu_l: l = 1,\dots,L)$, $\bm\alpha = (\alpha_{m,l}:m,l = 1,\dots,L)$, $\bm\beta = (\beta_{m,l}:m,l = 1,\dots,L)$, $\bm\gamma = (\gamma_{m,l}:m,l = 1,\dots,L)$. 

\subsubsection{Prior distributions}

For $m,l\in\{1,\dots,L\}$, we assume (independent) gamma priors for $\mu_l$, $\alpha_{m,l}$ and $\beta_{m,l}$, and inverse gamma priors for $\gamma^2_{m,l}$:
\begin{align*}
& \mu_{l}\sim\ga (a_{l,1},b_{l,1}), 
&& \alpha_{m,l}\sim\text{Trunc}\ga (a_{m,l,2},b_{m,l,2}),\, \\
& \beta_{m,l}\sim\ga (a_{m,l,3},b_{m,l,3}), 
&& \gamma_{m,l}^2\sim \ig(a_{m,l,4},b_{m,l,4}).
\end{align*}
where the truncation of the Gamma prior distribution on $\alpha_{m,l}$ imposes that $P(\alpha_{m,l} < 1)$ = 1. In the end of Supplement \ref{supp_sec:multi_HP}, we discuss an alternative prior on the parameters $\alpha$ that imposes a multivariate truncation, and makes these parameters a priori dependent.

For $m,l \in \{1,\dots,L\}$, we choose $a_{l,1} = a_{m,l,2} = 1$, $b_{m,l,1} = b_{m,l,2} = 0.1$, and $a_{m,l,4} = b_{m,l,4} = 0.001$  throughout. We set $a_{m,l,3} = 1$ and take $b_{m,l,3} = 0.1$ for the simulations and $b_{m,l,3} = 1$ for our application.
Therefore, the priors for the parameters of the multivariate Hawkes Process match the ones in Section 3.5 for the spatio-temporal Hawkes process.

\subsubsection{Conditional distribution of latent exact data}

Given the latent branching structure and model parameters, the distribution of exact data (under the exponential-Gaussian specification of the excitation function) is
\begin{align*}
    p(\bm{x}|\bm{Y},\theta) 
    &=\prod_{l = 1}^L \exp(-\mu_{l} T)(\mu_{l}|W|^{-1})^{|I_l|}\\
    &\hspace{10pt} \times\prod_{m = 1}^L\prod_{i=1}^{n_m}\prod_{l = 1}^L\Big\{\exp\big \{ -\alpha_{m,l} \{ 1-\exp[-\beta_{m,l}(T-t_{m,i})]\} \big\}\\
    &\hspace{10pt} \times\prod_{t_{l,j}\in O_{m,i}}\frac{\alpha_{m,l}\beta_{m,l}}{\sqrt{2\pi\gamma_{m,l}^2}}\exp\Big[-\beta_{m,l}(t_{l,j}-t_{m,i})-\frac{\|\sss_{l,j}-\sss_{m,i}\|^2}{2\gamma_{m,l}^2}\Big] \Big\},
\end{align*}
where $I_l$ denotes the set of immigrants in process $l$ and $O_{m,i}$ denotes the set of offspring of event $i$ in process $m$.

\subsubsection{MCMC updates}

We use Gibbs sampling for each of the parameter in $\theta$ and  the latent branching structure $\bm Y$.
Specifically, we iteratively sample from the parameters according to the distributions below:
\begin{align*}
 \mu_{l}|\bm{x},\theta_{-\mu_{l}},\bm{Y}, \bm N &\sim \ga \big(a_{l,1}+|I_l|,b_{l,1}+T \big) \\
 \alpha_{m,l}|\bm{x},\theta_{-\alpha_{m,l}},\bm{Y}, \bm N &\sim \text{Trunc}\ga \big(a_{m,l,2}+\sum_{i=1}^{n_m}|O_{m,i}|,\\&\hspace{60pt}b_{m,l,2}+n_m-\sum_{i=1}^{n_m} \exp(-\beta_{m,l}(T-t_{m,i}))\big)\\
 \gamma^2_{m,l}|\bm{x},\theta_{-\gamma_{m,l}},\bm{Y}, \bm N &\sim \ig\big(a_{m,l,4}+\sum_{i=1}^{n_m} |O_{m,i}|,\\&\hspace{60pt}b_{m,l,4}+\frac{1}{2}\sum_{i=1}^{n_m}\sum_{t_{l,j}\in O_{m,i}}\|\sss_{l,j}-\sss_{m,i}\|^2\big).
\end{align*}

We sample $Y_{l,i}$ from its full conditional distribution which is a multinomial distribution given by
$$
    P( Y_{l,i} = (m,j) \mid \theta,\bm{x},\bm{N}) \propto \begin{cases} 
      \mu_{l}|W|^{-1}, \\ \hspace{60pt} \text{for } m = j = 0, \quad \text{and} \\[5pt]
      \dfrac{\alpha_{m,l}\beta_{m,l}}{\sqrt{2\pi\gamma_{m,l}^2}}\exp(-\beta_{m,l}(t_{l,i}-t_{m,j})-\frac{\|\sss_{l,i}-\sss_{m,j}\|^2}{2\gamma_{m,l}^2}), \\ \hspace{60pt} \text{for } m>0,j>0, t_{m,k}<t_{l,i}
   \end{cases}
$$

Let $\bm t,\bm s$ denote vectors of the exact time and spatial marks of all events respectively. Since there are no conjugate priors for $\bm t,\bm s$ and $\beta_{m,l}$, we use element-wise Metropolis-Hasting to sample from their full conditionals. Specifically, we propose value $\beta_{m,l}'$ from a normal distribution with mean $\beta_{m,l}^c$ and standard deviation $\sigma_{\beta_{m,l}}$. We accept the move with probability
\begin{align*}
H_{\beta_{m,l}}&=\min\Big(1,\Big(\frac{\beta_{m,l}'}{\beta_{m,l}^c}\Big)^{a_{m,l,3}-1}\exp(b_{m,l,3}(\beta_{m,l}^c-\beta_{m,l}'))\\
&\hspace{50pt}\times\prod_{i = 1}^{n_m}\Big\{\exp\Big(\alpha_{m,l}\exp(-\beta_{m,l}'(T-t_{m,i}))-\alpha_{m,l}\exp(-\beta_{m,l}^c(T-t_{m,i}))\Big) \\
& \hspace{50pt}\times\prod_{j\in O_{m,i}}\frac{\beta_{m,l}'}{\beta_{m,l}^c}\exp\Big(-(\beta_{m,l}'-\beta_{m,l}^c)(t_{l,j}-t_{m,i})\Big)\Big\}\Big)
\end{align*}
For $i = 1,2\dots, n_m$, $t_{m,i}'$ is drawn from a continuous uniform distribution that satisfies the following restrictions: $t_{m,i}'$ is within the time bin of event $i$ in process $m$, it occurs after its parent event (if it has one, $Y_{m,i} \neq (0,0)$), and before its offspring (if any). The proposed value of $t_{m,i}$ is accepted with probability
\begin{align*}
H_{t_{m,i}} = \min\Big(1,&\exp\Big(\alpha_{m,l}\beta_{m,l}\exp(-\beta_{m,l}(T-t_{m,i}^c))
-\alpha_{m,l}\beta_{m,l}\exp(-\beta_{m,l}(T-t_{m,i}')))\\
&-(|O_{m,i}|+1{\{Y_{m,i}\neq (0,0)\}})\beta_{m,l}(t_{m,i}'-t_{m,i}^c)\Big)\Big).
\end{align*}
Similarly, $s_{m,i}$ is drawn from a uniform distribution on the space bin containing $s_{m,i}$ and the proposed move is accepted with probability
\begin{align*}
&H_{\sss_{m,i}} = \min \Big(1, \ \exp\Big(\frac{-\|\sss_{m,i}'-\sss_{\textrm{pa}}\|^2+\|\sss_{m,i}^c-\sss_{\textrm{pa}}\|^2}{2\gamma_{q,l}^2}1{\{Y_{m,i}\neq (0,0)\}}+ \\
& \hspace{90pt} +\sum_{j\in O_{m,i}}\frac{-\|\sss_{m,i}'-s_{l,j}\|^2+\|\sss_{m,i}^c-\sss_{l,j}\|^2}{2\gamma_{m,l}^2}\Big)\Big),
\end{align*}
where $q$ is the process that contains the parent of $t_{m,i}.$

\subsubsection{Alternative prior for $\alpha$}

Note that under the independent truncated Gamma priors for the $\alpha$ parameters described in the beginning of this section, it is possible to obtain posterior samples of the $\alpha$ matrix for which the stationarity condition is not satisfied. An alternative specification on $\alpha$ could impose a multivariate truncation on these parameters that ensures stationarity. Specifically, the prior on the collection of $\{\alpha_{m,l}\}$ could be the product of the independent truncated gamma priors $\alpha_{m,l} \sim \text{Trunc}\,\text{Gamma}(a_{m,l,2}, b_{m,l,2})$ subject to the additional constraint that the spectral radius of the $\alpha$ matrix is less than 1.

With this joint prior, separate updates for each $\alpha_{m,l}$ is complicated. At each iteration, the conditional posterior of $\alpha_{m,l}$ would include a truncation of its values over a range that satisfies that the spectral radius of the matrix $\alpha$ is less than 1. Even though this is theoretically possible, finding the allowed range for each of the $\alpha_{m,l}$ at each iteration would increase computation cost.
Alternatively, one could update all of the $\alpha$'s simultaneously. The full conditional distribution of the $\alpha$'s becomes the product of the full conditional distributions of each $\alpha_{m,l}$ as shown in the previous paragraph, combined with the constraint that the spectral radius of the $\alpha$ matrix remains less than 1. Performing this update is possible by drawing from the non-truncated version of this conditional posterior distribution until a matrix $\alpha$ that satisfies the condition is drawn. This can again increase computational cost since multiple draws from a multivariate distribution might be required until one satisfies the truncation condition, and the spectral radius would have to be computed for each one of them.

Instead, we investigate whether the posterior samples under our independent prior choice satisfy stationarity of the multivariate process by studying the proportion of posterior samples of the matrix $\alpha$ that have spectral radius below 1.

\subsection{MCMC updates for alternative specifications}\label{a: subsec: MCMC}
In this section, we present the details of MCMC updates for spatio-temporal Hawkes processes under specifications considered in the manuscripts. We first present the MCMC updates for the Lomax specification considered in simulations and then introduce the details of MCMC updates for the data-driven specification considered in the application.

\paragraph{Lomax kernel}
In the main manuscript, we introduce the sampling scheming for Hawkes process under exponential Gaussian specification. If the time component is Lomax instead of exponential density, then we need to update the unknown parameters $c,p$ defined in Table 1 and modify the Metropolis ratios for latent time and labels. We use independent gamma prior for $c$ and $p-1,$ i.e.
$$c\sim \ga\big(a_c,b_c\big)\hspace{0.5cm}\text{and}\hspace{0.5cm} p-1\sim\ga\big(a_p,b_p\big).$$ For all the simulations, we set $a_c = a_p = 1$ and $b_c = b_p = 0.1.$
In each iteration, We draw new values of $c$ and $p$ from normal distributions centered as the current values of $c$ and $p$ respectively, and accept the new values with probability $H_c$ and $H_p$ shown as follows. 
\begin{align*}
H_{c} &= 
\min \Big(1,\ \Big(\frac{c'}{c^c}\Big)^{a_c-1}\exp(b_c(c^c-c'))\\
& \hspace{60pt} \times\prod_{j = 1}^n\Big\{\exp\Big(\alpha\exp(-(1+\frac{T-t_j}{c'})^{-p+1})-\alpha\exp(-(1+\frac{T-t_j}{c^c})^{-p+1})\Big) \\
& \hspace{60pt} \times \prod_{i\in O_j}\frac{t_i-t_j+c^c}{t_i-t_j+c'}\Big\}\Big)
\end{align*}

\begin{align*}
H_{p} &= 
\min \Big(1,\ \Big(\frac{p'-1}{p^c-1}\Big)^{a_p-1}\exp(b_p(p^c-p'))\\
& \hspace{60pt} \times\prod_{j = 1}^n\Big\{\exp\Big(\alpha\exp(-(1+\frac{T-t_j}{c})^{-p'+1})-\alpha\exp(-(1+\frac{T-t_j}{c})^{-p^c+1})\Big) \\
& \hspace{60pt} \times \prod_{i\in O_j}\frac{(p'-1)c^{p'-p^c}}{(p^c-1)(t_i-t_j+c)^{p'-p^c}}\Big\}\Big)
\end{align*}
Moreover, the Metropolis ratio for latent time $t_i$ becomes
\begin{align*}
&H_{t_i} = \min\Big(1,\ \exp\Big(\alpha\exp(-\frac{T-t_i'+c}{c})-\alpha\exp(-\frac{T-t_i^c+c}{c}))\Big)\\
& \hspace{85pt}\times\prod_{j\in O_i}\big(\frac{t_j-t_i'+c}{t_j-t_i^c+c}\big)^p 1\{Y_i>0\}\big(\frac{t_i'-t_{Y_i}+c}{t_i^c-t_{Y_i}+c}\big)^p\Big).
\end{align*}
and the full posterior distribution of branching label $Y_i$ in this canse is
$$
\begin{aligned}
      &P( Y_i = j \mid \bm \theta,\bm{x},\bm{N}) \\\propto &\begin{cases} 
      \mu|W|^{-1}, & \text{for } j= 0, \quad \text{and} \\
      \dfrac{\alpha(p-1)c^{p-1}}{2\pi\gamma^2(t_i-t_j+c)^p} \exp \left(-\dfrac{\|\bm{s}_i-\sss_j\|^2}{2\gamma^2} \right)1\{t_i>t_j\}, & \text{for } j=1,2,...,n.
   \end{cases}  
\end{aligned}
$$
\begin{align*}
&H_{t_i} = \min\Big(1,\ \exp\Big(\alpha\exp(-\frac{T-t_i'+c}{c})-\alpha\exp(-\frac{T-t_i^c+c}{c}))\Big)\\
& \hspace{85pt}\times\prod_{j\in O_i}\big(\frac{t_j-t_i'+c}{t_j-t_i^c+c}\big)^p 1\{Y_i>0\}\big(\frac{t_i'-t_{Y_i}+c}{t_i^c-t_{Y_i}+c}\big)^p\Big).
\end{align*}

\paragraph{Data-driven specification}
In this application, we choose a data-driven distributions for the immigrant location distribution instead of a uniform distribution over the observed window $W.$ Specifically, we assume
\begin{equation*}
    f(\sss) = \frac{\rho}{C_1}\exp\Big\{-\frac{d^2_{c_1}(\sss)}{2\gamma^2_{c_1}}-\frac{d^2_{r}(\sss)}{2\gamma^2_{r}}\Big\}+\frac{1 - \rho}{C_2}\exp\Big\{-\frac{d^2_{c_2}(\sss)}{2\gamma^2_{c_2}}-\frac{d^2_{r}(\sss)}{2\gamma^2_{r}}\Big\},   
\end{equation*} 
where $\mu,\eta\ge 0.$
In this specification, we have some additional parameters, namely $\eta,\rho,\gamma_r,\gamma_{c_1},\gamma_{c_2}.$ Moreover, we have introduce latent variable $\bm z$ to simply the mixture distribution. We add steps to the sampling scheme introduce in Section 3.5 to update the addition parameters and latent variables. 
We can use Gibbs sampling for all the additional parameters and latent variables associated with $f(\sss).$
The prior distributions for the additional variance parameters are specified to be independent conjugate priors:
$
\gamma_r^2\sim \ig(a_r,b_r), 
\gamma_{c_1}^2\sim \ig(a_{c_1},b_{c_1}),$  $\gamma_{c_2}^2\sim \ig(a_{c_2},b_{c_2}),$ and
$\rho\sim \textrm{Beta}(1,1).$

We can use Gibbs sampling for all the additional parameters and latent variables. The full posterior distributions are 
\begin{align*}
    &z_i|\mathbf{N},\bm x,\bm Y,\bm \theta,\rho\sim \textrm{Bernoulli}\Big(\frac{\rho f(\bm{s}_i|z_i = 1)}{\rho f(\bm{s}_i|z_i = 1)+(1-\rho)f(\bm{s}_i|z_i = 0)}\Big) \quad \text{for}\  i\in I\\
    &\gamma_{r}^2|\mathbf{N},\bm x,\bm Y,\bm z\sim\ig \Big(a_{r}+|I|,b_r+\frac{1}{2}\sum_{i\in I} d^2_{r}(\bm{s}_i)\Big)\\
    &\gamma_{c_1}^2|\mathbf{N},\bm x,\bm Y,\bm z\sim\ig\Big(a_{c_1}+\sum_{i\in I} 1{\{z_i=1\}},b_{c_2}+\frac{1}{2}\sum_{i\in I}d^2_{c_1}(\bm{s}_i)\Big)\\
    &\gamma_{c_2}^2|\mathbf{N},\bm x,\bm Y,\bm z\sim\ig\Big(a_{c_2}+\sum_{i\in I}1{\{z_i=0\}},b_{c_2}+\frac{1}{2}\sum_{i\in I}d^2_{c_2}(\bm{s}_i)\Big), \quad \text{and} \\ 
    &\rho|\mathbf{N},\bm x,\bm Y,\bm z\sim \mathrm{Beta}\Big(a_\rho+\sum_{i\in I}1{\{z_i=1\}},b_\rho+\sum_{i\in I}1{\{z_i=0\}}\Big).
\end{align*}
The full posterior distribution of latent time $t_i$, and the Metropolis ratio $H_{\sss_i}$ for updating the latent time $\sss_i$ change in this specification. In this case, we have
$$
\begin{aligned}
      &P( Y_i = j \mid \bm \theta,\bm{x},\bm{N}) \\\propto &\begin{cases} 
      \mu f(\sss), & \text{for } j= 0, \quad \text{and} \\
      \dfrac{\alpha\beta}{2\pi\gamma^2} \exp \left( - \beta(t_i-t_j) -\dfrac{\|\bm{s}_i-\sss_j\|^2}{2\gamma^2} \right)1\{t_i>t_j\}, & \text{for } j=1,2,...,n.
   \end{cases}  
\end{aligned}
$$

\begin{align*}
&H_{\bm{s}_i} = \min\Big(1, \ \frac{f(\bm{s}_i'-\bm{s}_{\textrm{pa}})}{f(\bm{s}_i^c-\bm{s}_{\textrm{pa}})}1{\{Y_i>0\}}\exp\Big(\sum_{j\in O_i}\frac{-\|\bm{s}_i'-\bm{s}_{j}\|^2+\|\bm{s}_i^c-\bm{s}_{j}\|^2}{2\gamma^2}\Big)\Big).
\end{align*}
In applications, we also extend the constant intensity $\mu$ into a time-dependent intensity function $\mu\exp(-\eta t)$ where $\mu,\eta>0.$ We use the sample prior for $\mu$ as introduced in Section 3.5. For $\eta$, we adopt an independent Gamma prior, i.e $\eta\sim\ga\big(a_\eta,b_\eta\big).$ Under this specification, the full posterior of $\mu$ and latent branching label $Y_i$ are different from these presented the main manuscript. The updated posterior distribution is
$$\mu\mid \bm x,\ttheta_{-\mu},\bm Y,\bm N\sim \ga\big(|I|+a_1,\frac{1}{\eta}(1-\exp(-\eta T))+b_1\big),$$
For the parameter $\eta$, we employ the Metropolis-Hasting updates, in which we sample a new values of $\eta$ from a normal distribution centered at the current values of $\eta$, and accept the new values with probability
\begin{align*}
H_{\beta} &= 
\min \Big(1,\ \Big(\frac{\beta'}{\beta^c}\Big)^{a_\eta-1}\exp(b_\eta(\eta^c-\eta'))\\
& \hspace{60pt} \times\exp\Big(\frac{\mu}{\eta'}\exp(-\eta'(T))-\frac{\mu}{\eta^c}\exp(-\eta^c(T))\Big) \\
& \hspace{60pt} \times \prod_{i\in I}\frac{\eta'}{\eta^c}\exp\Big(-(\eta'-\eta^c)(T-t_i)\Big)\Big).
\end{align*}
Moreover, we need to update the posterior distribution and  Metropolis ratios for the following quantities.
$$
\begin{aligned}
      &P( Y_i = j \mid \bm \theta,\bm{x},\bm{N}) \\\propto &\begin{cases} 
      \mu\exp(-\eta t_i) f(\sss), & \text{for } j= 0, \quad \text{and} \\
      \dfrac{\alpha\beta}{2\pi\gamma^2} \exp \left( - \beta(t_i-t_j) -\dfrac{\|\bm{s}_i-\sss_j\|^2}{2\gamma^2} \right)1\{t_i>t_j\}, & \text{for } j=1,2,...,n.
   \end{cases}  
\end{aligned}
$$
\begin{align*}
&H_{t_i} = \min\Big(1,\ \exp\Big(\alpha\exp(-\beta(T-t_i'))-\alpha\exp(-\beta(T-t_i^c)))\\
& \hspace{85pt}+(|O_i|-1{\{Y_i>0\}})\beta(t_i'-t_i^c)\Big)\\
& \hspace{85pt}+1\{Y_i=0\}\exp(-\eta t_i'+\eta t_i^c)\Big).
\end{align*}

\section{Derivation
of the Metropolis ratios}\label{a: sec: acceptance}

In this section, we present details on the derivation of acceptance ratios for Metropolis-Hasting algorithm that are shown in Section 3.5. For parameter $\beta$, the ratio of the full posterior distribution evaluated at the proposed and current values of $\beta$ is 
\begin{align*}
&\frac{p(\beta'|\ttheta_{{-(\beta)}},\bm{x},\bm Y)}{p(\beta^{c}|\ttheta_{{-(\beta)}},\bm{x},\bm Y)}\\
 = &\frac{p(\bm{x},\bm{Y}|\ttheta_{-(\beta)},\beta')p(\beta')}{p(\bm{x},\bm{Y}|\ttheta_{-(\beta)},\beta^{c})p(\beta^{c
 })}\\
  = &\frac{\prod_{j=1}^n\Big\{\exp\Big(-\alpha\big(1- \beta'\exp(-\beta'(T-t_j))\big)\Big)\prod_{i\in O_j}\alpha \beta'(-\beta'(t_i-t_j))\Big\}{\beta'}^{a_3-1}\exp(-b_3\beta')}{\prod_{j=1}^n\Big\{\exp\Big(-\alpha\big(1- \beta^{c}\exp(-\beta^{c}(T-t_j))\big)\Big)\prod_{i\in O_j}\alpha \beta^{c}(-\beta^{c}(t_i-t_j))\Big\}{\beta^{c}}^{a_3-1}\exp(-b_3\beta^{c})}\\
\end{align*}
After simplifying the fraction, we have
\begin{align*}
    H_{\beta} &= 
\min \Big(1,\ \Big(\frac{\beta'}{\beta^c}\Big)^{a_3-1}\exp(b_3(\beta^c-\beta'))\\
& \hspace{60pt} \times\prod_{j = 1}^n\Big\{\exp\Big(\alpha\exp(-\beta'(T-t_j))-\alpha\exp(-\beta^c(T-t_j))\Big) \\
& \hspace{60pt} \times \prod_{i\in O_j}\frac{\beta'}{\beta^c}\exp\Big(-(\beta'-\beta^c)(t_i-t_j)\Big)\Big\}\Big).
\end{align*}
Next we present for derivation of the acceptance rates for latent time $t_i$ and location $\sss_i$, where $1\leq i\leq n$ and $n$ is the total number of events. Note that 
\begin{align*}
 p(\bm x,\bm Y|\ttheta) =  C_{t_i}\exp(\alpha\exp(-\beta(T-t_i))+(|O_i|-1\{Y_i>0\})\beta t_i)
\end{align*} 
where $C_{t_i}$ is a term that does not depend on $t_i.$ Then 
\begin{align*}
    \frac{p(t_i'|\ttheta,\bm{x}_{-(t_i)},\bm Y)}{p(t_i^c|\ttheta,\bm{x}_{-(t_i)},\bm Y)}&=
    \frac{p(t_i',\bm{x}_{-(t_i)},\bm Y)}{p(t_i^c,\bm{x}_{-(t_i)},\bm Y)}\\
    &=\frac{\exp(\alpha\exp(-\beta(T-t_i'))+(|O_i|-1\{Y_i>0\})\beta t_i')}{\exp(\alpha\exp(-\beta(T-t_i^c))+(|O_i|-1\{Y_i>0\})\beta t_i^c)}
\end{align*}
Since the proposal distribution of $t_i$ is symmetric, we have
\begin{align*}
    &H_{t_i} = \min\Big(1,\ \exp\Big(\alpha\exp(-\beta(T-t_i'))-\alpha\exp(-\beta(T-t_i^c)))\\
& \hspace{85pt}+(|O_i|-1{\{Y_i>0\}})\beta(t_i'-t_i^c)\Big)\Big).
\end{align*}
Similarly, 
\begin{align*}
 p(\bm x,\bm Y|\ttheta) =  C_{\sss_i}\exp\Big(\frac{-\|\bm{s}_i-\bm{s}_{\textrm{pa}}\|^2}{2\gamma^2}1{\{Y_i>0\}}+\sum_{j\in O_i}\frac{-\|\bm{s}_i-\bm{s}_{j}\|^2}{2\gamma^2}\Big)\Big)
\end{align*} 
where $C_{\sss_i}$ is a term that does not depend on $\sss_i$, and $\bm{s}_{\textrm{pa}}$ is the location of the parent of event $i$.
Then 
\begin{align*}
    \frac{p(\sss_i'|\ttheta,\bm{x}_{-(\sss_i)},\bm Y)}{p(t_i^c|\ttheta,\bm{x}_{-(t_i)},\bm Y)}&=
    \frac{p(\sss_i',\bm{x}_{-(\sss_i)},\bm Y)}{p(\sss_i^c,\bm{x}_{-(\sss_i)},\bm Y)}\\
    &=\frac{\exp\Big(\frac{-\|\bm{s}_i'-\bm{s}_{\textrm{pa}}\|^2}{2\gamma^2}1{\{Y_i>0\}}+\sum_{j\in O_i}\frac{-\|\bm{s}_i'-\bm{s}_{j}\|^2}{2\gamma^2}\Big)\Big)}{\exp\Big(\frac{-\|\bm{s}_i^c-\bm{s}_{\textrm{pa}}\|^2}{2\gamma^2}1{\{Y_i>0\}}+\sum_{j\in O_i}\frac{-\|\bm{s}_i^c\-\bm{s}_{j}\|^2}{2\gamma^2}\Big)\Big)}
\end{align*}
Thus,
\begin{align*}
&H_{\bm{s}_i} = \min\Big(1, \ \exp\Big(\frac{-\|\bm{s}_i'-\bm{s}_{\textrm{pa}}\|^2+\|\bm{s}_i^c-\bm{s}_{\textrm{pa}}\|^2}{2\gamma^2}1{\{Y_i>0\}}\\
&\hspace{85pt}+\sum_{j\in O_i}\frac{-\|\bm{s}_i'-\bm{s}_{j}\|^2+\|\bm{s}_i^c-\bm{s}_{j}\|^2}{2\gamma^2}\Big)\Big).
\end{align*}

\section{Theoretical Proofs of Identifiability}\label{appendix: proofs}
\subsection{Identifiability for aggregated temporal models}
In this section, we show all the proofs for the identifiability results of temporal Hawkes processes. In order to prove Theorem 3.1, let us first show a more general identifiability result for aggregated temporal models.

\begin{proof}[Proof of Theorem 3.1]
Let $\Delta_1^t,\dots,\Delta_K^t$ be sizes of aggregation bins. Let $\ttheta^h = (\alpha, \ttheta^g).$
We need to show that if $P(N_1 = n_1,\dots,N_k = n_K;\ttheta_1) =P(N_1 = n_1,\dots,N_k = n_K;\ttheta_2)$ for all $(n_1,\dots,n_K)\in\mathbb{N}^k$, then $\ttheta_1 = \ttheta_2$. We follow proof by contradiction.
Suppose that there exist $\ttheta_1\neq \ttheta_2$ such that $P(N_1 = n_1,\dots,N_k = n_K;\ttheta_1) =P(N_1 = n_1,\dots,N_k = n_K;\ttheta_2)$ for all $(n_1,\dots, n_K)\in\mathbb{N}^k.$ Define $\delta(0)=0$ and $\delta(k_0) = \sum_{k=1}^{k_0}\Delta^t_k$ for $k_0\in\mathbb{Z}^+$. Note that $\delta(0),\dots,\delta(K)$ are the end points of the $B_1^t,\dots,B_K^t.$ So
$$P(N_1=0,\dots,N_k = 0;\ttheta) = \exp\big({-\int_0^{\delta(k)} \mu(t;\ttheta^\mu)\dd t}\big),\mathrm{\  for\   all\  } k.$$ By assumption, we have
$$\exp\Big({-\int_0^{\delta(k)} \mu(t;{\ttheta^\mu_1})\dd t}\Big) = \exp\Big({-\int_0^{\delta(k)} \mu(t;{\ttheta^\mu_2})\dd t}\Big),$$ and thus $${\int_0^{\delta(k)} \mu(t;{\ttheta^\mu_1})\dd t} = {\int_0^{\delta(k)} \mu(t;{\ttheta^\mu_2})\dd t}$$ for all $k$. So $${\int_{\delta(k-1)}^{\delta(k)} \mu(t;{\ttheta^\mu_1})\dd t} = {\int_{\delta(k-1)}^{\delta(k)} \mu(t;{\ttheta^\mu_2})\dd t}$$ for all $k$. By assumption (a), $\ttheta^\mu_1 = \ttheta^\mu_2.$ Since $\ttheta_1\neq \ttheta_2$, we have $\ttheta^h_1\neq \ttheta^h_2.$ Consider intervals of the form $\big[T-\delta(k),T-\delta(k-1)\big)$ for $1\leq k\leq K.$ By assumption (b), there exist $k_0$ such that $[T-\delta(k_0),T-\delta(k))$ does not contain zeros of $h(t;{\ttheta^h_1})-h(t;{\ttheta^h_2})$. WLOG, suppose $h(t;{\ttheta}^h_1)>h(t;{\ttheta}^h_2)$ for $t\in [T-\delta(k_0),T-\delta(k_0-1)).$
Then
$$\int_{T-\delta(k_0)}^{T-\delta(k_0-1)}\exp\big(-h_(x;{\ttheta^h_1})\big)\dd x< \int_{T-\delta(k_0)}^{T-\delta(k_0-1)}\exp\big(-h(x;{\ttheta^h_2})\big)\dd x$$
Note that
\begin{align*}
   & P(N_1=0,\dots,N_{k_0-1}=0,N_{k_0} = 1,N_{k_0+1} = 0,\dots,N_{K} = 0;{\ttheta})\\
   =& \int_{\delta(k_0-1)}^{\delta(k_0)} \lambda^*(t_1;\ttheta)\exp\Big(\int_0^{T} -\lambda^*(t;\ttheta) \dd t \Big) \dd t_1\\
   =& \int_{\delta(k_0-1)}^{\delta(k_0)} \mu(t_1;{\ttheta}^\mu) \exp\big({-\mu(t_1;{\ttheta^\mu})}-h(T-t_1;{\ttheta^h})\big)\dd t_1\\
   =& \int_{T-\delta(k_0)}^{T-\delta(k_0-1)} \mu(t_1;{\ttheta}^\mu) \exp\Big(-\mu(t_1;{\ttheta}^\mu)-h(x;{\ttheta}^h)\big)\dd x
\end{align*}
Thus, 
\begin{align*}
& P(N_1=0,\dots,N_{k_0-1}=0,N_{k_0} = 1,N_{k_0+1} = 0,\dots,N_{K} = 0;{\ttheta_1}) \\
& \hspace{20pt} < P(N_1=0,\dots,N_{k_0-1}=0,N_{k_0} = 1,N_{k_0+1} = 0,\dots,N_{K} = 0;{\ttheta_2}),    
\end{align*}
a contradiction.
\end{proof}

\begin{lemma}[Properties of exponential offspring kernel]\label{lemma1}
  Let $h(t;{\ttheta}) = \alpha(1- e^{-\beta t})$ where $\theta = (\alpha,\beta)\in (0,1)\times(0,\infty).$ If $\ttheta_1\neq \ttheta_2$, then $h(t;{\ttheta_1})-h(t;{\ttheta_2})$ has at most 1 root on $(0,\infty).$
\end{lemma}

\begin{proof}
By contradiction, there exists $\ttheta_1\neq\ttheta_2$ such that $h(t;{\ttheta_1})-h(t;{\ttheta_2})$ has st least two positive roots. Since $h(0;{\ttheta_1})-h(0;{\ttheta_2}) = 0$, the derivative $h'(t;{\ttheta_1})-h'(t;{\ttheta_2})$ changes sign at least twice on $(0,\infty).$ Note that
    $$\frac{h'(t;{\ttheta_1})}{h'(t;{\ttheta_2})}=\frac{\alpha_1\beta_1}{\alpha_2\beta_2}e^{(\beta_2-\beta_1)t}$$ is a monotonic function for $t\in(0,\infty).$ So $h'(t;{\ttheta_1})-h'(t;{\ttheta_2})$ is also monotone on $(0, \infty).$ Thus, $h'(t;{\ttheta_1})-h'(t;{\ttheta_2})$ changes sign at most once, which leads to a contradiction. Hence, $h_{\ttheta_1}-h_{\ttheta_2}$ has at most 1 root on $(0,\infty).$
\end{proof}

\begin{lemma}[Properties of Lomax offspring kernel]\label{lemma2}
  Let $h(t;{\ttheta}) = \alpha-\alpha\Big(\frac{c}{c+x}\Big)^{p-1}$ where $\theta = (\alpha,c,p)\in (0,1)\times(0,\infty)\times(1,\infty).$ If $\ttheta_1\neq \ttheta_2$, then $h(t;{\ttheta_1})-h(t;{\ttheta_2})$ has at most 2 roots on $(0,\infty).$
\end{lemma}

\begin{proof}
    By contradiction, there exists $\ttheta_1\neq\ttheta_2$ such that $h(t;{\ttheta_1})-h(t;{\ttheta_2})$ has at least three positive roots. Since $h(0;{\ttheta_1})-h(0;{\ttheta_2}) = 0$, the derivative $h'(t;{\ttheta_1})-h'(t;{\ttheta_2})$ changes sign at least three times on $(0,\infty).$ Observe that $h'(t;{\ttheta_1})-h'(t;{\ttheta_2})$ and $\log h'(t;{\ttheta_1})-\log h'(t;{\ttheta_2})$ have the same sign. So the function $\log h'(t;{\ttheta_1})-\log h'(t;{\ttheta_2})$ changes sign at least three times on $(0,\infty).$  Note that
    $$\dt\Big(\log h'(t;{\ttheta_1})-\log h'(t;{\ttheta_2})\Big) = \frac{\alpha_1 p_1}{c_1+x}-\frac{\alpha_2 p_2}{c_2+x}.$$ We have the following three cases:
    \begin{case}
        Suppose $\alpha_1p_1 = \alpha_2p_2$ and $c_1 = c_2$. Then $\dt\Big(\log h'(t;{\ttheta_1})-\log h'(t;{\ttheta_2})\Big) = 0$ for all $t\in (0,\infty).$ So $\log h'(t;{\ttheta_1})-\log h'(t;{\ttheta_2})$ is constant. So $\log h'(t;{\ttheta_1})-\log h'(t;{\ttheta_2})$ does not change sign on $(0,\infty)$, a contradiction.
    \end{case}

    \begin{case}
        Suppose $\alpha_1p_1 = \alpha_2p_2$ and $c_1\neq c_2$. WLOG, assume $c_1<c_2.$ Then $\dt\Big(\log h'(t;{\ttheta_1})-\log h'(t;{\ttheta_2})\Big)>0$ for $t\in(0,\infty).$ So $\log h'(t;{\ttheta_1})-\log h'(t;{\ttheta_2})$ is increasing on $(0,\infty)$. Thus, $\log h'(t;{\ttheta_1})-\log h'(t;{\ttheta_2})$ changes sign at most once on $(0,\infty)$, a contradiction.
    \end{case}

    \begin{case}
        Suppose $\alpha_1p_1 \neq \alpha_2p_2$. Then $t_0 = \frac{\alpha_1p_1c_2-\alpha_2p_2c_1}{\alpha_2p_2-\alpha_1p_1}$ is the unique solution for $\dt\Big(\log h'(t;{\ttheta_1})-\log h'(t;{\ttheta_2})\Big) = 0$, and the function $\dt\Big(\log h'(t;{\ttheta_1})-\log h'(t;{\ttheta_2})\Big)$ changes sign at $t_0.$ WLOG, suppose $\dt\Big(\log h'(t;{\ttheta_1})-\log h'(t;{\ttheta_2})\Big)<0$ for $t<t_0$ and $\dt\Big(\log h'(t;{\ttheta_1})-\log h'(t;{\ttheta_2})\Big)>0$ for $t>t_0.$ Then $\log h'(t;{\ttheta_1})-\log h'(t;{\ttheta_2})$ is decreasing for $t<t_0$ and increasing for $t>t_0.$ Thus, $\log h'(t;{\ttheta_1})-\log h'(t;{\ttheta_2})$ changes sign at most twice on $(-\infty,\infty)$, a contradiction.
    \end{case}
\end{proof}

\begin{proof}[Proof of Proposition 3.1]
    It follows directly from Theorem 3.1, \cref{lemma1} and \cref{lemma2}.
\end{proof}

\subsection{Identifiability for aggregated spatio-temporal models}

In this section, we show the identifiability results for a spatio-temporal Hawkes process based on aggregated data. 

Note that for a given $t\in[0,T)$, $f(\sss\mid t)$ and $g_2(\sss\mid t)$ are density functions on $\mathbb{R}^2.$ According to \cite{Rasmussen2013bayesian}, a spatio-temporal Hawkes process can be treated as a marked temporal process with intensity $\lambda_g^*(t) = \mu(t)+\sum_{i:t_i<t}\alpha g_1(t-t_i)$ and mark distribution $$\gamma^*(s\mid t) = \frac{\mu(t) f(s\mid t)}{\lambda_g^*(t)}+\frac{\sum_{i:t_i<t}g_1(t-t_i)g_2(s-s_i\mid t-t_i)}{\lambda_g^*(t)}.$$ 

Conceiving the spatio-temporal Hawkes process as a marked process forms the basis for showing identifiability of the spatio-temporal Hawkes process parameters. By reducing the aggregated spatio-temporal data to aggregated temporal data by ignoring the location information, we apply Theorem 3.1 to show that the parameters in $\lambda_g^*$ are identifiable. Then, we use the spatial information in the aggregated data across bins to show that the parameters of the mark distribution are identifiable.

\begin{lemma}[Properties of log linear background intensity function]\label{lemma3}
  Let $\mu(t;{\ttheta}) = \mu\exp(-\eta t)$ where $\ttheta = (\mu,\eta)\in (0,\infty)\times (0,\infty).$ If $\ttheta_1\neq \ttheta_2$, then $\mu(t;{\ttheta_1})-\mu(t;{\ttheta_2})$ has at most 1 root on $(0,\infty).$
\end{lemma}
\begin{proof}
Let $\ttheta_1 = (\mu_1,\eta_1)\neq (\mu_2,\eta_2) = \ttheta_2.$ Note that $\mu'(t;\ttheta) = \mu\eta\exp(\eta t ).$ So $\frac{\mu'(t;\ttheta_1)}{\mu'(t;\ttheta_2)} = \frac{\mu_1\eta_1}{\mu_2\eta_2}\exp\big(-(\eta_1-\eta_2)t\big).$ Thus, $\frac{\mu'(t;\ttheta_1)}{\mu'(t;\ttheta_2)}$ is a monotonic function for $t\in(0,\infty)$, and so is $\mu'(t;\ttheta_1)-\mu'(t;\ttheta_2).$ So $\mu(t;\ttheta_1)-\mu(t;\ttheta_2)$ is either increasing or decreasing on $(0,\infty)$ and thus $\mu(t;\ttheta_1)-\mu(t;\ttheta_2)$ has at most one root on $(0,\infty).$
\end{proof}

\begin{proof}[Proof of Theorem 3.2]
Let $\ttheta_1,\ttheta_2\in\Theta$. Assume
$$P(N_{1,1}=n_{1,1},\dots,N_{K,R} = n_{k_1,k_2};{\ttheta_1}) = P(N_{1,1}=n_{1,1},\dots,N_{K,R} = n_{K,R};{\ttheta_2})$$
for all $\{n_{k,r}:1\leq k\leq K,1\leq r\leq R\}\subset\mathbb{N}.$  We need to show that $\ttheta_1 = \ttheta_2.$
    
Note that according to \cite{Rasmussen2013bayesian} the underlying spatio-temporal Hawkes process can be treated as a marked temporal process with ground intensity $\lambda_g^*(t;\ttheta) = \mu(t;\ttheta^\mu)+\sum_{i:t_i<t}\alpha g_1(t-t^*;\ttheta^{g_1})$ and mark distribution $$\gamma^*(s\mid t;\ttheta) = \frac{\mu(t;\ttheta^\mu) f(\sss\mid t;\ttheta^f)}{\lambda_g^*(t;\ttheta)}+\frac{\sum_{i:t_i<t}\alpha g_1(t-t_i;\ttheta^{g_1})g_2(\sss-\sss_i\mid t-t_i;\ttheta^{g_2})}{\lambda_g^*(t;\ttheta)}.$$
Let $\ttheta^h = (\alpha,\ttheta^{g_1}).$ We first show that if $\mu(t;\ttheta^\mu) = \mu\exp(-\eta)$, then condition (a) in Theorem 3.1 holds. Let $\ttheta^\mu_1\neq\ttheta^\mu_2.$ By \cref{lemma3} and $K\ge 3$, there exist one temporal bin $B_k^t$ such that $B_k^t$ does not contain a root of $\mu(t;\ttheta^\mu_1)-\mu(t;\ttheta^\mu_2)$. Since $\mu(t;\ttheta)$ is continuous, we have $\mu(t;\ttheta^\mu_1)>\mu(t;\ttheta^\mu_2)$ or $\mu(t;\ttheta^\mu_1)<\mu(t;\ttheta^\mu_2)$ for all $t\in B_k^t.$ Thus,
$$\int_{B^t_k} \mu(t;\ttheta^\mu_1)\dd t\neq \int_{B^t_k} \mu(t;\ttheta^\mu_2)\dd t,$$ and condition (a) in Theorem 3.1 holds. By similar arguments in Proposition 3.1, we have $\ttheta^\mu_1=\ttheta^\mu_2$ and $\ttheta^h_1 = \ttheta^h_2$. If $\ttheta^f_1\neq \ttheta^f_2$, then by assumption (c), there exists $k,r$ such that
$$\int_{B_r^{\sss}} f(\sss\mid t_1;{\ttheta^f_1})\dd \sss \neq \int_{B_r^{\sss}} f(\sss\mid t_1;{\ttheta^f_2})\dd \sss$$ for all $t_1\in B^t_k$.  Then $$
\begin{aligned}
  &P(N_{k,r}=1,\sum_{k=1}^K\sum_{r = 1}^R N_{k,r}=1;{\ttheta_1}) =\int_{B_k^t}\int_{B_r^{\sss}}p(t_1)f(\sss\mid t_1;{\ttheta^f_1})\dd\sss\dd t_1  \\&\neq \int_{B_k^t}\int_{B_r^{\sss}}p(t_1)f(\sss\mid t_1;{\ttheta^f_2})\dd\sss\dd t_1 = P(N_{k,r}=1,\sum_{k=1}^K\sum_{r = 1}^R N_{k,r}=1;{\ttheta_2}),  
\end{aligned}
$$ a contradiction. So $\ttheta^f_1=\ttheta^f_2.$ If $\ttheta^{g_2}_1\neq \ttheta^{g_2}_2$, by assumption (d), there exist $1\leq k\leq k'\leq K$ and $1\leq r,r' \leq R$ such that $$\int_{B_r^{\sss}}\int_{B_{r'}^{\sss}} {g_2}(\sss'-\sss\mid t_2-t_1;{\ttheta^{g_2}_1})\dd \sss\dd \sss' \neq \int_{B_r^{\sss}}\int_{B_{r'}^{\sss}} {g_2}(\sss'-\sss\mid t_2-t_1;{\ttheta^{g_2}_2})\dd \sss\dd \sss',$$ for all $t_1\in B^t_{k}$ and $t_2\in B^t_{k'}$. So 
$$\int_{B_r^{\sss}}\int_{B_{r'}^{\sss}} \gamma^*(\sss'-\sss\mid t_2-t_1;{\ttheta_1})\dd \sss\dd \sss' \neq \int_{B_r^{\sss}}\int_{B_{r'}^{\sss}} \gamma^*(\sss'-\sss\mid t_2-t_1;{\ttheta_2})\dd \sss\dd \sss',$$ for all $t_1\in B^t_{k}$ and $t_2\in B^t_{k'}.$
Thus,
$$
\begin{aligned}
    &P_{\ttheta_1}(N_{k,r} = 1,N_{k',r'} = 1, \sum_{k=1}^K\sum_{r=1}^R N_{k,r} = 2)\\=& \int_{B^t_k}\int_{B^{t}_{k'}}\int_{B_r^{\sss}}\int_{B_{r'}^{\sss}} p(t_1,t_2)f(s'\mid t_1;\ttheta^{f}_1)\gamma^*(\sss'-\sss\mid t_2-t_1;{\ttheta_1})\dd \sss\dd \sss'\dd t_2\dd t_1\\
    \neq &\int_{B^t_k}\int_{B^{t}_{k'}}\int_{B_r^{\sss}}\int_{B_{r'}^{\sss}} p(t_1,t_2)f(s'\mid t_1;\ttheta^f_2)\gamma^*(\sss'-\sss\mid t_2-t_1;{\ttheta_2})\dd \sss\dd \sss'\dd t_2\dd t_1\\
    =&P_{\ttheta_2}(N_{k,r} = 1,N_{k',r'} = 1, \sum_{k=1}^K\sum_{r=1}^R N_{k,r} = 2),
\end{aligned}
$$
a contradiction. So $\ttheta^{g_2}_1 = \ttheta^{g_2}_2.$ Hence $\ttheta_1=\ttheta_2$.
\end{proof}

\begin{lemma}\label{lemma: normal}
    Let $f_{\gamma}(x) = \frac{1}{2\pi\gamma^2}\exp(-\frac{x^2}{2\gamma^2})$. Suppose that $\gamma_1<\gamma_2$. Then there exists $x_0\in(0,\infty)$ such that $f_{\gamma_1}(x)<f_{\gamma_2}(x)$ for $x>x_0$ and $f_{\gamma_1}(x)>f_{\gamma_2}(x)$ for $x<x_0.$
\end{lemma}

\begin{proof}
    Note that $$\frac{f_1(x)}{f_2(x)} = \frac{\gamma_2}{\gamma_1}\exp(-\frac{x^2}{2\gamma_1^2}+\frac{x^2}{2\gamma_2^2}).$$ Since $-\frac{1}{2\gamma_1^2}+\frac{1}{2\gamma_2^2}<0$, we have $\exp(-\frac{x^2}{2\gamma_1^2}+\frac{x^2}{2\gamma_2^2})$ decreasing to 0. So there exist $x_0\in (0,\infty)$ such that $\exp(-\frac{x^2}{2\gamma_1^2}+\frac{x^2}{2\gamma_2^2})<\frac{\gamma_2}{\gamma_1}$ for $x>x_0$ and $\exp(-\frac{x^2}{2\gamma_1^2}+\frac{x^2}{2\gamma_2^2})>\frac{\gamma_2}{\gamma_1}$ for $x<x_0$. Thus, $\frac{f_1(x)}{f_2(x)}<1$ for all $x>x_0.$ Hence $f_1(x)<f_2(x)$ for all $x>x_0$ and $f_1(x)>f_2(x)$ for all $x<x_0.$
\end{proof}

\begin{proof}[Proof of Proposition 3.2]
Suppose all assumptions hold. We need to verify all the four conditions in Theorem 3.2.\\
\textbf{condition (a) and (b)}. By similar arguments in proofs of Proposition 3.1, assumption (a) and (d) imply that condition (a) and (b) in Theorem 3.2 are satisfied. \\
\textbf{condition (c)}. If $\ttheta^f=\varnothing$, then there is nothing to prove. If not, then $$f(\sss) = \frac{\rho}{C_1}\exp\Big\{-\frac{d_1^2(\sss)}{2\gamma^2_{1}}-\frac{d_3^2(\sss)}{2\gamma^2_{3}}\Big\}+\frac{1 - \rho}{C_2}\exp\Big\{-\frac{d_2^2(\sss)}{2\gamma^2_{2}}-\frac{d_3^2(\sss)}{2\gamma^2_{3}}\Big\},$$ where $C_1,C_2$ are normalizing constants, and there exist $r,r'$ such that for any choice of the order (greater or less), $d_1(\sss)$, $d_2(\sss)$, and $d_3(\sss)$ are related to $d_1(\sss')$, $d_2(\sss')$, and $d_3(\sss')$ according to the chosen order for all $\sss\in B^{\sss}_r$, $\sss'\in B^{\sss}_{r'}$. Let $\ttheta^f_1 = (\rho_1,\gamma_1,\gamma_2,\gamma_3)\neq (\rho_1',\gamma_1',\gamma_2',\gamma_3')\neq \ttheta^f_2$. Suppose $\gamma_1<\gamma_1'$, $\gamma_2<\gamma_2'$ and $\gamma_3<\gamma_3'$. By contradiction, suppose for all $1\leq k\leq K$ and $1\leq r \leq R$, 
\begin{equation}\label{a:eq1}
  \int_{B_r^{\sss}} f(\sss\mid t_1;\ttheta^{f}_1)\dd \sss = \int_{B_r^{\sss}} f(\sss\mid t_1;\ttheta^{f}_2)\dd \sss,  
\end{equation}
for all $t_1\in B^t_{k}.$  By assumption (b), there exist $k,k'$ such that $d_1(\sss)<d_1(\sss')$,  $d_2(\sss)<d_2(\sss')$, and $d_3(\sss)<d_3(\sss')$ for all $\sss\in B^{\sss}_r$ and $\sss'\in B^{\sss}_{r'}.$ Let $1\leq k\leq K$ and $t_1\in B^t_{k}$, by \cref{a:eq1}, there exist $\sss_0\in B^{\sss}_r$ such that 
$$  f(\sss_0\mid t_1;\ttheta^{f}_1) =  f(\sss_0\mid t_1;\ttheta^{f}_2).$$ Moreover, since $\gamma_1<\gamma_1'$, $\gamma_2<\gamma_2'$ and $\gamma_3<\gamma_3'$, it is easy to show that
$$  f(\sss_0\mid t_1;\ttheta^{f}_1)-f(\sss'\mid t_1;\ttheta^{f}_1) > f(\sss_0\mid t_1;\ttheta^{f}_2)-f(\sss'\mid t_1;\ttheta^{f}_2)$$ for all $\sss'\in B_{r'}^{\sss}.$ Thus,
$$ f(\sss'\mid t_1;\ttheta^{f}_1) <  f(\sss'\mid t_1;\ttheta^{f}_2)$$ for all $\sss'\in B_{r'}^{\sss}.$ So
$$  \int_{B_{r'}^{\sss}} f(\sss\mid t_1;\ttheta^{f}_1)\dd \sss < \int_{B_{r'}^{\sss}} f(\sss\mid t_1;\ttheta^{f}_2)\dd \sss,$$ a contradiction. So assumption (c) in Theorem 3.2 holds. The argument for other order relationships of parameters in $\ttheta^f_1$ and $\ttheta^f_2$ are very similar and thus omitted.\\
\textbf{condition (d)}. Let $\theta^{g_1}_1 = \gamma\neq \gamma'=\theta^{g_2}_2.$ By \cref{lemma: normal}, there exists $x_0\in(0,\infty)$ such that such that $f_{\gamma}(x)<f_{\gamma'}(x)$ for $x>x_0$ and $f_{\gamma}(x)>f_{\gamma'}(x)$ for $x<x_0.$ where $f_{\gamma}(x) = \frac{1}{2\pi\gamma^2}\exp(-\frac{x^2}{2\gamma^2}).$ Since $R\ge 3$, the number of rows or the number of columns of the spatial grid is greater than 3. WLOG, suppose the number of rows is greater than 3. Consider the spatial bin $B_1^{\sss}$. Let $x,x',y\in\mathbb{R}$ with $(x,y),(x',y)\in B_1^{\sss}$. Let $\sss = (x,y)$ and $\sss' = (x',y).$ If the side length of $B_1^{\sss}$ is less than $x_0$, then   $g_2(||\sss'-\sss||;\theta^{g_2}_1)> g_2(||\sss'-\sss||;\theta^{g_2}_2).$ Thus, $$\int_{B_1^{\sss}}\int_{B_{1}^{\sss}} {g_2}(||\sss'-\sss||;{\ttheta^{g_2}_1})\dd \sss\dd \sss' > \int_{B_1^{\sss}}\int_{B_{1}^{\sss}} {g_2}(||\sss'-\sss||;{\ttheta^{g_2}_2})\dd \sss\dd \sss'.$$ If the side length of $B_1^{\sss}$ is greater than $x_0$, then there exists $1<r\leq R$ such that $||\sss'-\sss||>x_0$ for all $(x,y)\in B_1^{\sss}$ and $(x',y)\in B_r^{\sss}.$ Then   $g_2(||\sss'-\sss||;\theta^{g_2}_1)< g_2(||\sss'-\sss||;\theta^{g_2}_2).$ Thus, $$\int_{B_1^{\sss}}\int_{B_{1}^{\sss}} {g_2}(||\sss'-\sss||;{\ttheta^{g_2}_1})\dd \sss\dd \sss' < \int_{B_1^{\sss}}\int_{B_{1}^{\sss}} {g_2}(||\sss'-\sss||;{\ttheta^{g_2}_2})\dd \sss\dd \sss'.$$ In both cases, condition 4 holds.\\
Hence, parameters of the Hawkes process are identifiable based on the aggregated data.
\end{proof}

\section{Compare different methods of updating latent times}\label{appendix: updates}
In this section, we assess the performance of the following three different updating methods which are introduced in Section 3.4:
\begin{itemize}
    \item [1.] one-at-a-time update
    \item [2.] block update by generations
    \item [3.] block update by clusters 
\end{itemize}

\subsection{Algorithm for sampling from proposal distribution of method 3}
In this section we introduce the algorithm for sample from the proposal distribution of method 3 mentioned in the manuscript under the exponential-Gaussian specification considered in simulations.

Let $C_i$ be a cluster formed by the immigrant $t_i.$ Let $T_{C_i} = \{t_j: j\in C_i\}.$ Re-index the elements in $T_{C_i}$ such that $T_{C_i}=\{t_1,\dots,t_n\}$ and $t_k<t_l$ whenever $1\leq k<l\leq n.$ Let $Y_{C_i}=\{Y_1,\dots,Y_n\}$ be the corresponding branching structure for $T_{C_i}$. For $t_j\in T_{C_i}$, suppose that $t_j$ is in the temporal bin $B^t_j$.
\begin{algorithm}
\caption{Sample from $p(T_{C_i}'\mid Y_{C_i},\theta)$ with desired restrictions}\label{alg:cap}
\begin{algorithmic}[1]
\State Sample $t_1'$ from a continuous uniform distribution over $B^t_1.$

    \For{$j \gets 1$ to $n$}                    
       \State $O_j = \{k: Y_k=j\}$
       \While{$O_j\neq \{\}$}
       \For{$k \in O_j$}
        \State Sample $D_k$ from a truncated exponential distribution with parameter$\beta$ \\\hspace{1.5cm} and support $B^t_k-t_j=\{t-t_j:t\in B_K\}$
        \State $t_k'\gets t_j+D_k$
       \EndFor
       \EndWhile
    \EndFor
\State $T_{C_i}' = \{t_1'\dots,t_n'\}$
\State \Return{$T_{C_i}'$}
\end{algorithmic}
\end{algorithm}

We can update the locations by cluster in a similar way when they are available. Let $S_{C_i} = \{\bm{s}_j:j\in C_i\}$ and re-index the locations so that $S_{C_i} = \{\bm{s}_1,\dots,\bm{s}_n\}$ matches $T_{C_i}.$ Suppose that $\bm{s}_j\in B^{\bm{s}}_j.$ Then we can use \cref{alg:cap2} to sample from $p(S_{C_i}'\mid Y_{C_i},\ttheta).$

\begin{algorithm}
\caption{Sample from $p(S_{C_i}'\mid Y_{C_i},\ttheta)$ with desired restrictions}\label{alg:cap2}
\begin{algorithmic}[1]
\State Sample $\sss_1'$ from a continuous uniform distribution over $B^{\sss}_1.$

    \For{$j \gets 1$ to $n$}                    
       \State $O_j = \{k: Y_k=j\}$
       \While{$O_j\neq \{\}$}
       \For{$k \in O_j$}
        \State Sample $E_k$ from a truncated bivariate normal distribution $N(\bm 0,\gamma I_2)$ \\\hspace{1.5cm} with support $B^{\bm{s}}_k-\bm{s}_j=\{\bm{s}-\bm{s}_j:\sss\in B^{\sss}_k\}$
        \State $\sss_k'\gets {\sss}_j+E_k$
       \EndFor
       \EndWhile
    \EndFor
\State $S_{C_i}' = \{\sss_1'\dots,\sss_n'\}$
\State \Return{$S_{C_i}'$}
\end{algorithmic}
\end{algorithm}

\subsection{Simulations comparing the efficiency of different algorithms for sampling the exact data}

\begin{table}[!t]
\centering
\caption{Results of the temporal simulation. The table shows the average Gelman-Rubin statistic (``Rhat''), average effective sample size over computing time in seconds (``ESS/t''), across 400 simulations, for temporal aggregations equal to 1.5 ($\Delta^t = 1.5$), and parameter sets 1 and 2.} 

\label{tab:sim_update_03}
\scalebox{0.95}{
\begin{tabular}{ c c c c c c c c c } 
\toprule
\multicolumn{5}{c}{Parameter set 1 -- (0.3, 0.7, 1)}  \\
\toprule
&Parameter  &  One-at-a-time & by generation & by cluster \\   
\midrule
\multirow{3}*{Rhat} 
&  $\mu$ &  1.0095  &  1.0117  &  1.007 \\
&  $\alpha$ & 1.0076  &  1.0088  &  1.005 \\
&  $\beta$ &  1.0364  &  1.0616  &  1.0295 \\

\hline
\multirow{3}*{ESS/t} 
&  $\mu$ &  1.3955  &  1.4332  &  1.0764 \\ 
&  $\alpha$  &  1.8642  &  1.9191  &  1.413 \\ 
&  $\beta$ &  0.2627  &  0.2817  &  0.2078 \\

\bottomrule
\end{tabular}
}

\vspace{10pt}
\scalebox{0.95}{
\begin{tabular}{ c c c c c c c c c } 
\multicolumn{5}{c}{Parameter set 2 -- (0.5, 0.5, 1)}  \\
\toprule
&Parameter  &  One-at-a-time & by generation & by cluster \\   
\midrule
\multirow{3}*{Rhat} 
&  $\mu$  &  1.038\phantom{0}  &  1.0357  &  1.0255  \\
&  $\alpha$  &  1.0409  &  1.0378  &  1.0266  \\
&  $\beta$  &  1.1307  &  1.1401  &  1.1007  \\

\hline
\multirow{3}*{ESS/t} 
&  $\mu$  &  0.6312  &  0.6582  &  0.4256  \\
&  $\alpha$  &  0.6157  &  0.6419  &  0.419  \\
&  $\beta$  &  0.1252  &  0.1384  &  0.0826 \\

\bottomrule
\end{tabular}
}
\end{table}

In each simulation, we first generate a temporal Hawkes process on $[0,500]$ and aggregate it by $\Delta^t = 1.5$.
For each of the updating method, we run two MCMC chains of 4,000 iterations. The first 2000 samples are discarded as burn-in samples. We use the Rubin statistic (Rhat) for analyzing the convergence and use effective sample size over computing time (ESS/t) as a metric for analyzing the efficiency. For generated Hawkes process, we consider two parameter sets:
\begin{itemize}
    \item[1.] $(\mu,\alpha,\beta)= (0.3,0.7,1)$
    \item[2.] $(\mu,\alpha,\beta)= (0.5,0.5,1)$
\end{itemize}
According to \cref{tab:sim_update_03}, 
the first and second methods have comparable performance. The third method has the smallest Rhat, but it also has the smallest ESS/t. Therefore, even though updating the exact data by cluster leads to better MCMC convergence for the same number of iterations compared to updating one-at-a-time or by generation, that comes with an increase in the computational time of each iteration, and a lower effective sample size by measure of time.

\section{Additional simulation results}\label{appendix:add_sim}

In addition to the simulations presented in the manuscript, we also investigated the performance of our method when the branching ratio (relative relationship between $\mu$ and $\alpha$) is more extreme, and alternative specifications for the excitation function.

\subsection{Compare Bayesian method with MC-EM method}\label{appendix:MCEM}
\begin{figure}[!t]
\centering
\includegraphics[width = \textwidth]{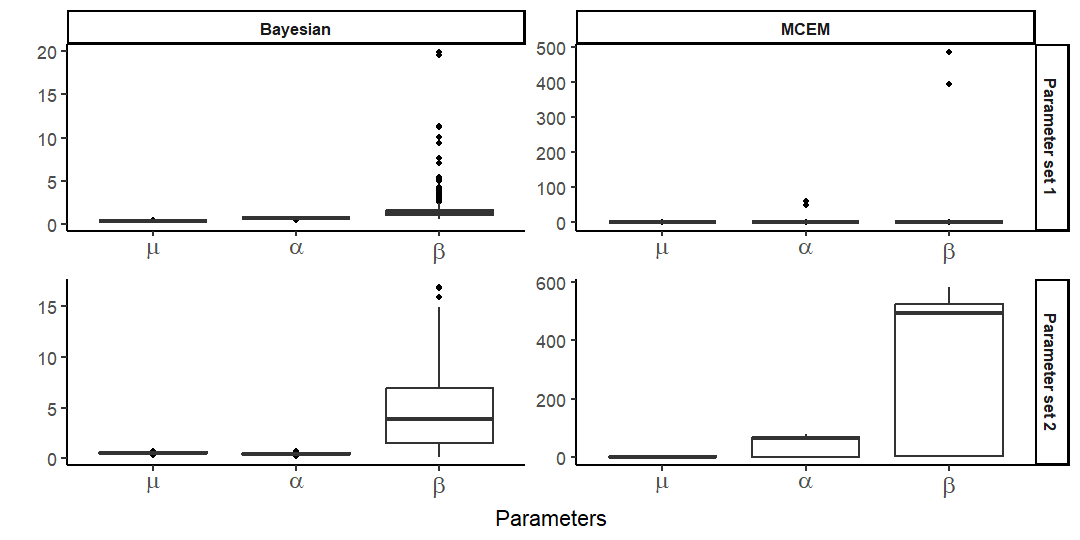}
\caption{Boxplot for estimates of $\mu$,$\alpha$ and $\beta$ using our Bayesian approach (left) and the MC-EM approach (right). The first row corresponds to datasets generated with parameter set 1: $(\mu,\alpha,\beta)=(0.3,0.7,1)$, and aggregation size $\Delta^t=4$. The second row corresponds to datasets generated with parameter set 2: $(\mu,\alpha,\beta)=(0.5,0.5,1)$, and aggregation size $\Delta^t=4$.}
\label{fig: compare_MCEM}
\end{figure}

\begin{figure}[!t]
\centering
\includegraphics[width = \textwidth]{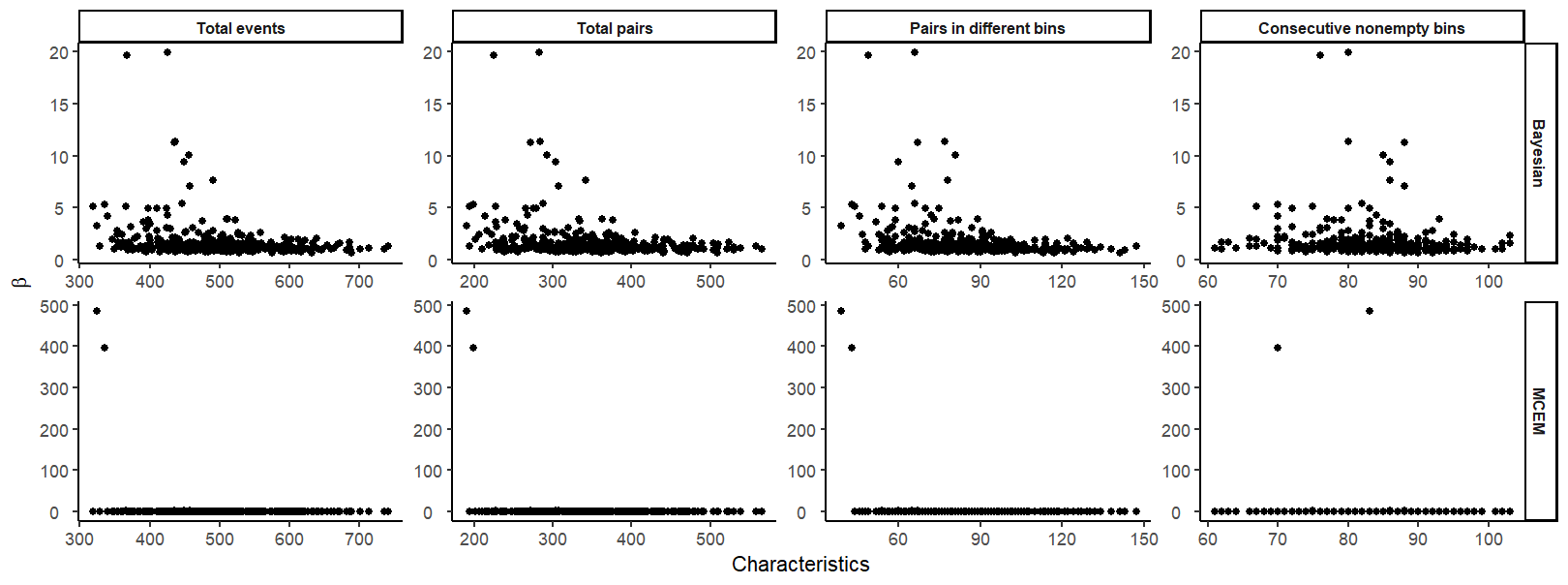}
\caption{Estimates of the $\beta$ parameter plotted against various characteristics of the generated dataset. The first row shows the estimates gained by the Bayesian method, and the second row corresponds to estimates using the MC-EM method. The plot includes 400 simulations, with the true parameter set as $(\mu,\alpha,\beta)=(0.3,0.7,1)$ and aggregation size $\Delta^t=4$. Characteristics considered are total events, total parent-offspring pairs, total parent-offspring pairs in different bins, and the number of nonempty consecutive bins.} 
\label{fig:compare_char_03}
\end{figure}

\begin{figure}[!t]
\centering
\includegraphics[width = \textwidth]{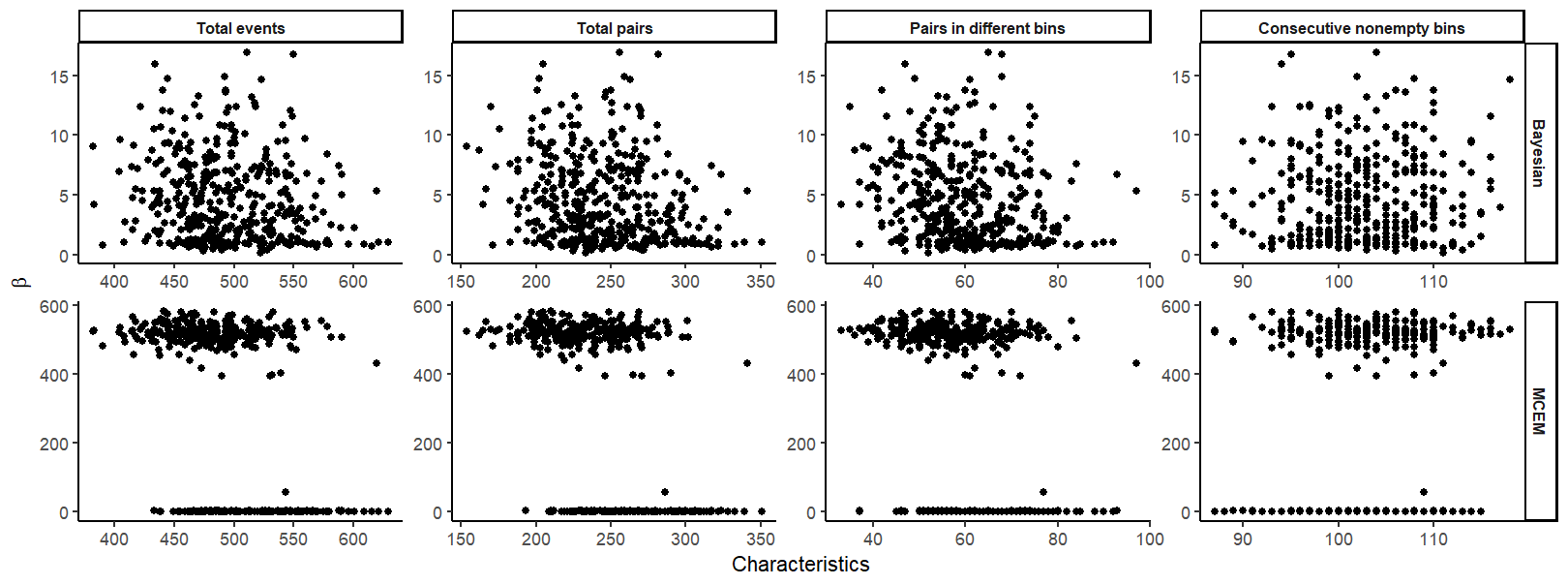}
\caption{Estimates of the $\beta$ parameter plotted against various characteristics of the generated dataset. The first row shows the estimates gained by the Bayesian method, and the second row corresponds to estimates using the MC-EM method. The plot includes 400 simulations, with the true parameter set as $(\mu,\alpha,\beta)=(0.5,0.5,1)$ and aggregation size $\Delta^t=4$. Characteristics considered are total events, total parent-offspring pairs, total parent-offspring pairs in different bins, and the number of nonempty consecutive bins.} 
\label{fig:compare_char_05}
\end{figure}

In this section, we compare the performance of the proposed Bayesian method with the MC-EM method. We generate dataset with two parameter sets specified in Supplement \ref{appendix: updates} with aggregation size $\Delta^t=4$, and then record the estimates gained by  each method. \cref{fig: compare_MCEM} shows the boxplots of estimates for each method. We observe that MC-EM estimates of $\alpha$ and $\beta$ are more variable than the Bayesian estimates.  Specifically, the results from MC-EM are very unstable for parameter set 2 and $\Delta^t=4$, including unreasonably large estimates of $\beta.$ There are some outlier in $\beta$ estimates from both method, which indicates potential converging issues. We plot the estimates of $\beta$ against characteristic of the dataset (\cref{fig:compare_char_03} and \cref{fig:compare_char_05}), and found that it is challenging to learn the $\beta$ parameter in datasets with less total events and total parent-offspring pairs.

\subsection{Simulation results for temporal data}\label{sec: sim_t}

\begin{table}[p]
\centering
\caption{Results of the temporal simulation under additional parameter sets shown in (a), (b) and (c). The table shows the average posterior mean (``Estimate''), average 95\% credible interval length (``CI length''), and coverage rate of the 95\% credible interval (``Coverage'') for parameters $\mu, \alpha$ and $\beta$ of the Hawkes process, across 400 simulated data sets, for the exact data ($\Delta^t = 0$), and temporal aggregations of different sizes. \\ }

\label{tab:sim_temporal_mu}

\subfloat[$(\mu,\alpha,\beta) = (0.1,0.9,1)$]{
\resizebox{0.37\textwidth}{!}{
\begin{tabular}{ c c c c c } 
\toprule
&$\Delta^t$  &  Estimate & CI length & Coverage \\   
\midrule
\multirow{7}*{$\mu$} 
& 0  & 0.1067  & 0.0781  & 0.9475  \\
& 0.5  & 0.1067  & 0.0781  & 0.95\phantom{00}  \\
& 1  & 0.1068  & 0.0784  & 0.9475  \\
& 2  & 0.1069  & 0.0788  & 0.9525  \\
& 3  & 0.1075  & 0.0798  & 0.9525  \\
& 4  & 0.108  & 0.081  & 0.9575  \\
& 5  & 0.1089  & 0.0823  & 0.9525  \\
\hline
\multirow{7}*{$\alpha$} 
& 0  & 0.8713  & 0.1827  & 0.9225  \\
& 0.5  & 0.8713  & 0.1827  & 0.925  \\
& 1  & 0.8712  & 0.183\phantom{0}  & 0.9225  \\
& 2  & 0.871\phantom{0}  & 0.1834  & 0.925  \\
& 3  & 0.8699  & 0.1839  & 0.9175  \\
& 4  & 0.8693  & 0.185\phantom{0}  & 0.9225  \\
& 5  & 0.868\phantom{0}  & 0.1857  & 0.9125  \\
\hline
\multirow{7}*{$\beta$} 
& 0  & 1.0299  & 0.4936  & 0.95\phantom{00}  \\
& 0.5  & 1.0307  & 0.5011  & 0.95\phantom{00}  \\
& 1  & 1.0311  & 0.517\phantom{0}   & 0.955\phantom{0}  \\
& 2  & 1.0432  & 0.5685  & 0.9625  \\
& 3  & 1.0625  & 0.66\phantom{00}   & 0.96\phantom{00}  \\
& 4  & 1.1021  & 0.8445  & 0.9375  \\
& 5  & 1.1868  & 1.2812  & 0.955\phantom{0}  \\
\bottomrule
\end{tabular}
}}
%
\subfloat[$(\mu,\alpha,\beta) = (0.7,0.3,1)$]{
\resizebox{0.37\textwidth}{!}{
\begin{tabular}{ c c c c c } 
\toprule
&$\Delta^t$  &  Estimate & CI length & Coverage \\   
\midrule
\multirow{7}*{$\mu$} 
& 0  & 0.7236  & 0.3113  & 0.95\phantom{00}  \\
& 0.5  & 0.7273  & 0.3141  & 0.9225  \\
& 1  & 0.7334  & 0.3157  & 0.925\phantom{0}  \\
& 2  & 0.7416  & 0.3158  & 0.9125  \\
& 3  & 0.7465  & 0.3003  & 0.91\phantom{00}  \\
& 4  & 0.743\phantom{0}  & 0.2965  & 0.9325  \\
& 5  & 0.7394  & 0.2981  & 0.94\phantom{00} \\
\hline
\multirow{7}*{$\alpha$} 
& 0  & 0.2774  & 0.2892  & 0.945\phantom{0}  \\
& 0.5  & 0.2738  & 0.291  & 0.915\phantom{0}  \\
& 1  & 0.2674  & 0.2917  & 0.9175  \\
& 2  & 0.2592  & 0.2901  & 0.9\phantom{000} \\
& 3  & 0.2542  & 0.2734  & 0.91\phantom{00}  \\
& 4  & 0.2577  & 0.2704  & 0.92 \phantom{00} \\
& 5  & 0.2613  & 0.2732  & 0.95 \phantom{00} \\
\hline
\multirow{7}*{$\beta$} 
& 0  & 1.5022  & 3.3037  & 0.965\phantom{0}  \\
& 0.5  & 1.838  & 5.3629  & 0.935 \phantom{0} \\
& 1  & 2.6731  & 10.227  & 0.93 \phantom{00} \\
& 2  & 4.8999  & 20.316  & 0.89 \phantom{00} \\
& 3  & 7.1738  & 26.8803  & 0.8025  \\
& 4  & 8.4524  & 30.5144  & 0.755 \phantom{0} \\
& 5  & 9.1575  & 32.1314  & 0.79\phantom{00}  \\
\bottomrule
\end{tabular}
}}
%

\subfloat[$(\mu,\alpha,\beta) = (0.9,0.1,1)$]{
\resizebox{0.37\textwidth}{!}{
\begin{tabular}{ c c c c c } 
\toprule
&$\Delta^t$  &  Estimate & CI length & Coverage \\   
\midrule
\multirow{7}*{$\mu$} 
& 0  & 0.905  & 0.3211  & 0.97\phantom{00}  \\
& 0.5  & 0.9087  & 0.3108  & 0.965 \phantom{0} \\
& 1  & 0.9079  & 0.3032  & 0.98 \phantom{00} \\
& 2  & 0.9018  & 0.2931  & 0.9875  \\
& 3  & 0.894  & 0.2924  & 0.99 \phantom{00} \\
& 4  & 0.8862  & 0.2955  & 0.975\phantom{0}  \\
& 5  & 0.8794  & 0.3028  & 0.975 \phantom{0} \\
\hline
\multirow{7}*{$\alpha$} 
& 0  & 0.0985  & 0.2641  & 0.96 \phantom{00} \\
& 0.5  & 0.0943  & 0.2523  & 0.9775  \\
& 1  & 0.095  & 0.2456  & 0.9875  \\
& 2  & 0.101\phantom{0}  & 0.2377  & 0.995 \phantom{0} \\
& 3  & 0.1087  & 0.2396  & 0.9975  \\
& 4  & 0.1166  & 0.2455  & 0.98\phantom{00}  \\
& 5  & 0.1233  & 0.2561  & 0.9775  \\
\hline
\multirow{7}*{$\beta$} 
& 0  & 4.6946  & 19.3281  & 0.9575  \\
& 0.5  & 6.142  & 25.7129  & 0.945\phantom{0}  \\
& 1  & 7.2787  & 29.9935  & 0.9375  \\
& 2  & 8.4626  & 32.7962  & 0.93\phantom{00}  \\
& 3  & 9.2211  & 34.3572  & 0.9275  \\
& 4  & 9.3212  & 34.1318  & 0.9325  \\
& 5  & 9.5876  & 34.9262  & 0.935\phantom{0}  \\
\bottomrule
\end{tabular}
}}
\end{table}

We run additional simulations for more challenging cases where data are generated from a very small or large branching ratio $\alpha$. In these cases, $\mu$ is chosen to be $1-\alpha$, so the expected total number of events are around 500 for all the cases we considered. The results are summarized in \cref{tab:sim_temporal_mu}. Estimation of $\mu$ and $\alpha$ is closed to unbiased in all cases, with adequate inferential performance and coverage near 95\% throughout. We notice that the estimation for $\beta$ that drives the excitation density function deteriorates when $\alpha$ is small. However, that is true even when the exact data are known ($\Delta^t = 0$) which illustrates that this is not an issue of the proposed approach, rather than the estimability of $\beta$ with a small number of offspring. That said, for all aggregation sizes, the confidence intervals remain wide to manifest limited information on $\beta$, and coverage of the 95\% credible intervals is close to 95\%. 

\cref{tab: sim_t_betas} shows simulation results for temporal data with different values of $\beta$, while holding other parameters constant. We observe that the lengths of the credible intervals increase as the true value of $\beta$ increases. When $\beta$ is small, it takes a short time for an event to trigger another event. Thus, these results indicate that Hawkes processes with short excitation effects are more prone to aggregation. As a result, it is more challenging to recover the Hawkes process parameters from aggregated data, where most events occur in some bins while others remain empty.

\cref{tab:temporal_data_summary} shows the summary for datasets generated under all specifications considered in the temporal simulations. The parameter $\alpha$ is shown to control the total number of offspring, with larger values of $\alpha$ resulting in more parent-offspring pairs. On the other hand, the number of empty bins and the number of parent-offspring pairs in different bins are influenced by the parameter $\beta$. Specifically, when 
$\beta$ is small, there are fewer empty bins and a larger number of offspring pairs.

\begin{table}[!t]
\caption{Simulation results for temporal data with different value of $\beta$. The table shows the average posterior mean (``Estimate''), average 95\% credible interval length (``CI length''), and coverage rate of the 95\% credible interval (``Coverage'') for parameters $\mu, \alpha$ and $\beta$, $\gamma$ of the Hawkes process, across 400 simulated data sets with aggregation size $\Delta^t = 3$. Parameter $\mu$ and $\alpha$ are set to be 0.3 and 0.7 for all simulations, and $\beta = 0.5,1,3,5.$}

\label{tab: sim_t_betas} 

\begin{minipage}{\linewidth}
    \centering

    \medskip
\scalebox{0.86}{
\begin{tabular}{ c c c c c } 
\toprule
&True $\beta$  &  Estimate & CI length & Coverage \\   
\midrule
\multirow{5}*{$\mu$} 
& 0.5  & 0.3245  & 0.2256  & 0.935\phantom{0}  \\
& 1  & 0.3169  & 0.1756  & 0.9496  \\
& 3  & 0.3124  & 0.1406  & 0.965\phantom{0}  \\
& 5  & 0.3087  & 0.131\phantom{0}  & 0.965\phantom{0}  \\
& 7  & 0.3046  & 0.1274  & 0.965\phantom{0}  \\
\hline
\multirow{5}*{$\alpha$} 
& 0.5  & 0.669  & 0.2586  & 0.93\phantom{00}  \\
& 1  & 0.6794  & 0.2101  & 0.927\phantom{0}  \\
& 3  & 0.6815  & 0.1805  & 0.945\phantom{0}  \\
& 5  & 0.6851  & 0.1733  & 0.95\phantom{00}  \\
& 7  & 0.6891  & 0.1711  & 0.955\phantom{0}  \\
\hline
\multirow{5}*{$\beta$} 
& 0.5  & 0.5899  & 0.6073  & 0.935\phantom{0} \\
& 1  & 1.1935  & 1.284\phantom{0}  & 0.9345  \\
& 3  & 5.4116  & 11.1737  & 0.895\phantom{0}  \\
& 5  & 8.9448  & 17.3601  & 0.93\phantom{00} \\
& 7  & 10.6478  & 20.0971  & 0.96\phantom{00}  \\
\bottomrule
\end{tabular}
}
\end{minipage}\hfill
\end{table}

\begin{table}[!t]
\centering
\caption{Summary of the simulated data sets for the temporal simulations. The table shows average percentage of empty bins, average maximum count across different parameter sets, average number of parent offspring pairs in different bins, and average total number of parent offspring pairs} 

\label{tab:temporal_data_summary}
\scalebox{0.95}{
\begin{tabular}{ c c c c c c c c} 
\toprule
$\mu$  &  $\alpha$ & $\beta$ & $\Delta^t$ & Empty bins (\%) & Max count & Different pairs & Total pairs\\   
\midrule
0.1  & 0.9  & 1  & 5  & 44.39  & 49.96  & 82.585  & 420.3575 \\
0.3  & 0.7  & 1  & 5  & 14.36  & 27.55  & 67.2975  & 343.335 \\
0.5  & 0.5  & 1  & 5  & 5.45  & 18.635  & 48.365  & 244.44 \\
0.7  & 0.3  & 1  & 5  & 2.38  & 14.6  & 29.0425  & 147.6025 \\
0.9  & 0.1  & 1  & 5  & 0.97  & 12.21  & 9.7375  & 48.7425 \\
0.3  & 0.7  & 0.5  & 3  & 19.53  & 15.1725  & 175.18  & 339.5025 \\
0.3  & 0.7  & 1  & 3  & 26.6  & 20.2925  & 107.905  & 343.335 \\
0.3  & 0.7  & 3  & 3  & 35.02  & 30.02  & 37.9625  & 343.2425 \\
0.3  & 0.7  & 5  & 3  & 37.1  & 33.9025  & 22.63  & 343.5825 \\
0.3  & 0.7  & 7  & 3  & 38.08  & 35.675  & 16.3725  & 343.335 \\
\bottomrule
\end{tabular}
}

\end{table}

\subsection{Simulation results for spatio-temporal data}\label{sec: sim_st}

\begin{figure}[!b]
\centering
\includegraphics[width = 0.95\textwidth]{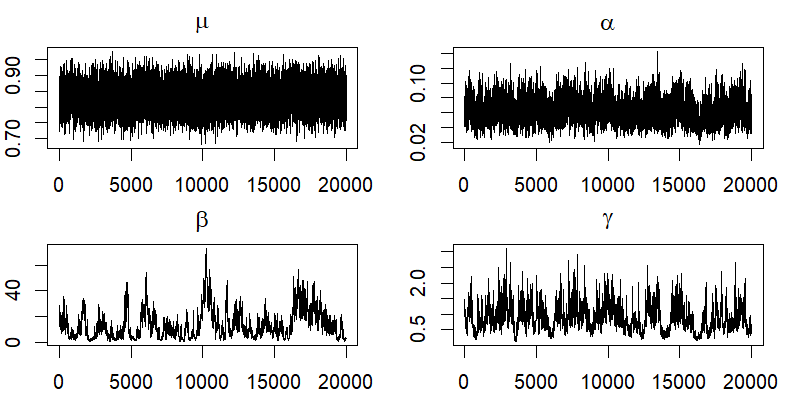}
\caption{Trace plots for a randomly chosen MCMC chain of the aggregated spatio-temporal model, when the data are generated from parameter set $(\mu,\alpha,\beta,\gamma) = (0.9,0.1,1,1)$ with aggregation sizes $\Delta^t = 5$, $\Delta^s=5$}.
\label{fig:st_traceplot}
\end{figure}

We considered spatio-temporal simulations with various choices of $\mu$, and $\alpha$. Specifically, for $\alpha = 1 - \mu$, we considered $\mu = 0.1, 0.3, 0.5, 0.7, 0.9$, as we did in the temporal simulations. The parameters $\beta, \gamma$ were set to 1 throughout.

First, we should mention that we investigated MCMC convergence using the Rhat statistic and by investigating traceplots for the model parameters of randomly chosen simulated data sets. In \cref{fig:st_traceplot} we show traceplots for the model parameters of a randomly chosen simulated data set from the spatio-temporal Hawkes process with parameters $(\mu,\alpha,\beta,\gamma) = (0.9,0.1,1,1)$, and the coarsest aggregation size $\Delta^t = \Delta^s = 5$. We see that even in this challenging setting, the traceplots do not indicate any sign of lack of convergence.


\begin{table}[p]
\caption{The result of the spatio-temporal simulation. The table shows the average posterior mean (``Estimate''), average 95\% credible interval length (``CI length''), and coverage rate of the 95\% credible interval (``Coverage'') for parameters $\mu, \alpha$ and $\beta$, $\gamma$ of the Hawkes process, across 400 simulated data sets. Parameter set: $(\mu,\alpha,\beta,\gamma) = (0.1,0.9,1,1)$}
\label{tab:sim_st_mu01}

\begin{minipage}{\linewidth}
    \centering

    \medskip
\scalebox{0.86}{
\begin{tabular}{c c c c c c c c c c} 
\toprule
\multicolumn{5}{c}{$\mu$} &\multicolumn{5}{c}{$\alpha$} \\
\toprule
$\Delta^t$ &$\Delta^s$  &  Estimate & CI length & Coverage &$\Delta^t$ &$\Delta^s$  &  Estimate & CI length & Coverage\\   
\midrule 
\multirow{4}*{ 0 } & 0  & 0.1016  & 0.0559  & 0.925\phantom{0}   & \multirow{4}*{ 0 } & 0  & 0.8775  & 0.1737  & 0.95\phantom{00}   \\
& 1  & 0.1016  & 0.0559  & 0.92\phantom{00}   & & 1  & 0.8775  & 0.1738  & 0.95\phantom{00}   \\
& 3  & 0.1017  & 0.056  & 0.925\phantom{0}   & & 3  & 0.8775  & 0.1738  & 0.95\phantom{00}   \\
& 5  & 0.1017  & 0.0562  & 0.92\phantom{00}   & & 5  & 0.8774  & 0.174  & 0.95\phantom{00}   \\
\hline
\multirow{4}*{ 1 } & 0  & 0.1016  & 0.0559  & 0.92\phantom{00}   & \multirow{4}*{ 1 } & 0  & 0.8775  & 0.1739  & 0.95\phantom{00}   \\
& 1  & 0.1016  & 0.0559  & 0.925\phantom{0}   & & 1  & 0.8775  & 0.1738  & 0.95\phantom{00}   \\
& 3  & 0.1017  & 0.056  & 0.93\phantom{00}   & & 3  & 0.8774  & 0.1738  & 0.95\phantom{00}   \\
& 5  & 0.1017  & 0.0562  & 0.92\phantom{00}   & & 5  & 0.8774  & 0.1738  & 0.95\phantom{00}   \\
\hline
\multirow{4}*{ 3 } & 0  & 0.1019  & 0.0561  & 0.925\phantom{0}   & \multirow{4}*{ 3 } & 0  & 0.8771  & 0.174  & 0.945  \\
& 1  & 0.1019  & 0.056  & 0.93\phantom{00}   & & 1  & 0.8772  & 0.174  & 0.945\phantom{0}   \\
& 3  & 0.1021  & 0.0562  & 0.93\phantom{00}   & & 3  & 0.877  & 0.1739  & 0.945\phantom{0}   \\
& 5  & 0.1022  & 0.0564  & 0.925\phantom{0}   & & 5  & 0.8768  & 0.174  & 0.94\phantom{00}   \\
\hline
\multirow{4}*{ 5 } & 0  & 0.1023  & 0.0563  & 0.9246  & \multirow{4}*{ 5 } & 0  & 0.8776  & 0.1737  & 0.9497  \\
& 1  & 0.1023  & 0.0563  & 0.9246  & & 1  & 0.8775  & 0.1737  & 0.9447  \\
& 3  & 0.1025  & 0.0565  & 0.9246  & & 3  & 0.8773  & 0.1739  & 0.9447  \\
& 5  & 0.1028  & 0.0568  & 0.9246  & & 5  & 0.877  & 0.174  & 0.9447  \\

\toprule
\multicolumn{5}{c}{$\beta$} &\multicolumn{5}{c}{$\gamma$} \\
\toprule
$\Delta^t$ &$\Delta^s$  &  Estimate & CI length & Coverage &$\Delta^t$ &$\Delta^s$  &  Estimate & CI length & Coverage\\  
\midrule 
\multirow{4}*{ 0 } & 0  & 1.0189  & 0.2857  & 0.95\phantom{00}   & \multirow{4}*{ 0 } & 0  & 1.0026  & 0.1398  & 0.92  \\
& 1  & 1.0187  & 0.2898  & 0.96\phantom{00}   & & 1  & 1.0007  & 0.1544  & 0.94\phantom{00}  \\
& 3  & 1.0242  & 0.3056  & 0.93\phantom{00}   & & 3  & 0.9853  & 0.2185  & 0.935\phantom{0}   \\
& 5  & 1.0278  & 0.3208  & 0.93\phantom{00}   & & 5  & 0.9603  & 0.3076  & 0.86\phantom{00}  \\
\hline
\multirow{4}*{ 1 } & 0  & 1.0214  & 0.2989  & 0.94\phantom{00}   & \multirow{4}*{ 1 } & 0  & 1.0016  & 0.1441  & 0.915  \\
& 1  & 1.022  & 0.3026  & 0.935\phantom{0}   & & 1  & 1.0012  & 0.158  & 0.94\phantom{00}   \\
& 3  & 1.0264  & 0.317  & 0.93\phantom{00}   & & 3  & 0.9853  & 0.2196  & 0.925\phantom{0}   \\
& 5  & 1.0312  & 0.3318  & 0.955\phantom{0}   & & 5  & 0.9596  & 0.3097  & 0.84\phantom{00}   \\
\hline
\multirow{4}*{ 3 } & 0  & 1.0449  & 0.3624  & 0.92\phantom{00}   & \multirow{4}*{ 3 } & 0  & 1.0018  & 0.1508  & 0.93  \\
& 1  & 1.0463  & 0.3654  & 0.935\phantom{0}   & & 1  & 1.0001  & 0.1633  & 0.945\phantom{0}   \\
& 3  & 1.053  & 0.3805  & 0.91\phantom{00}   & & 3  & 0.9868  & 0.2234  & 0.925\phantom{0}   \\
& 5  & 1.0614  & 0.3961  & 0.93\phantom{00}   & & 5  & 0.9624  & 0.31  & 0.855\phantom{0}   \\
\hline
\multirow{4}*{ 5 } & 0  & 1.0707  & 0.4583  & 0.9196  & \multirow{4}*{ 5 } & 0  & 1.0014  & 0.1551  & 0.9347  \\
& 1  & 1.0744  & 0.4665  & 0.9095  & & 1  & 0.9992  & 0.1675  & 0.9598  \\
& 3  & 1.0815  & 0.4839  & 0.9146  & & 3  & 0.9874  & 0.2266  & 0.9146  \\
& 5  & 1.094  & 0.504  & 0.8995  & & 5  & 0.9659  & 0.3112  & 0.8693  \\

\bottomrule
\end{tabular}
}
\end{minipage}\hfill
\end{table}

\begin{table}[p]
\caption{The result of the spatio-temporal simulation. The table shows the average posterior mean (``Estimate''), average 95\% credible interval length (``CI length''), and coverage rate of the 95\% credible interval (``Coverage'') for parameters $\mu, \alpha$ and $\beta$, $\gamma$ of the Hawkes process, across 400 simulated data sets. Parameter set: $(\mu,\alpha,\beta,\gamma) = (0.3,0.7,1,1)$}
\label{tab:sim_st_mu03}

\begin{minipage}{\linewidth}
    \centering

    \medskip
\scalebox{0.86}{
\begin{tabular}{c c c c c c c c c c} 
\toprule
\multicolumn{5}{c}{$\mu$} &\multicolumn{5}{c}{$\alpha$} \\
\toprule
$\Delta^t$ &$\Delta^s$  &  Estimate & CI length & Coverage &$\Delta^t$ &$\Delta^s$  &  Estimate & CI length & Coverage\\   
\midrule 
\multirow{4}*{ 0 } & 0  & 0.3019  & 0.0967  & 0.935\phantom{0}  & \multirow{4}*{ 0 } & 0  & 0.6931  & 0.1486  & 0.915\phantom{0}  \\
& 1  & 0.3021  & 0.0968  & 0.93\phantom{00}  & & 1  & 0.6929  & 0.1487  & 0.915\phantom{0}  \\
& 3  & 0.3022  & 0.0972  & 0.93\phantom{00}  & & 3  & 0.6928  & 0.1488  & 0.925\phantom{0}  \\
& 5  & 0.3031  & 0.0978  & 0.925\phantom{0}  & & 5  & 0.6918  & 0.1492  & 0.915\phantom{0}  \\
\hline
\multirow{4}*{ 1 } & 0  & 0.302  & 0.0968  & 0.93\phantom{00}  & \multirow{4}*{ 1 } & 0  & 0.693  & 0.1487  & 0.915  \\
& 1  & 0.3021  & 0.0969  & 0.93\phantom{00}  & & 1  & 0.6929  & 0.1486  & 0.915\phantom{0}  \\
& 3  & 0.3023  & 0.0972  & 0.925\phantom{0}  & & 3  & 0.6927  & 0.1489  & 0.925\phantom{0}  \\
& 5  & 0.3034  & 0.0979  & 0.925\phantom{0}  & & 5  & 0.6916  & 0.1492  & 0.92\phantom{00}  \\
\hline
\multirow{4}*{ 3 } & 0  & 0.3025  & 0.097  & 0.93\phantom{00}  & \multirow{4}*{ 3 } & 0  & 0.6926  & 0.1487  & 0.905\phantom{0}  \\
& 1  & 0.3026  & 0.0971  & 0.93\phantom{00}  & & 1  & 0.6925  & 0.1489  & 0.92\phantom{00}  \\
& 3  & 0.303  & 0.0975  & 0.925\phantom{0}  & & 3  & 0.692  & 0.149  & 0.915\phantom{0}  \\
& 5  & 0.3045  & 0.0984  & 0.93\phantom{00}  & & 5  & 0.6904  & 0.1493  & 0.92\phantom{00}  \\
\hline
\multirow{4}*{ 5 } & 0  & 0.3031  & 0.0974  & 0.93\phantom{0}  & \multirow{4}*{ 5 } & 0  & 0.6918  & 0.1489  & 0.91\phantom{00}  \\
& 1  & 0.3033  & 0.0974  & 0.925\phantom{0}  & & 1  & 0.6916  & 0.1489  & 0.9\phantom{000}  \\
& 3  & 0.3042  & 0.098  & 0.92\phantom{00}  & & 3  & 0.6907  & 0.1491  & 0.9\phantom{000}  \\
& 5  & 0.3059  & 0.0993  & 0.915\phantom{0}  & & 5  & 0.6888  & 0.1498  & 0.895\phantom{0}  \\

\toprule
\multicolumn{5}{c}{$\beta$} &\multicolumn{5}{c}{$\gamma$} \\
\toprule
$\Delta^t$ &$\Delta^s$  &  Estimate & CI length & Coverage &$\Delta^t$ &$\Delta^s$  &  Estimate & CI length & Coverage\\  
\midrule 
\multirow{4}*{ 0 } & 0  & 1.0172  & 0.2564  & 0.925\phantom{0}  & \multirow{4}*{ 0 } & 0  & 0.9986  & 0.129  & 0.945\phantom{0}  \\
& 1  & 1.0177  & 0.2588  & 0.915\phantom{0}  & & 1  & 0.9952  & 0.1441  & 0.915\phantom{0}  \\
& 3  & 1.0195  & 0.2681  & 0.93\phantom{00}  & & 3  & 0.9886  & 0.2213  & 0.93\phantom{00}  \\
& 5  & 1.0244  & 0.2769  & 0.905\phantom{0}  & & 5  & 0.9493  & 0.3386  & 0.9\phantom{000}  \\
\hline
\multirow{4}*{ 1 } & 0  & 1.0178  & 0.268  & 0.93\phantom{00}  & \multirow{4}*{ 1 } & 0  & 0.998  & 0.1322  & 0.96\phantom{00}  \\
& 1  & 1.0179  & 0.2703  & 0.92\phantom{00}  & & 1  & 0.9954  & 0.1467  & 0.93\phantom{00}  \\
& 3  & 1.0206  & 0.279  & 0.905\phantom{0}  & & 3  & 0.9884  & 0.2224  & 0.92\phantom{00} \\
& 5  & 1.0258  & 0.2886  & 0.905\phantom{0}  & & 5  & 0.9493  & 0.3396  & 0.905\phantom{0} \\
\hline
\multirow{4}*{ 3 } & 0  & 1.0363  & 0.3356  & 0.925  & \multirow{4}*{ 3 } & 0  & 0.9979  & 0.1366  & 0.955\phantom{0}  \\
& 1  & 1.0375  & 0.3374  & 0.905\phantom{0}  & & 1  & 0.9956  & 0.1505  & 0.93\phantom{00}  \\
& 3  & 1.0426  & 0.3482  & 0.915\phantom{0}  & & 3  & 0.9884  & 0.2258  & 0.94\phantom{00} \\
& 5  & 1.0541  & 0.3638  & 0.89\phantom{00}  & & 5  & 0.9505  & 0.3401  & 0.915\phantom{0}  \\
\hline
\multirow{4}*{ 5 } & 0  & 1.0785  & 0.4793  & 0.925\phantom{0}  & \multirow{4}*{ 5 } & 0  & 1.0001  & 0.1397  & 0.94\phantom{00}  \\
& 1  & 1.0792  & 0.4794  & 0.93\phantom{00}  & & 1  & 0.9954  & 0.1534  & 0.93\phantom{00}  \\
& 3  & 1.0919  & 0.5102  & 0.925\phantom{0}  & & 3  & 0.9873  & 0.2289  & 0.93\phantom{00}  \\
& 5  & 1.1092  & 0.5476  & 0.9\phantom{000}  & & 5  & 0.957  & 0.3463  & 0.915\phantom{0}  \\

\bottomrule
\end{tabular}
}
\end{minipage}\hfill
\end{table}


\begin{table}[p]
\caption{Simulation results for spatio-temporal data. The table shows the average posterior mean (``Estimate''), average 95\% credible interval length (``CI length''), and coverage rate of the 95\% credible interval (``Coverage'') for parameters $\mu, \alpha$ and $\beta$, $\gamma$ of the Hawkes process, across 400 simulated data sets. Parameter set: $(\mu,\alpha,\beta,\gamma) = (0.5,0.5,1,1)$ }
\label{tab:sim_st_mu05}

\begin{minipage}{\linewidth}
    \centering

    \medskip
\scalebox{0.86}{
\begin{tabular}{c c c c c c c c c c} 
\toprule
\multicolumn{5}{c}{$\mu$}  &\multicolumn{5}{c}{$\alpha$}  \\
\toprule
$\Delta^t$ &$\Delta^s$  &  Estimate & CI length & Coverage &$\Delta^t$ &$\Delta^s$  &  Estimate & CI length & Coverage\\    
\midrule 
\multirow{4}*{ 0 } & 0  & 0.5014  & 0.1248  & 0.945\phantom{0}  & \multirow{4}*{ 0 } & 0  & 0.4951  & 0.1252  & 0.94\phantom{00}  \\
& 1  & 0.5015  & 0.1249  & 0.945\phantom{0}   & & 1  & 0.4951  & 0.1253  & 0.94\phantom{00}  \\
& 3  & 0.5021  & 0.1255  & 0.945\phantom{0}   & & 3  & 0.4945  & 0.1257  & 0.935\phantom{0}   \\
& 5  & 0.5028  & 0.1267  & 0.945\phantom{0}   & & 5  & 0.4938  & 0.1268  & 0.94\phantom{00}  \\
\hline
\multirow{4}*{ 1 } & 0  & 0.5015  & 0.1248  & 0.945\phantom{0}   & \multirow{4}*{ 1 } & 0  & 0.4951  & 0.1252  & 0.945\phantom{0}  \\
& 1  & 0.5015  & 0.125  & 0.945\phantom{0}   & & 1  & 0.4951  & 0.1254  & 0.94\phantom{00}  \\
& 3  & 0.5023  & 0.1256  & 0.945\phantom{0}   & & 3  & 0.4943  & 0.1259  & 0.935\phantom{0}   \\
& 5  & 0.5029  & 0.1268  & 0.945\phantom{0}   & & 5  & 0.4937  & 0.1269  & 0.94\phantom{00}  \\
\hline
\multirow{4}*{ 3 } & 0  & 0.5022  & 0.1251  & 0.945\phantom{0}   & \multirow{4}*{ 3 } & 0  & 0.4944  & 0.1254  & 0.945  \\
& 1  & 0.5024  & 0.1253  & 0.945\phantom{0}   & & 1  & 0.4943  & 0.1255  & 0.945\phantom{0}   \\
& 3  & 0.5032  & 0.1259  & 0.945\phantom{0}   & & 3  & 0.4934  & 0.1259  & 0.935\phantom{0}   \\
& 5  & 0.5042  & 0.1274  & 0.94\phantom{0}   & & 5  & 0.4924  & 0.1272  & 0.945\phantom{0}   \\
\hline
\multirow{4}*{ 5 } & 0  & 0.5031  & 0.1256  & 0.94\phantom{00}  & \multirow{4}*{ 5 } & 0  & 0.4935  & 0.1256  & 0.94\phantom{00}  \\
& 1  & 0.5033  & 0.1257  & 0.945\phantom{0}   & & 1  & 0.4932  & 0.1258  & 0.935\phantom{0}   \\
& 3  & 0.5048  & 0.1268  & 0.94\phantom{00}  & & 3  & 0.4917  & 0.1265  & 0.93\phantom{00}  \\
& 5  & 0.507  & 0.1288  & 0.945\phantom{0}   & & 5  & 0.4895  & 0.1281  & 0.945\phantom{0}   \\

\toprule
\multicolumn{5}{c}{$\beta$}  &\multicolumn{5}{c}{$\gamma$}  \\
\toprule
$\Delta^t$ &$\Delta^s$  &  Estimate & CI length & Coverage &$\Delta^t$ &$\Delta^s$  &  Estimate & CI length & Coverage\\ 
\midrule 
\multirow{4}*{ 0 } & 0  & 1.0238  & 0.2879  & 0.92\phantom{00}  & \multirow{4}*{ 0 } & 0  & 0.9953  & 0.1448  & 0.94\phantom{00}  \\
& 1  & 1.0246  & 0.2906  & 0.92\phantom{00}  & & 1  & 0.9933  & 0.1642  & 0.92\phantom{00}  \\
& 3  & 1.0276  & 0.3014  & 0.94\phantom{00}  & & 3  & 0.9753  & 0.276  & 0.925\phantom{0}  \\
& 5  & 1.03  & 0.3129  & 0.94\phantom{00}  & & 5  & 0.9388  & 0.4352  & 0.935\phantom{0}  \\
\hline
\multirow{4}*{ 1 } & 0  & 1.0229  & 0.304  & 0.94\phantom{00}  & \multirow{4}*{ 1 } & 0  & 0.9954  & 0.1476  & 0.95\phantom{00}  \\
& 1  & 1.0238  & 0.3058  & 0.935\phantom{0}  & & 1  & 0.9932  & 0.1665  & 0.915\phantom{0}  \\
& 3  & 1.0286  & 0.3182  & 0.93\phantom{00}  & & 3  & 0.9747  & 0.2774  & 0.925\phantom{0}  \\
& 5  & 1.0304  & 0.3297  & 0.95\phantom{00}  & & 5  & 0.9402  & 0.4337  & 0.945\phantom{0}  \\
\hline
\multirow{4}*{ 3 } & 0  & 1.062  & 0.4177  & 0.955  & \multirow{4}*{ 3 } & 0  & 0.9951  & 0.1514  & 0.93\phantom{00}  \\
& 1  & 1.0626  & 0.4214  & 0.94\phantom{00}  & & 1  & 0.9931  & 0.1702  & 0.93\phantom{00}  \\
& 3  & 1.069  & 0.4388  & 0.93\phantom{00}  & & 3  & 0.9754  & 0.2813  & 0.93\phantom{00}  \\
& 5  & 1.0779  & 0.4614  & 0.925\phantom{0}  & & 5  & 0.9423  & 0.4385  & 0.965\phantom{0}  \\
\hline
\multirow{4}*{ 5 } & 0  & 1.1243  & 0.6653  & 0.905\phantom{0}  & \multirow{4}*{ 5 } & 0  & 0.9943  & 0.1544  & 0.945\phantom{0}  \\
& 1  & 1.1286  & 0.6756  & 0.905\phantom{0}  & & 1  & 0.9919  & 0.1731  & 0.945\phantom{0}  \\
& 3  & 1.1517  & 0.742  & 0.895\phantom{0}  & & 3  & 0.9726  & 0.2853  & 0.91\phantom{00}  \\
& 5  & 1.1902  & 0.8625  & 0.87\phantom{00}  & & 5  & 0.9384  & 0.4439  & 0.95\phantom{00}  \\

\bottomrule
\end{tabular}
}
\end{minipage}\hfill
\end{table}


\begin{table}[p]
\caption{Simulation results for spatio-temporal data. The table shows the average posterior mean (``Estimate''), average 95\% credible interval length (``CI length''), and coverage rate of the 95\% credible interval (``Coverage'') for parameters $\mu, \alpha$ and $\beta$, $\gamma$ of the Hawkes process, across 400 simulated data sets. Parameter set: $(\mu,\alpha,\beta,\gamma) = (0.7,0.3,1,1)$ }
\label{tab:sim_st_mu07}

\begin{minipage}{\linewidth}
    \centering

    \medskip
\scalebox{0.86}{
\begin{tabular}{c c c c c c c c c c} 
\toprule
\multicolumn{5}{c}{$\mu$}  &\multicolumn{5}{c}{$\alpha$}  \\
\toprule
$\Delta^t$ &$\Delta^s$  &  Estimate & CI length & Coverage &$\Delta^t$ &$\Delta^s$  &  Estimate & CI length & Coverage\\    
\midrule 
\multirow{4}*{ 0 } & 0  & 0.7009  & 0.1477  & 0.94\phantom{00}  & \multirow{4}*{ 0 } & 0  & 0.2986  & 0.0977  & 0.92\phantom{00}  \\
& 1  & 0.7009  & 0.1479  & 0.945\phantom{0}  & & 1  & 0.2985  & 0.098  & 0.92\phantom{00}  \\
& 3  & 0.7015  & 0.1485  & 0.935\phantom{0}  & & 3  & 0.2979  & 0.0989  & 0.935\phantom{0}  \\
& 5  & 0.703  & 0.1503  & 0.935\phantom{0}  & & 5  & 0.2964  & 0.101  & 0.935\phantom{0}  \\
\hline
\multirow{4}*{ 1 } & 0  & 0.701  & 0.1478  & 0.94\phantom{00}  & \multirow{4}*{ 1 } & 0  & 0.2985  & 0.0978  & 0.925  \\
& 1  & 0.701  & 0.1478  & 0.945\phantom{0}  & & 1  & 0.2984  & 0.098  & 0.925\phantom{0}  \\
& 3  & 0.7017  & 0.1487  & 0.935\phantom{0}  & & 3  & 0.2977  & 0.099  & 0.93\phantom{00}  \\
& 5  & 0.7034  & 0.1505  & 0.935\phantom{0}  & & 5  & 0.296  & 0.1012  & 0.93\phantom{00}  \\
\hline
\multirow{4}*{ 3 } & 0  & 0.7014  & 0.148  & 0.94\phantom{00}  & \multirow{4}*{ 3 } & 0  & 0.2981  & 0.0981  & 0.93\phantom{00}  \\
& 1  & 0.7015  & 0.1482  & 0.94\phantom{00}  & & 1  & 0.2979  & 0.0983  & 0.925\phantom{0}  \\
& 3  & 0.7025  & 0.1492  & 0.945\phantom{0}  & & 3  & 0.2968  & 0.0994  & 0.925\phantom{0}  \\
& 5  & 0.705  & 0.1514  & 0.935\phantom{0}  & & 5  & 0.2943  & 0.102  & 0.935\phantom{0}  \\
\hline
\multirow{4}*{ 5 } & 0  & 0.7018  & 0.1484  & 0.94\phantom{00}  & \multirow{4}*{ 5 } & 0  & 0.2976  & 0.0985  & 0.92\phantom{00}  \\
& 1  & 0.702  & 0.1486  & 0.94\phantom{00}  & & 1  & 0.2973  & 0.0988  & 0.92\phantom{00}  \\
& 3  & 0.7043  & 0.1503  & 0.93\phantom{00}  & & 3  & 0.295  & 0.1005  & 0.915\phantom{0}  \\
& 5  & 0.7085  & 0.1534  & 0.9347  & & 5  & 0.2907  & 0.1042  & 0.9296  \\

\toprule
\multicolumn{5}{c}{$\beta$}  &\multicolumn{5}{c}{$\gamma$}  \\
\toprule
$\Delta^t$ &$\Delta^s$  &  Estimate & CI length & Coverage &$\Delta^t$ &$\Delta^s$  &  Estimate & CI length & Coverage\\ 
\midrule 
\multirow{4}*{ 0 } & 0  & 1.028  & 0.3717  & 0.93\phantom{00}  & \multirow{4}*{ 0 } & 0  & 1.0021  & 0.1884  & 0.92\phantom{00}   \\
& 1  & 1.0284  & 0.3757  & 0.945\phantom{0}  & & 1  & 0.9996  & 0.217  & 0.915\phantom{0}  \\
& 3  & 1.0321  & 0.3943  & 0.95\phantom{00}  & & 3  & 0.9684  & 0.391  & 0.93\phantom{00}   \\
& 5  & 1.0374  & 0.4143  & 0.935\phantom{0}  & & 5  & 0.9194  & 0.5991  & 0.9\phantom{000}   \\
\hline
\multirow{4}*{ 1 } & 0  & 1.0386  & 0.4039  & 0.925\phantom{0}  & \multirow{4}*{ 1 } & 0  & 1.0013  & 0.1908  & 0.93\phantom{00}  \\
& 1  & 1.0386  & 0.4071  & 0.935\phantom{0}  & & 1  & 0.9993  & 0.2195  & 0.93\phantom{00}  \\
& 3  & 1.0433  & 0.4281  & 0.94\phantom{00}  & & 3  & 0.968  & 0.3932  & 0.925\phantom{0}  \\
& 5  & 1.0507  & 0.4531  & 0.93\phantom{00}  & & 5  & 0.9175  & 0.6022  & 0.89 \phantom{00} \\
\hline
\multirow{4}*{ 3 } & 0  & 1.1017  & 0.6284  & 0.935\phantom{0}  & \multirow{4}*{ 3 } & 0  & 1.0028  & 0.1957  & 0.935\phantom{0}  \\
& 1  & 1.1027  & 0.6371  & 0.935\phantom{0}  & & 1  & 0.9991  & 0.224  & 0.94\phantom{00}  \\
& 3  & 1.1212  & 0.6879  & 0.93\phantom{00}  & & 3  & 0.9683  & 0.3999  & 0.92\phantom{00}  \\
& 5  & 1.1478  & 0.7559  & 0.91\phantom{00}  & & 5  & 0.9159  & 0.6144  & 0.89\phantom{00}  \\
\hline
\multirow{4}*{ 5 } & 0  & 1.2628  & 1.2986  & 0.905\phantom{0}  & \multirow{4}*{ 5 } & 0  & 1.0016  & 0.1987  & 0.945\phantom{0}  \\
& 1  & 1.3084  & 1.4659  & 0.895\phantom{0}  & & 1  & 0.9965  & 0.227  & 0.955\phantom{0}  \\
& 3  & 1.3851  & 1.8385  & 0.89\phantom{00}  & & 3  & 0.9625  & 0.4058  & 0.93\phantom{00}  \\
& 5  & 2.1876  & 4.6501  & 0.8543  & & 5  & 0.9178  & 0.6223  & 0.9095  \\

\bottomrule
\end{tabular}
}
\end{minipage}\hfill
\end{table}


\begin{table}[p]
\caption{Simulation results for spatio-temporal data. The table shows the average posterior mean (``Estimate''), average 95\% credible interval length (``CI length''), and coverage rate of the 95\% credible interval (``Coverage'') for parameters $\mu, \alpha$ and $\beta$, $\gamma$ of the Hawkes process, across 400 simulated data sets. Parameter set: $(\mu,\alpha,\beta,\gamma) = (0.9,0.1,1,1)$ }
\label{tab:sim_st_mu09}

\begin{minipage}{\linewidth}
    \centering

    \medskip
\scalebox{0.86}{
\begin{tabular}{c c c c c c c c c c} 
\toprule
\multicolumn{5}{c}{$\mu$}  &\multicolumn{5}{c}{$\alpha$}  \\
\toprule
$\Delta^t$ &$\Delta^s$  &  Estimate & CI length & Coverage &$\Delta^t$ &$\Delta^s$  &  Estimate & CI length & Coverage\\    
\midrule 
\multirow{4}*{ 0 } & 0  & 0.9007  & 0.1674  & 0.93\phantom{00}  & \multirow{4}*{ 0 } & 0  & 0.0996  & 0.0586  & 0.94\phantom{00}  \\
& 1  & 0.9009  & 0.1677  & 0.93\phantom{00}  & & 1  & 0.0993  & 0.059  & 0.945\phantom{0}   \\
& 3  & 0.9021  & 0.1687  & 0.935\phantom{0}   & & 3  & 0.0982  & 0.0616  & 0.94\phantom{00}  \\
& 5  & 0.9049  & 0.1704  & 0.93\phantom{00}  & & 5  & 0.0954  & 0.0649  & 0.94\phantom{00}  \\
\hline
\multirow{4}*{ 1 } & 0  & 0.901  & 0.1674  & 0.93\phantom{00}   & \multirow{4}*{ 1 } & 0  & 0.0993  & 0.0587  & 0.935\phantom{0}   \\
& 1  & 0.9013  & 0.1677  & 0.93\phantom{00}  & & 1  & 0.099  & 0.059  & 0.94\phantom{00}  \\
& 3  & 0.9025  & 0.1688  & 0.935\phantom{0}   & & 3  & 0.0978  & 0.0618  & 0.945\phantom{0}   \\
& 5  & 0.906  & 0.1708  & 0.935\phantom{0}   & & 5  & 0.0942  & 0.0653  & 0.93\phantom{00}  \\
\hline
\multirow{4}*{ 3 } & 0  & 0.9013  & 0.1679  & 0.93\phantom{00}  & \multirow{4}*{ 3 } & 0  & 0.0989  & 0.0595  & 0.925\phantom{0}   \\
& 1  & 0.9017  & 0.1681  & 0.93\phantom{00}  & & 1  & 0.0985  & 0.0599  & 0.93\phantom{00}  \\
& 3  & 0.9041  & 0.1696  & 0.93\phantom{00}  & & 3  & 0.0961  & 0.0631  & 0.92\phantom{00}  \\
& 5  & 0.909  & 0.1726  & 0.935\phantom{0}   & & 5  & 0.0912  & 0.0682  & 0.895\phantom{0}   \\
\hline
\multirow{4}*{ 5 } & 0  & 0.9029  & 0.1686  & 0.93\phantom{00}  & \multirow{4}*{ 5 } & 0  & 0.0973  & 0.0604  & 0.93\phantom{00}  \\
& 1  & 0.9038  & 0.1687  & 0.925\phantom{0}   & & 1  & 0.0964  & 0.0608  & 0.925\phantom{0}   \\
& 3  & 0.9079  & 0.1706  & 0.925\phantom{0}   & & 3  & 0.0923  & 0.0638  & 0.9\phantom{000}   \\
& 5  & 0.9132  & 0.1735  & 0.92\phantom{00}  & & 5  & 0.0869  & 0.0685  & 0.89\phantom{00}   \\

\toprule
\multicolumn{5}{c}{$\beta$}  &\multicolumn{5}{c}{$\gamma$}  \\
\toprule
$\Delta^t$ &$\Delta^s$  &  Estimate & CI length & Coverage &$\Delta^t$ &$\Delta^s$  &  Estimate & CI length & Coverage\\ 
\midrule 
\multirow{4}*{ 0 } & 0  & 1.0837  & 0.7479  & 0.9\phantom{000}   & \multirow{4}*{ 0 } & 0  & 1.0061  & 0.365  & 0.965\phantom{0}   \\
& 1  & 1.0863  & 0.7643  & 0.91\phantom{00}  & & 1  & 0.9984  & 0.4315  & 0.96\phantom{00}  \\
& 3  & 1.1099  & 0.8593  & 0.935\phantom{0}   & & 3  & 0.9549  & 0.8205  & 0.94\phantom{00}  \\
& 5  & 1.1406  & 0.9593  & 0.905\phantom{0}   & & 5  & 0.8121  & 1.1877  & 0.915\phantom{0}   \\
\hline
\multirow{4}*{ 1 } & 0  & 1.1221  & 0.892  & 0.925\phantom{0}   & \multirow{4}*{ 1 } & 0  & 1.0043  & 0.368  & 0.975\phantom{0}   \\
& 1  & 1.131  & 0.9219  & 0.915\phantom{0}   & & 1  & 0.9951  & 0.4341  & 0.97\phantom{00}  \\
& 3  & 1.1707  & 1.0832  & 0.92\phantom{00}  & & 3  & 0.9524  & 0.8245  & 0.935\phantom{0}   \\
& 5  & 1.2551  & 1.3819  & 0.89\phantom{00}  & & 5  & 0.7951  & 1.201  & 0.9\phantom{000}  \\
\hline
\multirow{4}*{ 3 } & 0  & 1.5252  & 2.8713  & 0.875\phantom{0}   & \multirow{4}*{ 3 } & 0  & 1.0079  & 0.3807  & 0.975\phantom{0}   \\
& 1  & 1.567  & 3.1375  & 0.875\phantom{0}   & & 1  & 0.9964  & 0.4478  & 0.965\phantom{0}   \\
& 3  & 2.1775  & 5.9797  & 0.85\phantom{00}  & & 3  & 0.9435  & 0.8484  & 0.945\phantom{0}   \\
& 5  & 3.3635  & 11.7478  & 0.825\phantom{0}   & & 5  & 0.8095  & 1.229  & 0.95\phantom{00}  \\
\hline
\multirow{4}*{ 5 } & 0  & 4.0148  & 13.977  & 0.81\phantom{00}  & \multirow{4}*{ 5 } & 0  & 1.011  & 0.3943  & 0.96\phantom{00}  \\
& 1  & 4.2957  & 14.9312  & 0.775\phantom{0}  & & 1  & 0.9951  & 0.464  & 0.965\phantom{0}   \\
& 3  & 6.8014  & 23.7393  & 0.72\phantom{00}  & & 3  & 0.9366  & 0.8834  & 0.945\phantom{0}   \\
& 5  & 9.528  & 32.5244  & 0.62\phantom{00}  & & 5  & 0.834  & 1.2793  & 0.965\phantom{0}   \\

\bottomrule
\end{tabular}
}
\end{minipage}\hfill
\end{table}

\begin{figure}[!b]
\centering
\includegraphics[width = \textwidth,trim=0 5 0 5, clip]{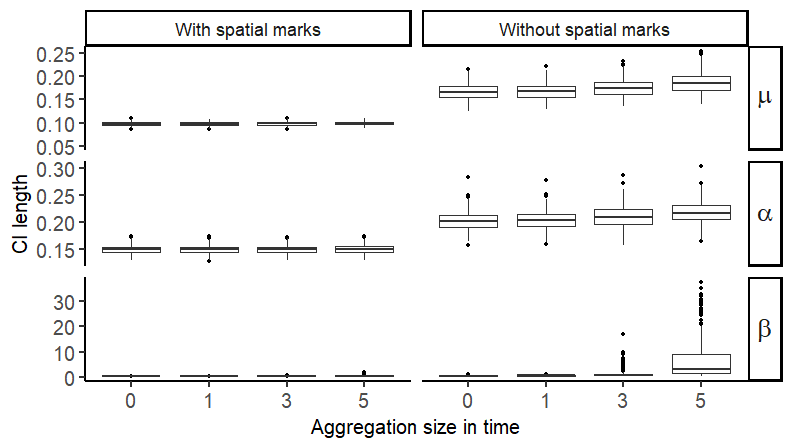}
\caption{Credible Interval Lengths from the Temporal and Spatio-temporal Models. Boxplots of the 95\% credible interval for $\mu$, $\alpha$, and $\beta$ across 400 simulated data sets under parameter set 1, for varying temporal aggregations $\Delta^t = 0,1,3,5$ (horizontal axis) and coarse spatial aggregation $\Delta^{\sss} = 5$ }
\label{fig:compare_t_st}
\end{figure}

\begin{figure}[!b]
\centering
\hspace{10pt}
\includegraphics[width = 0.43\textwidth,trim=10 5 230 8, clip]{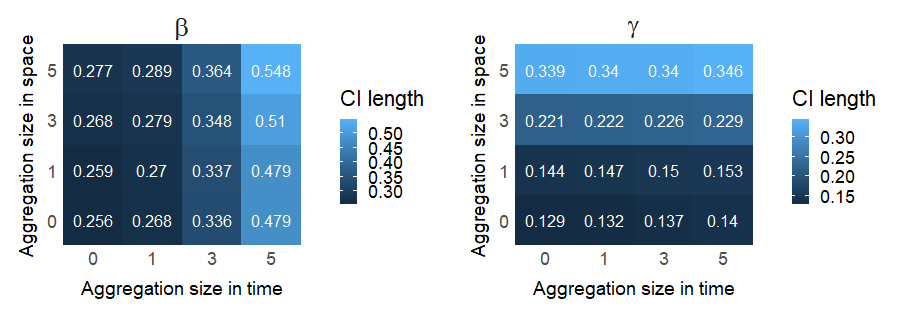}
\hspace{20pt}
\includegraphics[width = 0.43\textwidth,trim=230 5 10 8, clip]{heatmap_st_CI.png}
\caption{Average credible interval length for $\beta$ and $\gamma$ over 400 simulated spatio-temporal data sets under parameter set 1 and based on different aggregation sizes in time and space. }
\label{fig:heatmap_st}
\end{figure}

Tables \ref{tab:sim_st_mu01}, \ref{tab:sim_st_mu03}, \ref{tab:sim_st_mu05}, \ref{tab:sim_st_mu07} and \ref{tab:sim_st_mu09} show the spatio-temporal simulation results for $\mu = 0.1, 0.3, 0.5, 0.7$ and $0.9$, respectively. Tables \ref{tab:sim_st_mu03} and \ref{tab:sim_st_mu05} correspond to parameter sets 1 and 2 in the manuscript, respectively, and they compliment the results and text shown in Section 4.2 of the manuscript. We see that the estimator performs well for $\mu$, $\alpha$ and $\gamma$ across all scenarios. In nearly all scenarios and parameters, coverage of the 95\% credible intervals is close to 95\%.
Focusing on the most challenging scenario (\cref{tab:sim_st_mu09}), we see that there is some bias in estimating $\beta$ when the branching ratio is small ($\alpha = 0.1$). In this challenging setting estimating $\beta$ becomes harder with increasing length of credible intervals as the coarseness of temporal aggregation increases. However, coverage is above 80\% for all parameters in most cases.

Estimation of all parameters is improved when spatial information is available, compared to the situation with temporal data only. \cref{fig:compare_t_st} shows the boxplots of credible interval lengths of parameter $\mu, \alpha$ and $\beta$ when the aggregation size in time is 5. We can see that the length of the credible intervals in the spatio-temporal simulations are substantially lower than those in the temporal simulations for the common parameters ($\mu, \alpha, \beta$).

\subsection{Simulation results for multivariate spatio-temporal data}

\begin{figure}[!thb]
\centering

\includegraphics[width = \textwidth,trim=0 0 0 0, clip]{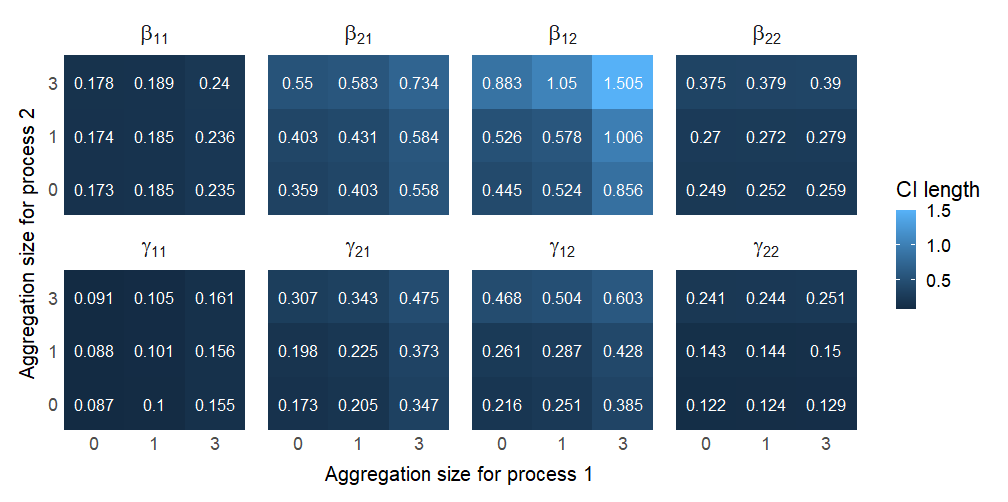}

\caption{Average credible interval length for $\beta_{m,l}$ and $\gamma_{m,l}$ where $m,l \in \{1,2\}$ over 400 simulated multiple spatio-temporal data sets and based on different aggregation sizes in each process.}
\label{fig:heatmap_st2}
\end{figure}

\begin{table}[!b]
\caption{The result of selected parameters ($\mu_1,\alpha_{21},\beta_{11},\gamma_{21}$) in the multivariate spatio-temporal simulation. True parameter: ($\mu_1,\alpha_{21},\beta_{11},\gamma_{21}$) = (0.3,0.3,1,1)}
\label{tab:sim_st2}

\begin{minipage}{\linewidth}
    \centering

    \medskip
\scalebox{0.86}{
\begin{tabular}{c c c c c c c c c c} 
\toprule
\multicolumn{5}{c}{$\mu_1$}  &\multicolumn{5}{c}{$\alpha_{21}$}  \\
\toprule
$\Delta_1$ &$\Delta_2$  &  Estimate & CI length & Coverage &$\Delta_1$ &$\Delta_2$  &  Estimate & CI length & Coverage\\   
\midrule 
\multirow{3}*{ 0 } & 0  & 0.3044  & 0.0983  & 0.9425  & \multirow{3}*{ 0 } & 0  & 0.2994  & 0.089  & 0.9475  \\
& 1  & 0.3047  & 0.1003  & 0.9525  & & 1  & 0.2992  & 0.0921  & 0.96\phantom{00}  \\
& 3  & 0.3057  & 0.1048  & 0.9525  & & 3  & 0.298\phantom{0}  & 0.1016  & 0.9625  \\
\hline
\multirow{3}*{ 1 } & 0  & 0.3051  & 0.1004  & 0.94\phantom{00}  & \multirow{3}*{ 1 } & 0  & 0.2991  & 0.0927  & 0.9525  \\
& 1  & 0.3053  & 0.1014  & 0.9575  & & 1  & 0.2989  & 0.0947  & 0.96\phantom{00}  \\
& 3  & 0.3065  & 0.1059  & 0.9575  & & 3  & 0.2973  & 0.1044  & 0.965\phantom{0}  \\
\hline
\multirow{3}*{ 3 } & 0  & 0.3069  & 0.105\phantom{0}  & 0.9375  & \multirow{3}*{ 3 } & 0  & 0.2978  & 0.1033  & 0.9575  \\
& 1  & 0.3075  & 0.1059  & 0.955\phantom{0}  & & 1  & 0.2969  & 0.1049  & 0.9625  \\
& 3  & 0.3089  & 0.1095  & 0.9524  & & 3  & 0.2951  & 0.1133  & 0.9524  \\

\toprule

\multicolumn{5}{c}{$\beta_{11}$}  &\multicolumn{5}{c}{$\gamma_{12}$}  \\
\toprule
$\Delta_1$ &$\Delta_2$  &  Estimate & CI length & Coverage &$\Delta_1$ &$\Delta_2$  &  Estimate & CI length & Coverage\\ 
\midrule 
\multirow{3}*{ 0 } & 0  & 1.0106  & 0.1729  & 0.9575  & \multirow{3}*{ 0 } & 0  & 0.9992  & 0.1728  & 0.95\phantom{00}  \\
& 1  & 1.0103  & 0.1742  & 0.9475  & & 1  & 0.999\phantom{0}  & 0.1979  & 0.94\phantom{00}  \\
& 3  & 1.0083  & 0.1777  & 0.96\phantom{00}  & & 3  & 0.9972  & 0.3065  & 0.935\phantom{0}  \\
\hline
\multirow{3}*{ 1 } & 0  & 1.0139  & 0.1846  & 0.955\phantom{0}  & \multirow{3}*{ 1 } & 0  & 0.9971  & 0.2052  & 0.965\phantom{0}  \\
& 1  & 1.0135  & 0.1854  & 0.9525  & & 1  & 0.9978  & 0.2252  & 0.9525  \\
& 3  & 1.0113  & 0.189\phantom{0}  & 0.9475  & & 3  & 0.9944  & 0.3434  & 0.9425  \\
\hline
\multirow{3}*{ 3 } & 0  & 1.0265  & 0.2346  & 0.9425  & \multirow{3}*{ 3 } & 0  & 0.9857  & 0.3468  & 0.945\phantom{0}  \\
& 1  & 1.0261  & 0.2359  & 0.9475  & & 1  & 0.983\phantom{0}  & 0.3726  & 0.9375  \\
& 3  & 1.0257  & 0.2404  & 0.9398  & & 3  & 0.9856  & 0.4751  & 0.9424  \\

\bottomrule
\end{tabular}
}
\end{minipage}\hfill
\end{table}

In \cref{tab:sim_st2} we show simulation results for data from a multivariate spatio-temporal Hawkes process. These results compliment the results and text shown in Section 4.3 of the manuscript. Since the number of parameters is large, we show results corresponding to a subset of them. We include parameters that represent background intensity ($\mu_1$), self-excitation ($\beta_{11}$), and cross-excitation ($\alpha_{21}, \gamma_{21}$). Results for other parameters are similar. 

We see that the posterior mean of all estimated parameters is, on average, near the true values, irrespective of aggregation size. Confidence intervals are wider under coarser aggregations. We see that the credible interval length for the parameters of the first component of the process $(\mu_1, \beta_{11})$ is not affected by the aggregation size for the second process $(\Delta^2)$. However, the credible interval length for the cross-excitation parameters ($\alpha_{21}, \gamma_{21}$) is affected by the aggregation size of both processes. Coverage of 95\% credible intervals is near the nominal level throughout.

\subsection{Simulation results for Hawkes processes with Lomax kernel}\label{appendix: Lomax_sim}

In this section, we run simulations for Hawkes process with Lomax kernel. In other words, we assume $g(t)$ in the conditional intensity function of temporal Hawkes process and $g_1(t)$ are probability density functions of two-parameter Lomax distribution, which is in the form: $\frac{(p-1)c^{p-1}}{(t+c)^p}$. The $g_2(\sss)$ function in the spatio-temporal is still assumed to be a bivariate normal density. For both temporal and spatio-temporal cases, we generate 400 data sets with chosen parameters and fit the model with exact and aggregated data with varying aggregation sizes.  We choose $\mu = 0.3,\alpha = 0.7$ for the temporal cases, and $\mu = 0.5,\alpha = 0.5,\gamma=1$ for the spatio-temporal case. The parameters in the Lomax kernel was set to $c = 10$, $p=12$ for both temporal and spatio-temporal cases. The mean and variance of the excitation times under this Lomax specification are both 1 (as for the exponential kernel with parameter 1 used in the other simulations).

\begin{figure}[p]
\centering
\includegraphics[width = 0.9\textwidth]{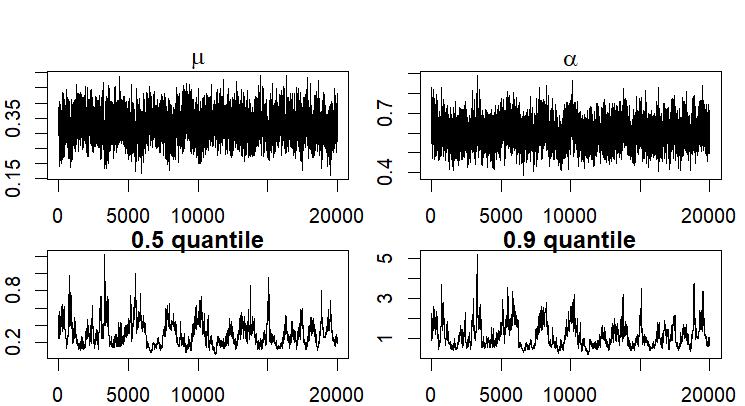}
\caption{Trace plots for a randomly chosen MCMC chain of the aggregated temporal model with Lomax kernel, when the data are generated from parameter set $(\mu,\alpha,c,p) = (0.3,0.7,10,13)$ with aggregation sizes $\Delta^t = 5$}
\label{fig:Lomax_t_traceplot}

\centering
\includegraphics[width = 0.95\textwidth]{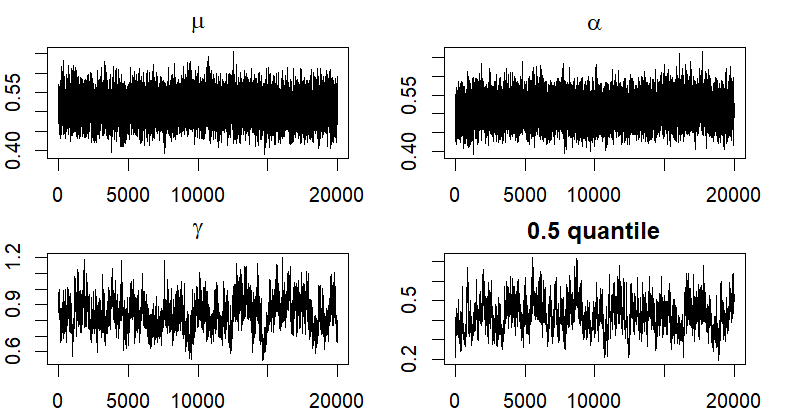}
\caption{Trace plots for a randomly chosen MCMC chain of the aggregated spatio-temporal model, when the data are generated from parameter set $(\mu,\alpha,c,p,\gamma) = (0.5,0.5,10,13,1)$ with aggregation sizes $\Delta^t = 5$, $\Delta^s=5$ }.
\label{fig:Lomax_st_traceplot}
\end{figure}

Although the model is theoretically identifiable, we find that the parameters of the Lomax excitation function $c$ and $p$ are not individually well identified, {\it even in the case of no aggregation}. Thus, instead of studying $c$ and $p$ individually, we study whether the excitation function is estimated appropriately by focusing on quantiles of the Lomax distribution. We show results for the median here, though other quantiles were also studied. Firstly, we analyze the convergence of the MCMC by investigating trace plots of randomly selected data sets. \cref{fig:Lomax_t_traceplot} show the trace plots for parameter $\mu$, $\alpha$ and quantiles of the Lomax function in the most challenging case ($\Delta^t = 5$) for temporal simulation, and \cref{fig:Lomax_st_traceplot} show the trace plot for parameter $\mu$, $\alpha$, $\gamma$ and the median of the Lomax function in the most challenging case ($\Delta^t = \Delta^s = 5$) for the spatio-temporal simulation, for a random data set. Both plots show relative good convergence of the posterior samples, without any signs of convergence issues.

The simulation results for temporal and spatio-temporal simulations are shown in \cref{tab:sim_temporal_Lomax} and \cref{tab:sim_st_Lomax} respectively. We observed that the posterior mean for the $\mu$, $\alpha$, $\gamma$, and quantiles of the Lomax function are generally close to the true values in all scenarios, indicating that the estimates are overall unbiased and the fitted Lomax kernel function is close to the truth. The bias increases a little in the temporal simulations as aggregation sizes get larger, while are almost unchanged in the spatial temporal simulations. The coverage rates for the model parameters $\mu, \alpha$ and $\gamma$ in the all simulations are close to 95\%. The coverage rate for $\beta$ and $\gamma$ drops a little the scenarios with course aggregation in both time and space, while the coverage maintains above 90\% throughout.  
Similar to simulations for exponential kernel function, the credible interval lengths increases as the aggregation size increases. Moreover, the credible interval lengths are generally smaller when the spatial information are available.


\begin{table}[!t]
\centering
\caption{Results of the temporal simulation with Lomax kernel. The table shows the average posterior mean (``Estimate''), average 95\% credible interval length (``CI length''), and coverage rate of the 95\% credible interval (``Coverage'') for parameters $\mu, \alpha$ of the Hawkes process and 0.5 quantile of the Lomax distribution, across 400 simulated data sets, for the exact data ($\Delta^t = 0$), temporal aggregations of different sizes. Parameter set: $(\mu,\alpha,c,p) = (0.3,0.7,10,12)$}
\label{tab:sim_temporal_Lomax}
\scalebox{0.95}{
\begin{tabular}{ c c c c c } 
\toprule
&$\Delta^t$  &  Estimate & CI length & Coverage \\   
\midrule
\multirow{7}*{$\mu$} 
& 0  & 0.3021  & 0.1743  & 0.9625  \\
& 0.5  & 0.3023  & 0.1752  & 0.9575  \\
& 1  & 0.3027  & 0.1764  & 0.965\phantom{0}  \\
& 2  & 0.3039  & 0.1785  & 0.955\phantom{0}   \\
& 3  & 0.3061  & 0.1825  & 0.9475  \\
& 4  & 0.3093  & 0.1875  & 0.9475  \\
& 5  & 0.3121  & 0.1917  & 0.94\phantom{00}   \\
\hline
\multirow{7}*{$\alpha$} 
& 0  & 0.6969  & 0.2111  & 0.96\phantom{00}   \\
& 0.5  & 0.6966  & 0.2119  & 0.95\phantom{00}  \\
& 1  & 0.6962  & 0.213  & 0.955\phantom{0}   \\
& 2  & 0.695  & 0.2149  & 0.9475  \\
& 3  & 0.6927  & 0.2182  & 0.9425  \\
& 4  & 0.6893  & 0.2223  & 0.94\phantom{00}   \\
& 5  & 0.6864  & 0.2262  & 0.9425  \\
\hline
\multirow{7}*{0.5 quantile} 
& 0  & 0.6762  & 0.4114  & 0.9575  \\
& 0.5  & 0.6758  & 0.4215  & 0.9575  \\
& 1  & 0.6741  & 0.4416  & 0.945\phantom{0}   \\
& 2  & 0.668  & 0.4917  & 0.95\phantom{00}   \\
& 3  & 0.6556  & 0.5619  & 0.9425  \\
& 4  & 0.6335  & 0.6531  & 0.94\phantom{00}   \\
& 5  & 0.6049  & 0.752  & 0.94\phantom{00}   \\
\bottomrule
\end{tabular}
}
\end{table}

%

\begin{table}[p]
\caption{Simulation results for spatio-temporal data with Lomax kernel. The table shows the average posterior mean (``Estimate''), average 95\% credible interval length (``CI length''), and coverage rate of the 95\% credible interval (``Coverage'') for parameters $\mu, \alpha$ and $\gamma$ of the Hawkes process, and 0.5 quantile of the Lomax distribution, across 400 simulated data sets. Parameter set: $(\mu,\alpha,c,p,\gamma) = (0.5,0.5,11,12,1)$ }
\label{tab:sim_st_Lomax}

    \centering

    \medskip
\scalebox{0.86}{
\begin{tabular}{c c c c c c c c c c} 
\toprule
\multicolumn{5}{c}{$\mu$}  &\multicolumn{5}{c}{$\alpha$}  \\
\toprule
$\Delta^t$ &$\Delta^s$  &  Estimate & CI length & Coverage &$\Delta^t$ &$\Delta^s$  &  Estimate & CI length & Coverage\\    
\midrule
\multirow{4}*{ 0 } & 0  & 0.5034  & 0.1252  & 0.955\phantom{0}  & \multirow{4}*{ 0 } & 0  & 0.4941  & 0.1251  & 0.9425  \\
& 1  & 0.5035  & 0.1253  & 0.955\phantom{0}  & & 1  & 0.4941  & 0.1252  & 0.9425  \\
& 3  & 0.5036  & 0.1258  & 0.95\phantom{00}  & & 3  & 0.4939  & 0.1257  & 0.945\phantom{0}  \\
& 5  & 0.5045  & 0.1271  & 0.9525  & & 5  & 0.493  & 0.1268  & 0.9375  \\
\hline
\multirow{4}*{ 1 } & 0  & 0.5034  & 0.1253  & 0.955\phantom{0}  & \multirow{4}*{ 1 } & 0  & 0.4941  & 0.1251  & 0.945\phantom{0}  \\
& 1  & 0.5035  & 0.1254  & 0.955\phantom{0}  & & 1  & 0.4941  & 0.1253  & 0.9425  \\
& 3  & 0.5037  & 0.126  & 0.95\phantom{00}  & & 3  & 0.4938  & 0.1259  & 0.945\phantom{0}  \\
& 5  & 0.5046  & 0.1273  & 0.955\phantom{0}  & & 5  & 0.4929  & 0.1269  & 0.9325  \\
\hline
\multirow{4}*{ 3 } & 0  & 0.5037  & 0.1254  & 0.9575  & \multirow{4}*{ 3 } & 0  & 0.4938  & 0.1253  & 0.9425  \\
& 1  & 0.5038  & 0.1256  & 0.955\phantom{0}  & & 1  & 0.4937  & 0.1254  & 0.9425  \\
& 3  & 0.5042  & 0.1263  & 0.95\phantom{00}  & & 3  & 0.4932  & 0.126\phantom{0}  & 0.94\phantom{00}  \\
& 5  & 0.5054  & 0.1278  & 0.9575  & & 5  & 0.4921  & 0.1273  & 0.935\phantom{0}  \\
\hline
\multirow{4}*{ 5 } & 0  & 0.5041  & 0.1257  & 0.9575  & \multirow{4}*{ 5 } & 0  & 0.4934  & 0.1255  & 0.945\phantom{0}  \\
& 1  & 0.5042  & 0.1259  & 0.9575  & & 1  & 0.4933  & 0.1257  & 0.94\phantom{00}  \\
& 3  & 0.5052  & 0.1267  & 0.9475  & & 3  & 0.4923  & 0.1264  & 0.9425  \\
& 5  & 0.5068  & 0.1288  & 0.9475  & & 5  & 0.4906  & 0.1281  & 0.935\phantom{0}  \\

\toprule
\multicolumn{5}{c}{$\gamma$}  &\multicolumn{5}{c}{0.5 quantile}  \\
\toprule
$\Delta^t$ &$\Delta^s$  &  Estimate & CI length & Coverage &$\Delta^t$ &$\Delta^s$  &  Estimate & CI length & Coverage\\    
\midrule
\multirow{4}*{ 0 } & 0  & 0.9962  & 0.1457  & 0.93\phantom{00}  & \multirow{4}*{ 0 } & 0  & 0.641  & 0.1993  & 0.94\phantom{00}  \\
& 1  & 0.9929  & 0.1649  & 0.93\phantom{00}  & & 1  & 0.6409  & 0.2002  & 0.9425  \\
& 3  & 0.9755  & 0.2772  & 0.94\phantom{00}  & & 3  & 0.6405  & 0.2056  & 0.9375  \\
& 5  & 0.9497  & 0.4348  & 0.9125  & & 5  & 0.6383  & 0.2108  & 0.94\phantom{00}  \\
\hline
\multirow{4}*{ 1 } & 0  & 0.996\phantom{0}  & 0.1485  & 0.9375  & \multirow{4}*{ 1 } & 0  & 0.6404  & 0.2124  & 0.9475  \\
& 1  & 0.9925  & 0.1672  & 0.9325  & & 1  & 0.64  & 0.2135  & 0.9475  \\
& 3  & 0.9743  & 0.2789  & 0.95\phantom{00}  & & 3  & 0.6397  & 0.2192  & 0.95\phantom{00}  \\
& 5  & 0.9498  & 0.4377  & 0.91\phantom{00}  & & 5  & 0.6369  & 0.2241  & 0.9325  \\
\hline
\multirow{4}*{ 3 } & 0  & 0.995\phantom{0}  & 0.1523  & 0.9225  & \multirow{4}*{ 3 } & 0  & 0.6328  & 0.2677  & 0.945\phantom{0}  \\
& 1  & 0.9919  & 0.1708  & 0.9325  & & 1  & 0.6322  & 0.268  & 0.9375  \\
& 3  & 0.9727  & 0.2828  & 0.935\phantom{0}  & & 3  & 0.6307  & 0.2747  & 0.94\phantom{00}  \\
& 5  & 0.9486  & 0.4437  & 0.9075  & & 5  & 0.6251  & 0.281  & 0.9375  \\
\hline
\multirow{4}*{ 5 } & 0  & 0.9948  & 0.155  & 0.9325  & \multirow{4}*{ 5 } & 0  & 0.6173  & 0.3379  & 0.93\phantom{00}  \\
& 1  & 0.9912  & 0.1736  & 0.9425  & & 1  & 0.6163  & 0.3404  & 0.9325  \\
& 3  & 0.9708  & 0.2864  & 0.935\phantom{0}  & & 3  & 0.6114  & 0.3472  & 0.9225  \\
& 5  & 0.9494  & 0.4494  & 0.9175  & & 5  & 0.6013  & 0.3624  & 0.9175  \\
\bottomrule
\end{tabular}
}
\medskip

\centering

\end{table}

\subsection{Computation efficiency}
\label{supp_sec:sim_temporal_computing}

In this section, we show the computing time of the proposed method under the model specifications in Section \ref{sec: sim_t} and Section \ref{sec: sim_st}. \cref{tab:sim_temporal_computing} shows the shows the minimum, median and maximum time (in seconds) for
running a single MCMC chain of 40,000 iterations, across 400 simulated data sets, for $\mu = 0.1,0.9$ , $\alpha = 1-\mu$ and $\beta = \gamma=1$. We observe that it takes longer time to fit the model on data sets with more offspring (larger $\alpha$) when the total number events is the same, but for both parameter sets, the computing time for getting 40,000 posterior samples when time and locations are both aggregated are about 1.5 times the computing time when time and locations are exact. 

To analyze the computing time for the aggregated temporal model relative to the size of the point pattern, we generate 50 data sets using $(\mu,\alpha,\beta) = (0.3,0.7,1)$ and $t = 100,200,300,400,500.$ Then we fit aggregated model with $\Delta^t = 1$ and record the time for getting 1,000 posteriors samples. Similarly, we generate 50 data sets from the spatio-temporal model with $\mu  = 0.3,\alpha = 0.7,\beta = 1,\gamma = 1$, and $t = 100,200,300,400,500$, and run 1,000 iterations for the aggregated data with $\Delta^t = \Delta^s = 1.$ The relationships between computing time and number of events in aggregated temporal and aggregated spatio-temporal model are shown in \cref{tab:sim_temporal_computing}. We observe a linear trend between the computing time and the size of the point pattern for both temporal and spatio-temporal models.


\begin{table}[!t]
\caption{The table shows the minimum, median and maximum time (in seconds) for running a single MCMC chain of 40,000 iterations, across 400 simulated data sets, for different parameter sets with $\beta=\gamma=1$}
\label{tab:sim_temporal_computing}
\centering
\scalebox{0.9}{
\begin{tabular}{ c c c c c c c c c c} 
\toprule
\multicolumn{5}{c}{$\mu=0.1,\alpha=0.9$} & \multicolumn{5}{c}{$\mu=0.9,\alpha=0.1$}\\
\midrule
$\Delta^t$  &  $\Delta^s$ & Min  & Median & Max & $\Delta^t$  &  $\Delta^s$ & Min  & Median & Max\\   
\midrule
\multirow{4}*{ 0 } & 0  & 220  & 885  & 4384  & \multirow{4}*{ 0 } & 0  & 335  & 615  & 959  \\
& 0.5  & 328  & 1313  & 6547  & & 0.5  & 487  & 939  & 1358  \\
& 1  & 325  & 1329  & 6610  & & 1  & 489  & 950  & 1354  \\
& 1.5  & 325  & 1310  & 6184  & & 1.5  & 491  & 948  & 1332  \\
\hline
\multirow{4}*{ 0.5 } & 0  & 340  & 1351  & 6563  & \multirow{4}*{ 0.5 } & 0  & 536  & 1061  & 1498  \\
& 0.5  & 441  & 1783  & 8223  & & 0.5  & 694  & 1376  & 1840  \\
& 1  & 442  & 1777  & 6715  & & 1  & 700  & 1378  & 1879  \\
& 1.5  & 440  & 1837  & 5699  & & 1.5  & 704  & 1379  & 1849  \\
\hline
\multirow{4}*{ 1 } & 0  & 333  & 1412  & 4360  & \multirow{4}*{ 1 } & 0  & 539  & 1059  & 1458  \\
& 0.5  & 441  & 1848  & 5757  & & 0.5  & 699  & 1366  & 1866  \\
& 1  & 438  & 1835  & 5833  & & 1  & 702  & 1367  & 1876  \\
& 1.5  & 438  & 1825  & 5696  & & 1.5  & 705  & 1365  & 1867  \\
\hline
\multirow{4}*{ 1.5 } & 0  & 334  & 1415  & 4386  & \multirow{4}*{ 1.5 } & 0  & 544  & 1058  & 1823  \\
& 0.5  & 421  & 1830  & 4970  & & 0.5  & 703  & 1360  & 1913  \\
& 1  & 431  & 1811  & 5133  & & 1  & 616  & 1336  & 1879  \\
& 1.5  & 289  & 1743  & 5069  & & 1.5  & 463  & 1303  & 1818  \\
\bottomrule
\end{tabular}
}
\end{table}

\begin{figure}[!t]
\centering
\includegraphics[width = \textwidth]{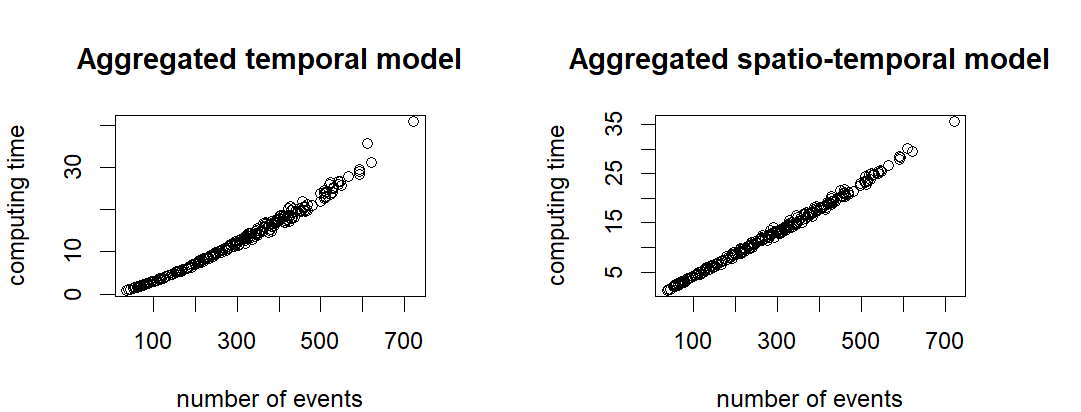}
\caption{Computing time (in seconds) of 1,000 posterior samples. The temporal data are generated using $(\mu,\alpha,\beta) = (0.3,0.7,1)$, with $t = 100,200,300,400,500.$, The spatio-temporal data are generated using $(\mu,\alpha,\beta,\gamma) = (0.3,0.7,1,1)$, with $t = 100,200,300,400,500.$  50 data sets are generated, for each value of $t$.} 
\end{figure}


\subsection{Analyzing the truncated categorical distribution}\label{a: subsec: truncation}
\begin{figure}[!t]
\centering
\includegraphics[width = 0.9\textwidth]{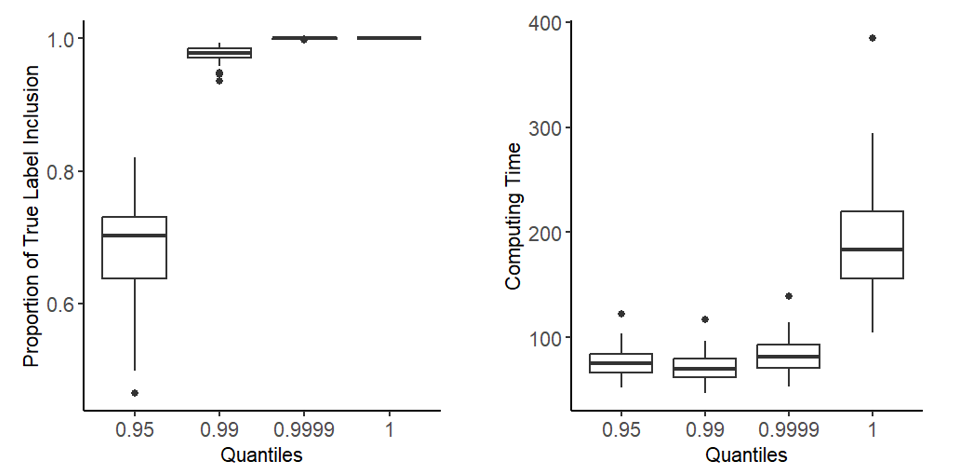}
\caption{Mean proportion of true label inclusion (left) and computing time in seconds of 5,000 posterior samples (right). The temporal data are generated using $(\mu,\alpha,\beta) = (0.3,0.7,1)$, $t = 500$ and exponential offspring kernel. The posterior distributions of branching labels are truncated based on quantiles 0.95,0.99,0.9999, and 1. Results are averaged over all 50 simulated datasets} 
\label{a: fig: truncated}
\end{figure}

In this section, we provided details about the truncated posterior distribution of the branching and show simulation results for performance of truncated posterior distribution based on different truncation ranges.
We truncate the posterior distribution of branching label based on the quantile of quantile of $g_1(t)$ function in order to reduce the computing time for large dataset. More specifically, when updating the branching label for event $i$,  We truncated out events happened before $t_j-q$, where $q$ is a chosen quantile of $g_1(t)$ with the parameter value in the current iteration. To analyze the performance of the truncated posterior distribution based on different quantiles, we generate 50 data sets using $(\mu,\alpha,\beta) = (0.3,0.7,1)$ and $t = 500$ with the exponential kernel. For each dataset, we run MCMC chains 0f 5,000 iterations and truncate the posterior distribution of each braching label using quantiles 0.95,0.99,0.9999,1, and record the whether the range of the truncated distribution include the true label for each event in each iteration and the computing time for running one chain in each case. When the time stamps of events are aggregated, the order of the events cannot be determined and thus we cannot identify the true branching label of an events if it is not the only event in the corresponding temporal bin. So we set the latent time to the true values and did not update them in these simulations. The results are shown in \cref{a: fig: truncated}. We observed that the computing times for MCMC chains utilizing truncated posteriors distribution are about 50\% smaller than the chains without truncation, and the computing time for truncation based on 0.9999,0.99,0.95 quantiles are comparable. For 0.9999 quantile, the range of the truncated distribution include the true label for almost all the iterations and all the time, while the mean proportion of true label inclusion for quantile 0.95 is relatively low. So we will use the 0.9999 quantile for the application.

For sensitivity analysis, we fit the model on datasets generated under parameter set 1 based on untruncated posterior distribution of the branching structure. The results are summarized in \cref{tab:temporal_sensitivity}. We observed that the estimates and credible intervals are almost identical for truncated and untruncated posterior across all parameters when $\Delta^t\le 3.$ When $\Delta^t$ is large, the credible intervals for untruncated posterior are slightly wider than those for truncated posterior.

\begin{table}[!htb]
\centering
\caption{Results of the temporal simulation utilizing untruncated and truncated(based on the 0.9999 quantile) posterior distribution of the branching structure. The table shows the average posterior mean (``Estimate''), average 95\% credible interval length (``CI length''), and coverage rate of the 95\% credible interval (``Coverage'') for parameters $\mu, \alpha$ and $\beta$ of the Hawkes process, across 400 simulated data sets for parameter set $(\mu,\alpha,\beta) = (0.3,0.7,1)$ and  temporal aggregations of different sizes. 95\% credible interval is determined by the 2.5 and 97.5 quantiles of the posterior samples.
}

\label{tab:temporal_sensitivity}
\scalebox{0.9}{
\begin{tabular}{ c c c c c c c c c } 
\toprule 
& \multicolumn{4}{c}{With truncation} & \multicolumn{4}{c}{Without truncation}  \\
\cmidrule(lr){2-5} \cmidrule(lr){6-9}
&$\Delta^t$  &  Estimate & CI length & Coverage &$\Delta^t$  &  Estimate & CI length & Coverage\\   
\midrule
\multirow{6}*{$\mu$} 
& 0  & 0.3115  & 0.1666  & 0.945\phantom{0} & 0  & 0.3123  & 0.1667  & 0.945\phantom{0}  \\
& 1  & 0.3122  & 0.1679  & 0.945\phantom{0} & 1  & 0.3131  & 0.1681  & 0.94\phantom{00}  \\
& 2  & 0.3139  & 0.171\phantom{0}  & 0.95\phantom{00} & 2  & 0.315\phantom{0}  & 0.1711  & 0.9475  \\
& 3  & 0.3169  & 0.1756  & 0.9496 & 3  & 0.318\phantom{0}  & 0.1763  & 0.95\phantom{00}  \\
& 4  & 0.3225  & 0.1822  & 0.915\phantom{0} & 4  & 0.3255  & 0.1846  & 0.9125  \\
& 5  & 0.3283  & 0.1866  & 0.8975 & 5  & 0.3343  & 0.1897  & 0.8725  \\
\hline
\multirow{6}*{$\alpha$} 
& 0  & 0.6854  & 0.2024  & 0.9375  & 0  & 0.6846  & 0.2025  & 0.94\phantom{00}  \\
& 1  & 0.6847  & 0.2036  & 0.945\phantom{0} & 1  & 0.6838  & 0.2036  & 0.94\phantom{00}  \\
& 2  & 0.6829  & 0.206  & 0.9375 & 2  & 0.6818  & 0.206\phantom{0}  & 0.935\phantom{0}  \\
& 3  & 0.6794  & 0.2101  & 0.927\phantom{0} & 3  & 0.6785  & 0.2106  & 0.925\phantom{0}  \\
& 4  & 0.6734  & 0.2152  & 0.8975 & 4  & 0.6704  & 0.2169  & 0.8875  \\
& 5  & 0.6672  & 0.2179  & 0.8675 & 5  & 0.6611  & 0.2198  & 0.845\phantom{0}  \\
\hline
\multirow{6}*{$\beta$} 
& 0  & 1.0567  & 0.5995  & 0.95\phantom{00} & 0  & 1.0625  & 0.6022  & 0.95\phantom{00}  \\
& 1  & 1.07\phantom{00}  & 0.6587  & 0.955\phantom{0} & 1  & 1.0774  & 0.6636  & 0.9525  \\
& 2  & 1.0991  & 0.7968  & 0.965 & 2  & 1.1097  & 0.8072  & 0.965\phantom{0}  \\
& 3  & 1.1935  & 1.284\phantom{0}  & 0.9345 & 3  & 1.2541  & 1.6604  & 0.92\phantom{00}  \\
& 4  & 1.6732  & 3.2711  & 0.9025 & 4  & 1.7943  & 4.5248  & 0.8775  \\
& 5  & 2.4718  & 6.3484  & 0.85\phantom{00} & 5  & 3.4574  & 10.5078  & 0.815\phantom{0}  \\
\bottomrule
\end{tabular}
}
\end{table}

\section{Trace plots in the analysis on insurgent attacks}

\begin{figure}[!b]
\centering

\includegraphics[width = 0.9\textwidth]{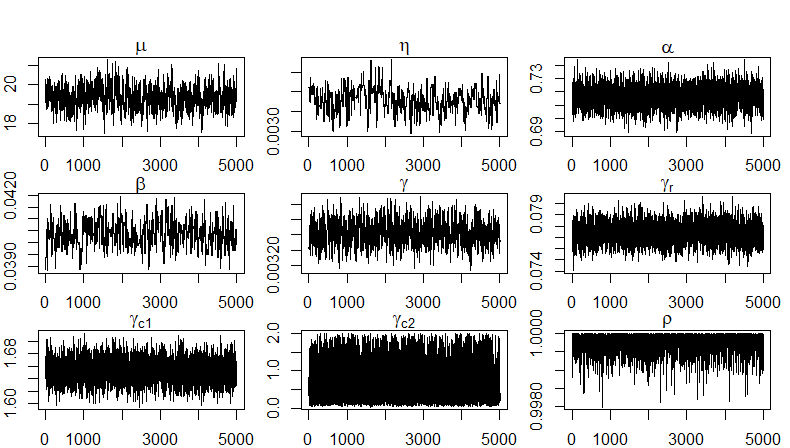}
\caption{Trace plots for parameters in the separate spatio-temporal on IED attacks}
\label{fig:IED_traceplot}

\includegraphics[width = 0.9\textwidth]{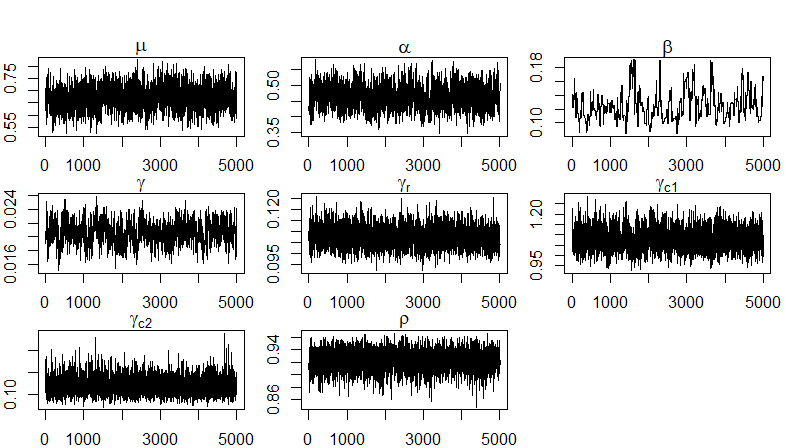}
\caption{Trace plots for parameters in the separate spatio-temporal on airstrikes}
\label{fig:AIR_traceplot}
\end{figure}

\begin{figure}[t]
\centering
\includegraphics[width = \textwidth]{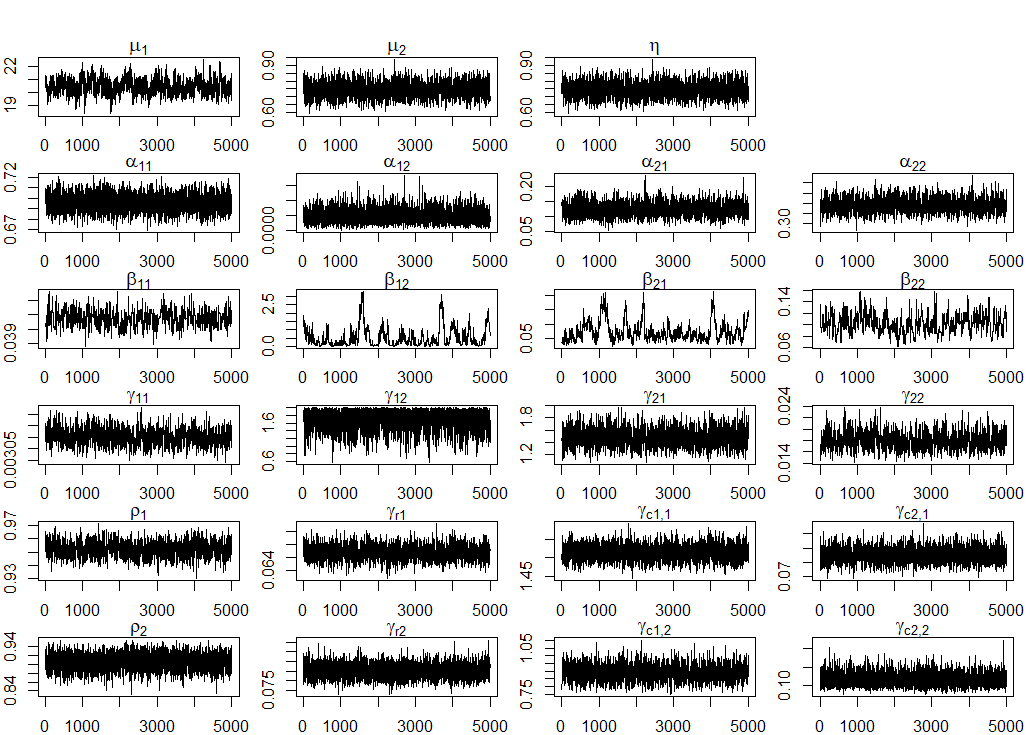}
\caption{Trace plots for parameters in the mutually exciting model based on IED attacks and airstrikes}
\label{fig:IEDAIR_traceplot}

\end{figure}

In this section, we show the trace plots in Section 5 in the main manuscript. We run 20,000, for both the joint and separate models and half of the samples are discared as burn-in samples, and there is no thinning. The trace plots of parameters in the separate spatio-temporal model on IED attack and airstrikes are shown in \cref{fig:IED_traceplot} and \cref{fig:AIR_traceplot} respectively.  The trace plots for parameters in the mutually exciting model on IED and airstrikes is presented in \cref{fig:IEDAIR_traceplot}. We can see that these trace plots indicate good mixing, and there is no observable trend in all the plots.

\section{Interpretation of $\gamma_{c_1}$, $\gamma_{c_2}$ and $\gamma_r$ in the empirical analysis}\label{a:sec:scale_parameters}
\begin{figure}[!ht]
\centering
\includegraphics[width = 0.9\textwidth]{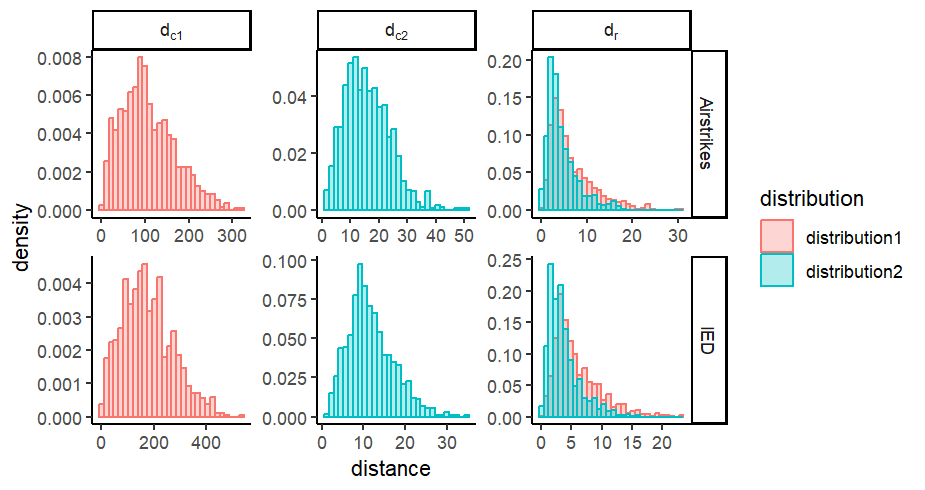}
\caption{Histogram of the distance (in kilometers) to the city Baghdad ($d_{c_1}$), closest distance to the city Mosul or Al Basrah ($d_{c_2}$), and closest distance to the road network ($d_r$) when a location is generated from the first and the second distribution in the estimated mixture distribution for the background intensity of airstrikes and IED attacks under the joint model. The $y$ axis is the probability density.}
\label{fig:distance_hist}
\end{figure}

In the empirical analysis on airstrikes and IED attacks, the immigrant location distribution is assume to be a mixture distribution where
$$f(\sss) = \frac{\rho}{C_1}\exp\Big\{-\frac{d_{c_1}^2(\sss)}{2\gamma^2_{c_1}}-\frac{d_r^2(\sss)}{2\gamma^2_{r}}\Big\}+\frac{1 - \rho}{C_2}\exp\Big\{-\frac{d_{c_2}^2(\sss)}{2\gamma^2_{c_2}}-\frac{d_r^2(\sss)}{2\gamma^2_{r}}\Big\}.$$
So the locations of immigrant events are generate from two distributions, where the first distribution is centered at Baghdad and the distribution is centered at cities Mosul and Al Basrah. The parameter $\rho$ is the probability for a location generated from the first distribution. It is difficult to interpret the values of $\gamma_{c_1}, \gamma_{c_2}$ and $\gamma_{r}$ since they influence the immigrant locations jointly. To have a better understanding of the magnitude of these values, we generated points from the estimated two distributions in the mixture distribution $f(\sss)$ for airstrikes and IED attacks under the joint model using the estimates in Table 3 in the main manuscript and record the $d_{c_1}$, $d_{c_2}$ and $d_{r}$ for each point. The results are shown in \cref{fig:distance_hist}. 

For airstrikes generated from the first distribution of the estimated mixture distributions, the distance to Baghdad is generally smaller than that for IED attacks. Specifically, the 0.9 quantile is 194 kilometers for airstrikes and 304 kilometers for IED attacks. Conversely, airstrikes generated from the second distribution have larger distances to the cities Mosul and Al Basrah, with a 0.9 quantile distance of 26.1 kilometers, compared to IED attacks, which have a 0.9 quantile distance of 19.3 kilometers. Furthermore, we observe that the distributions of distances to the road network for airstrikes generated from the first and second distributions are 13.4 and 9.12 kilometers, respectively. The distance to the road network for IED attacks has a 0.9 quantile of 10.9 kilometers for those generated from the first distribution and 7.39 kilometers for those from the second distribution.

\section{Analyze latent labels}\label{a:branching}

\subsection{Determine the label for each event}\label{a:subsec:event_label}
\begin{figure}
\centering

\subfloat[Airstrikes and triggered IED attacks]{
\includegraphics[width = 0.35\textwidth,trim = 300 65 280 60, clip]{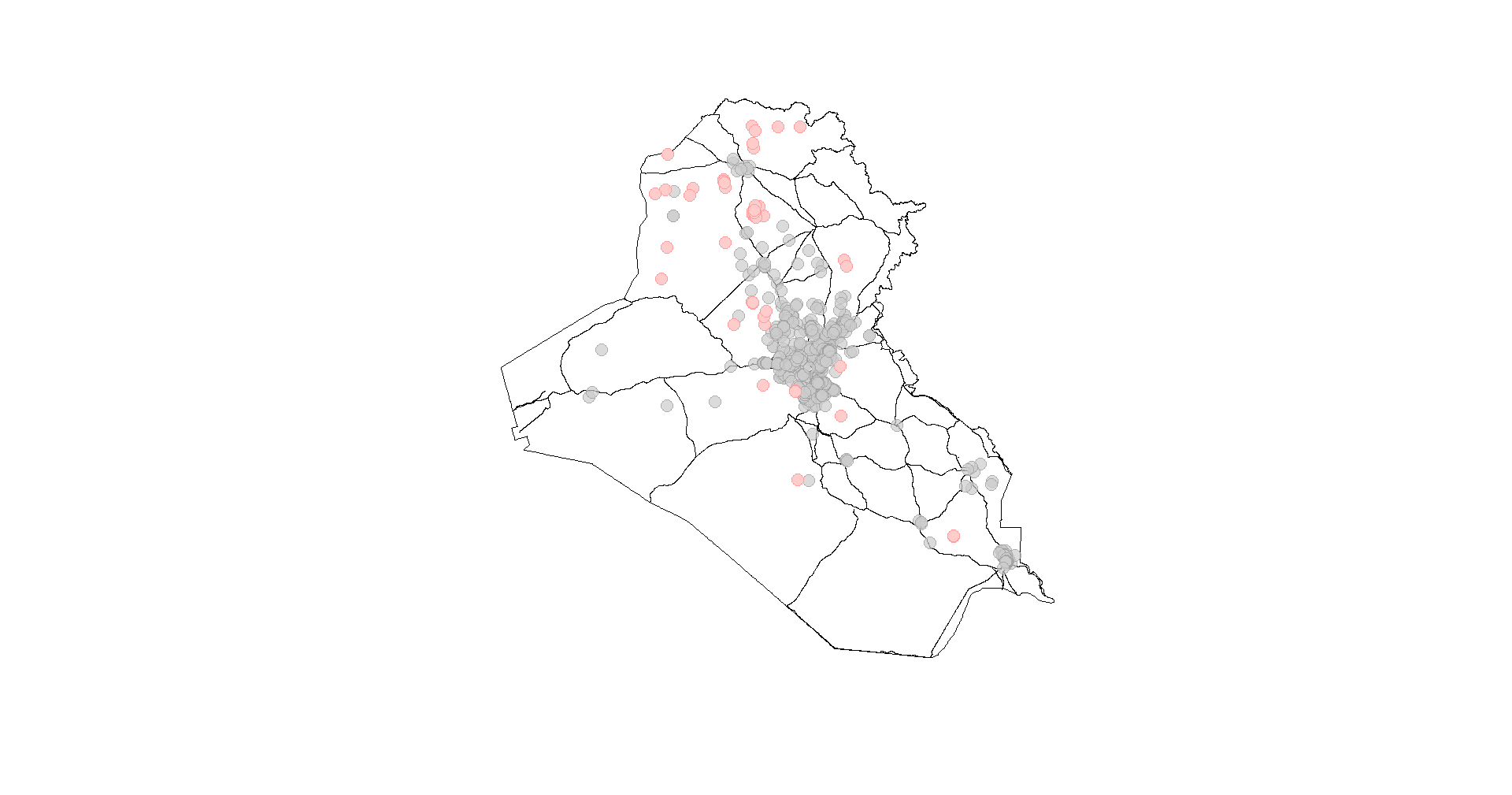}
\label{fig:IED_triggered}
}\hspace{2mm}  
\subfloat[Distance between parents and offspring for IED attacks]{
\includegraphics[width = 0.55\textwidth,trim = 0 0 0 0, clip]{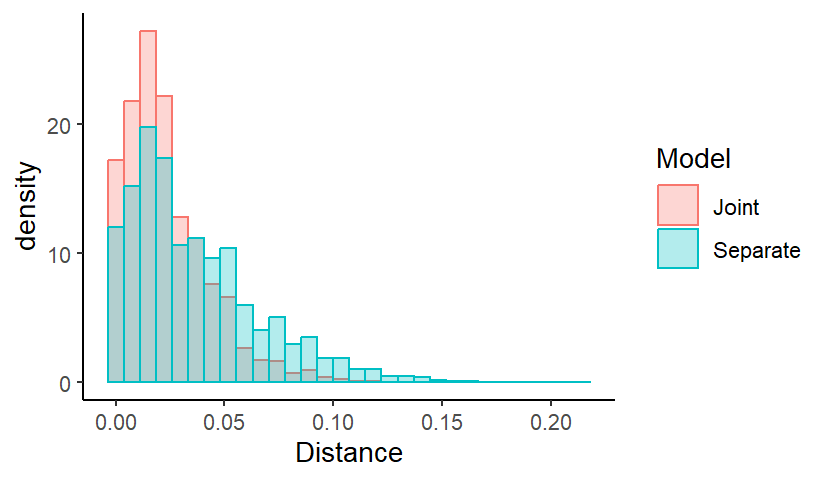}
\label{fig:Air_parent_offspring_dsitance}
}
\caption{Panel (a) shows locations of airstrikes and triggered IED attacks in the joint model. Panel (b) show the histogram of the distance of a parent and its offspring in airstrikes over all iterations, where the y-axis is the probability density.}
\end{figure}

Bases on posterior samples of the branching structure, we can analyze the branching label for each event and examine how labels changes after including the cross excitation effect in the joint model. We determine the branching label of a particular events by its most prevalent labels in all the posterior samples. For separate models on IED and airstrikes, there are 4 possible type of events: immigrants in the IED process ($I_1$), immigrants the airstrike process ($I_2$), offspring in the IED process ($O_{11}$) and offspring in in the airstrike process ($O_{22}$). After including the cross excitation, there are two more types of events: IED attacks triggered by airstrikes ($O_{21}$) and airstrikes triggered by IED attacks. \cref{tab: label} shows the number of events of each type in the separate and joint models. We observe that 50 immigrant events in the IED process in the separate model are classified as offspring of airstrikes in the joint model. Spatial locations of these IED attacks are shown in \cref{fig:IED_triggered}. Moreover, some offspring in the separate model become immigrants in the joint model. Furthermore, the posterior of branching labels enable us to analyze the distance of parent events and their offspring in both separate and joint model. \cref{fig:Air_parent_offspring_dsitance} shows histogram of the distances of all parent-offspring pairs over all iterations in airstrikes based on the separate and joint models. We observe that the distances between parent-offspring pairs tend to be smaller in the joint model than these in the separate model, aligning with the result that estimates for $\gamma$ for airstrikes becomes smaller in the joint model than in the separate model.

\subsection{Analyze parent-offspring pairs}\label{a:subsec: parent_offspring}
Using the posterior samples of the branching labels, we can count the number of parent-offspring pairs to assess the estimability of the parameters. Given that the exact timing for most IED attacks is available, we focus on parent-offspring pairs involving airstrikes. Specifically, we examine the following types of offspring: $O_{22}$, $O_{12}$, and $O_{21}$. In each iteration of the MCMC chain, we count the number of offspring with a parent on the same day and on different days for each offspring type. We then calculate the median of these values. The results are shown in \cref{tab:parent_offspring}. We observe that for both the separate and joint models, the number of events in $O_{22}$ with a parent in different days are relative large, showing no issues with estimability for parameters $\beta_{22}$. Similar result is observed for $O_{21}$ in the joint model. However, the number of events in $O_{21}$ with a parent in different days is small, indicating that the estimate for $\beta_{12}$ may not be reliable.

\begin{table}[ht]
\centering
\caption{Number of events with different labels in the separate and joint model on IED (process2) and airstrikes (process2). $I_1$ and $I_2$ present the immigrant events in the IED and airstrikes processes, respectively. $O_{m,l}$ represent events in process $l$ and is triggered by an event in process $m$, for $m,l\in\{1,2\}$. \vspace{5pt}}
\label{tab: label}
\begin{tabular}{|c|cccccc|c|}
  \hline
 \diagbox{Separate}{Joint}& $I_1$ & $I_2$ & $O_{11}$ & $O_{22}$ & $O_{12}$ & $O_{21}$ & Total  \\ 
  \hline
$I_1$ & 3,723 &   0 & 251 &   0 &   0 &  50 & 4,024 \\ 
$I_2$ &   0 & 140 &   0 &  84 &   1 &   0 & 225 \\ 
$O_{11}$  & 429 &   0 & 10,893 &   0 &   0 &   3 & 11,325 \\ 
$O_{22}$ &   0 & 196 &   0 & 173 &   3 &   0 & 372 \\ 
   \hline
   Total & 4,152 & 336 & 11,144 & 257 & 4 & 53 & \\
   \hline 
\end{tabular}
\end{table}

\begin{table}[!htb]
\centering
\caption{Posterior median of the number of offspring grouped by whether they are with their parent on the same day or on different days, comparing results from the separate and joint models. $O_{m,l}$ represent events in process l and is
triggered by an event in process m, for $m, l \in \{1, 2\}$where process1 is IED attacks and process2 is airstrikes.} 
\label{tab:parent_offspring}
\begin{tabular}{c c c c c} 
\toprule 
& \multicolumn{1}{c}{Separate} & \multicolumn{3}{c}{Joint}  \\
\cmidrule(lr){2-2} \cmidrule(lr){3-5}
& $O_{22}$ & $O_{12}$ & $O_{21}$ & $O_{22}$ \\  
\midrule
Same day & 39 & 2 & 3 & 26  \\
Different days & 260 & 5 & 71 & 209  \\
\bottomrule
\end{tabular}
\end{table}

\section{Forecasting the number of airstrikes and IED attacks}\label{appendix: prediction}

In this section, we show results that complement the text in Section 5.3. We assess the model performance by analyzing the ability of forecasting the number of events in the holdout period. 

\begin{figure}[p]

\centering

\includegraphics[width = 5cm]{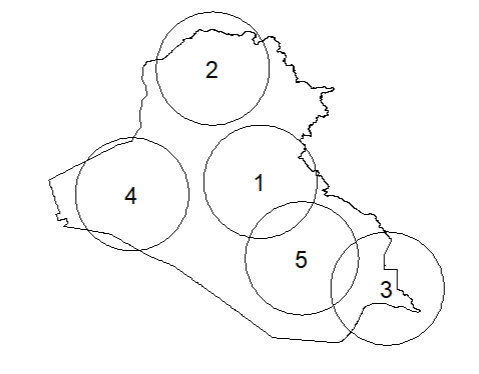}
\caption{The selected regions for forecasting}
\label{fig:region}

\subfloat[Airstrikes]{
\includegraphics[width = \textwidth,trim = 0 0 0 0, clip]{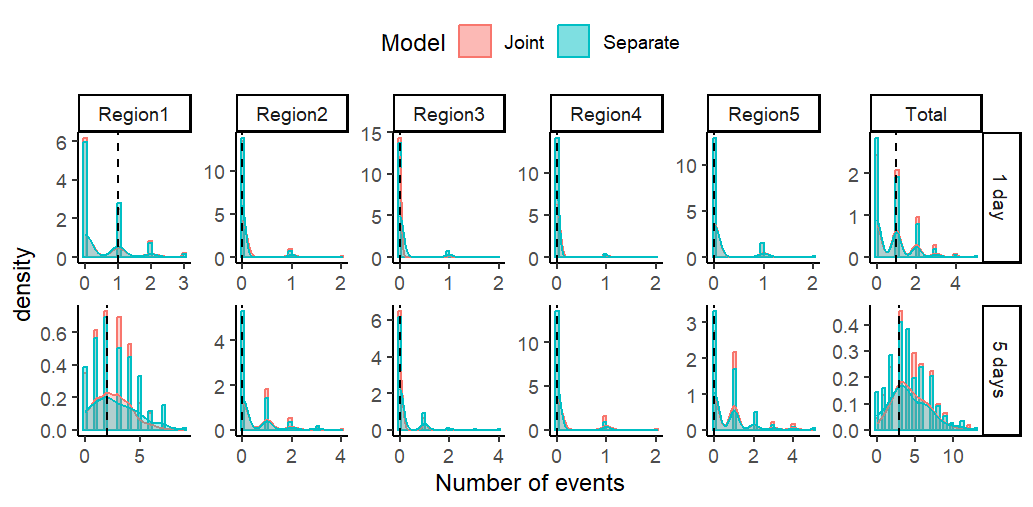}
\label{fig:AIR_forecast}
}  \\
\subfloat[IED]{
\includegraphics[width = \textwidth,trim = 0 0 0 30, clip]{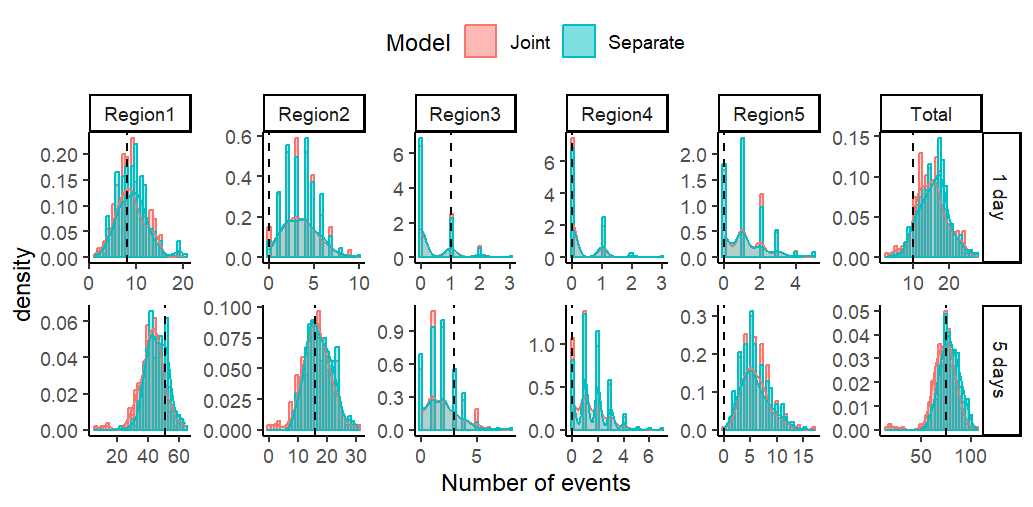}
\label{fig:IED_forecast}
}

\caption{Posterior predictive distribution of the number of airstrikes in (a) and IED
 attacks in (b) during one day and five days after June 30, 2008 for different regions using separate and joint models on airstrikes and IED attacks. The y-axis shows estimated probability densities. The true number of events is indicated by the black dashed line.
}
\label{fig:IEDAIR_forecast}
\end{figure}

We evaluate the predictive posterior distribution of number of airstrikes and IED attacks in five selected regions and in the full holdout time period. Those regions (\cref{fig:region}) are chosen based on the locations of major cities of Iraq.  We use these regions only for evaluating the performance of the model and the spatial dimension is not aggregated when we fit the model. In \cref{fig:IEDAIR_forecast}, we show the predictive posterior distributions of airstrikes and IED attacks in the selected regions using separate and joint spatio-temporal models. We observe that the predictive posterior distributions using separate model are very similar to those using the joint model, and most posterior modes are close to the true value. The number of events in region 5 seems to be underestimated for airstrikes and IED attacks and the number of events in region 2 potentially overestimated for airstrikes, but the estimates for the total number of events is relatively accurate.

\end{document}